\documentclass[11pt]{article}
\usepackage{amsfonts,amsmath,amssymb}
\usepackage{graphicx}
\newfont{\twelvemsb}{msbm10 scaled\magstep1}
\newfont{\eightmsb}{msbm8}
\newfam\msbfam
\textfont\msbfam=\twelvemsb \scriptfont\msbfam=\eightmsb
\catcode`\@=11
\def\Bbb{\ifmmode\let\next\Bbb@\else
\def\next{\errmessage{Use \string\Bbb\space only in math mode}}\fi\next}
\def\Bbb@#1{{\fam\msbfam{{#1}}}}
\newcommand{\be}{\begin{equation}}
\newcommand{\ee}{\end{equation}}
\newcommand{\ba}{\begin{eqnarray}}
\newcommand{\ea}{\end{eqnarray}}

\begin{document}
\sloppy
\renewcommand{\thefootnote}{\fnsymbol{footnote}}
\newpage
\setcounter{page}{1} \vspace{0.7cm}

\vspace*{1cm}
\begin{center}
{\bf \large  From classical Lax ODEs to quantum integrable theories: the moduli.}\\
\vspace{1.8cm} {\large Davide Fioravanti $^a$ and Marco Rossi $^b$
\footnote{E-mail: fioravanti@bo.infn.it, rossi@cs.infn.it}}\\
\vspace{.5cm} $^a${\em Sezione INFN di Bologna, Dipartimento di Fisica e Astronomia,
Universit\`a di Bologna\\
Via Irnerio 46, 40126 Bologna, Italy} \\
\vspace{.3cm} $^b${\em Dipartimento di Fisica dell'Universit\`a della Calabria and
INFN, Gruppo collegato di Cosenza\\
Arcavacata di Rende, 87036 Cosenza, Italy} \\
\end{center}
\renewcommand{\thefootnote}{\arabic{footnote}}
\setcounter{footnote}{0}
\begin{abstract}
The general idea of this paper is to start from a classical integrable (partial differential) equation which arises as a compatibility condition for a matrix linear differential problem. For definitiveness' sake, a generalised sinh-Gordon equation depending on $2N-1$ complex coefficients or moduli is considered. Then, the connexion coefficients (Wronskians) of different solutions to this problem satisfy, in the spirit of the Ordinary Differential Equation/Integrable Model correspondence, functional relations, which can be considered, -- if supplemented by suitable asymptotic behaviours --, as identifying a specific state of a quantum integrable model: in fact they are the eigenvalues of extensions of Baxter operators $Q$ and $T$, the transfer matrix. Moreover, Y-system and (implementing the asymptotic conditions) thermodynamic Bethe Ansatz equations originate from them, without any passage through the scattering theory, and providing an invariant parametrisation of the monodromy space. The crucial novelty is the modification of all the relations because of their dependence on the moduli. For zero momentum, they fully describe physically the quantum homogeneous sine-Gordon model, {\it i.e.} scattering amplitudes of gauge fields in $\mathcal{N} = 4$ SYM at strong coupling or their dual null polygonal light-like Wilson loops in $AdS_3$. As a direct consequence of the correspondence, two Zamolodchikov's conjectures, based on previous results, are also proven.

\end{abstract}
\vspace{1cm} %{\noindent PACS}: 11.30-j; 02.40.-k; 03.50.-z
\newpage

\section{Introduction}

Shortly declared, the ODE/IM (Ordinary differential equation/Integrable Model) correspondence may be considered as a way to move from the analysis of the connexion coefficients of some specific ODE to identify them with eigenvalues of  some (non-local) operators of a specific quantum IM. In fact, these eigenvalues/operators are generating functions of the local and non-local charges when Taylor-Laurent expanded around some specific value of the spectral parameter ({\it i.e.} rapidity) of the quantum IM: for this reason and others to come, these operators are crucial in quantum integrability theory. More precisely, for some fixed ODE its connexion and monodromy coefficients coincide with the eigenvalues of these operators in a specific IM, changing 'a little' the former with the state (see below).

Initially derived for (differential) stationary Schr\"{o}dinger equations with monomial potentials and angular momentum term as reproducing the ground state of conformal minimal models \cite {ODE-IM1, ODE-IM3, ODE-IM2}, it was extended to excited states \cite {BLZ-excited-ode-im, DF-excited-ode-im} and then to a massive case, namely it was established from the classical Sinh-Gordon Lax problem to the quantum sine-Gordon field theory on a cylinder \cite {LZ}. On one hand, the addition (to the potential) of terms with apparent (double) poles provided a description of excited states of CFT; on the other, the introduction of a classical integrable evolution in the form a Lax pair of matrix linear differential operators (for the equation of motion in the form of a zero-curvature condition) lead to a massive theory, more precisely the ground state of quantum sine-Gordon theory: a specific solution of the classical theory, respecting the CFT case discrete symmetries, plays the r\^ole of (complicated) Schr\"{o}dinger potential (excitations can be dealt with the same general principle as in the conformal scenario).  

Very importantly, the ODE enjoys many more methods of analysis and solution than the (non linear) functional or integral equations satisfied by the eigenvalues, albeit the hidden price to pay consists in the mysterious fact that the differential variable seems to play no r\^ole for the IM ({\it cf.} \cite{FRletter} for some tentative explanations). In fact, an ODE permits, for instance, different kinds of expansions as some parameter can be small and large, so that also some strong/weak dualities can be studied. Relevantly, in \cite{F-G-pure-Liouville} the asymptotic semiclassical expansion $\frac{\hslash}{\Lambda}=e^{-\theta} \rightarrow 0$ was computed at all orders; it is very usual in integrability since it can be computed by other means as the physical rapidity $\theta\rightarrow -\infty$ \cite{FRaugust}. Moreover, the leading order corresponds to the Seiberg-Witten contribution while the others correspond to the Nekrasov-Shatashvili quantisation corrections, with the proper differences between gauge periods and integrable charges (for all this matters, see the detailed explanations in \cite{F-G-pure-Liouville}). Yet, the ODE analysis gives probably the most effective way, also in the off-critical case \cite{LZ}. On the other side, the {\it instanton} expansion $\frac{\Lambda}{\hslash}=e^{\theta} \rightarrow 0$ was computed at all orders in \cite{FRaugust} by an efficient algorithm and is conjectured to be convergent, in that a gauge quantity, the prepotential, derivable from it \cite{FRaugust}, enjoy indeed a combinatoric instanton expansion \cite{Nek-Ok} which is allegedly convergent \cite{bon-tanz}. On the contrary, it is rather new and very profitable in the integrability literature and realm as an example of stimulated cross-fertilisation.

In more detail, the non-local operators -- whose eigenvalues are these connection coefficients -- are the proper extensions of the celebrated Baxter's $Q$- and $T$-matrices 
\cite{BLZ12, FStan,FR03}. 
All their eigenvalues satisfy primarily three kind of functional equations: the $QQ$-system, the $TQ$-system and the $TT$-system, enumerated in the logic of their following derivation. Eventually, the $T$s construct the $Y$-functions which close a $Y$-system and, importantly, can be converted into Thermodynamic Bethe Ansatz (TBA) integral equations. In all these equations the role of spectral parameter is played by the energy of the original Schr\"{o}dinger problem.

The extension of these results to the case of generic polynomial potentials is non-trivial both in the {\it static} case and in the Lax {\it dynamical} one. The $Q$, $T$ and $Y$-functions depend not only on the energy, but also on the coefficients $c_n$  of the potential, playing the r\^ole of 'moduli' of the theory. Then, the functional relations satisfied by them can (and actually do) involve functions computed on 
different sets of 'moduli'. To be concrete, these different sets of moduli contain coefficients multiplied by a phase, for instance $c_n \rightarrow c_n e^{\frac {i\pi n}{N}}$, where $2N$ is the degree of the potential. This is indeed a technical complication we want to analyse in this work. Other than generalising the functional relations that appear in integrable models, the content of this article firmly establishes the functional relations that we used in the paper \cite {FRS}, which is devoted to computing $AdS_3$ scattering amplitudes by using a set of non-linear integral equations. Moreover, it gives a concrete realisation of the starting hypothesis used in the letter \cite {FRletter} and in the forthcoming publication \cite {FRinverse}, which both deal with the 'inverse' problem. {\it i.e.} with the derivation of a differential operator from a set of functional relations. Finally, it provides several examples of TBA equations, which are the starting points for explicit evaluations of $Y$,$T$ and $Q$-functions.

\medskip

The summary of this paper is as follows. We start in Section \ref {funrel} by considering a general differential Lax pair arising in the context of scattering amplitudes in $\mathcal{N} = 4$ SYM at strong coupling, dual to Wilson loops in $AdS_3$. A set of complex numbers, the 'moduli', are present to characterise scattering states (or dual Wilson loops). The Lax pair forms a linear problem of two coupled Partial differential Equations (PDEs) and their compatibility zero curvature condition defines a classical theory, a modified version of the sinh-Gordon equation. Concentrating on connection coefficients ($Q$ functions) and the monodromy ($T$-functions) appearing in formul{\ae} relating different solutions of the two PDEs, we derive the functional equations they satisfy: $QQ$-, $TQ$- , $T$- and $Y$-systems of Homogeneous sine-Gordon theory.
Particular attention is paid to the two symmetries of the PDEs (named $\hat \Pi$ and $\hat \Omega$) which are crucial for deriving these functional relations. Importantly, the functional relations we derive are not closed, as they involve functions that depend on different sets of moduli. We show how to derive closed functional relations by redefining the $T$ and $Y$ functions in a specific way. The $T$- and $Y$- systems obtained in this way can be called 'universal', as they are independent of the coupling constants of the quantum theory. 
In Section \ref {QT}, we provide a more detailed analysis of the $Q$-functions. We prove that they can be obtained by specialising a particular solution (a Jost wave function) to the origin. Using a change of variables, $z\rightarrow w$, that transforms the modified into the ordinary sinh-Gordon equation, the limits at large rapidity of the $Q$ functions is computed.  
%We discuss a physical interpretation of their zeros.
With this result in Section \ref {l0N32case} we translate the $Y$-system into TBA equations, by paying particular attention to the particular simpler case, when only two moduli are present. TBA equations coincide with those of Homogeneous sine-Gordon theory, thus giving meaning to the ODE/IM correspondence, where now the ODE part has been necessarily substituted by a Lax linear differential problem.
Section \ref {conf-lim} treats the 'conformal' limit in which the Lax pair reduces to a single ODE. In this limit the natural parameters (moduli) are not the naive limit of the (off-critical) moduli, which leads to a modification, with respect to the off-critical case, of the $T$- and $Y$- systems, and then of TBA equations, when $T$ and $Y$ are expressed as functions of the conformal case parameters. 
In following Section \ref {lzero} we demonstrate how the $l\rightarrow 0$ limit of the $Y$-system yields realisations of solutions of a system of TBA-like equations in terms of a peculiar integral kernel as brilliantly proven in \cite {FS,CFIV} with supersymmetric field theories. Actually, this proves the two famous Zamolodchikov's conjectures \cite{Zam-poly} as deriving purely and naturally as a consequence of massive ODE/IM correspondence (in the inverse direction \cite{FRletter}). Two appendices, Appendix \ref {app1} and Appendix \ref {app3}, contain technical results concerning the $T$-functions and the $Y$-functions, respectively. Appendix \ref {lp} concerns two particular cases in which the linear problem has radial symmetry. In these circumstances important connections with the Painlev\'e III$_3$ equation are highlighted. Appendix \ref {app4} shows how TBA equations discussed in this paper can be equivalently written by using two different kernels.

\section{Derivation of all functional relations for {\it generalised} amplitudes}
\setcounter{equation}{0}
\label{funrel}

All the relevant functional equations satisfied by the various connexion coefficients will be derived here for the linear differential problem introduced in \cite {GMN} and so extending this work and \cite{FRS}. In particular, this procedure opens the way to computing the amplitudes in $\mathcal{N} = 4$ SYM at strong coupling, which are dual to null polygonal light-like Wilson loops in $AdS_3$; on the IM side this corresponds to a massive integrable quantum field theories (in $1+1$ dimensional space-time), the homogeneous Sine-Gordon quantum field theory.

\medskip

The starting point is a classical integrable system, in specific a real function $\eta(z,{\bar z}; \vec{c})$ of the complex variables $z$ and of its complex conjugate $\bar z$, which satisfies the classical modified sinh-Gordon equation\footnote{We discuss below the suitable asymptotic conditions on $\eta(z,{\bar z}; \vec{c})$.}
\be
\partial _z \partial _{\bar z} \eta -e^{2\eta}+p(z,\vec{c})\bar p(z, \vec{c})e^{-2\eta}=0 \, .
\label {cl-sinh}
\ee
Modification comes from a given polynomial $p(z,\vec{c})= z^{2N} + \sum \limits _{n=0}^{2N-2}c_n z^n$ ($c_{2N}=1$, $c_{2N-1}=0$ without loss of generality by using scaling and translation of $z$ (and $c_0<0$)) depending on a $2N-1$-ple of complex coefficients $\vec{c}=(c_0,...,c_{2N-2})$ (the bar means complex conjugation), with asymptotic condition (realising later the quantum ground state)
\be
\eta=l \ln z\bar z +O(1) \, , \label{etaasy}
\ee
as $|z|\rightarrow 0$, for $|l|<\frac {1}{2}$. This solution generalises the seminal paper \cite{LZ} with the presence of the moduli $\vec{c}$ (besides $c_0$): the case of \cite{LZ}, which in the IM side describes the sine-Gordon model, can be recovered by setting to zero the moduli $c_1,c_2,..., c_{2N-2}$ and by letting $N$ to be a positive real number. It generalises also the $AdS_3$ string problem representing the Wilson loops \cite{AM-Sinh-G-1,YSA,HISS} which is regular in $z=0$, {\it i.e.} $l=0$: to have a parallel with quantum integrable models we may think of the series of homogeneous sine-Gordon ones \cite {HSG1,HSG2}. In this sense, it may be called {\it generalised} amplitudes. We may recover (\ref{cl-sinh}) as zero curvature or compatibility condition of the associated linear 
\be
 \mathcal{D}_z\Psi =0,\,\,\,  { \mathcal{D}}_{\bar z} \Psi =0 \,,
\label{ass-lin-prob}
\ee
with these forms of the differential operators in the $su(2)$ fundamental representation\footnote{To be more precise we shall be writing about a specific gradation of a Kac-Moody algebra; but we want to reckon these generalisations, though important, not too far from our present specific treatment (see for instance \cite{A22-Marian}).}
\ba
 \mathcal{D}_z &=&\partial _z + \frac{1}{2}\partial _z \eta \sigma ^3 -e^{\theta}\left [ \sigma ^+ e^{\eta}+\sigma ^- p(z,\vec{c}) e^{-\eta}  \right ] \, ,
\label {D}\\
 { \mathcal{D}}_{\bar z} &=&\partial _{\bar z} - \frac{1}{2}\partial _{\bar z} \eta \sigma ^3 -e^{-\theta}\left [ \sigma ^- e^{\eta}+\sigma ^+\bar p(z,\vec{c}) e^{-\eta} \right ] \label {barD} \, , \\
\sigma ^3&=&\begin{pmatrix} 1 & 0 \\ 0 & -1 \end{pmatrix} \, , \quad \sigma ^{+}=\frac {\sigma _1+i\sigma _2}{2}=\begin{pmatrix} 0 & 1 \\ 0 & 0 \end{pmatrix} \, , \quad  \quad \sigma ^{-}=\frac {\sigma _1-i\sigma _2}{2}=\begin{pmatrix} 0 & 0 \\ 1 & 0 \end{pmatrix} \, .
\ea
They are invariant with respect to the operation (symmetry)
\be
\hat \Omega : \quad z \rightarrow z e^{\frac{i\pi}{N}} \, , \quad \theta \rightarrow \theta - \frac{i\pi}{N} \, , \quad \vec{c} \rightarrow \vec{c}^R \, , \quad \vec{c}^R =(c_0,c_1e^{-\frac{i\pi}{N}}, ..., c_n e^{-\frac{i\pi n}{N}}, ..., c_{2N-2}e^{\frac{2i\pi}{N}}) \, ,
\label{Omega-symm}
\ee
although, as we will see, the solution $\Psi$ is not. In fact, this operation leaves invariant the polynomial $p$ ($p(ze^{\frac{i\pi}{N}}, \vec{c}^R)=p(z, \vec{c})$) and then the scalar $\eta$. However, it corresponds to a change of sheet and then $\sqrt{p(ze^{\frac{i\pi}{N}}, \vec{c}^R)}=-\sqrt{p(z, \vec{c})}$. The ancestor of this generalised symmetry involving the moduli appeared in \cite{Suzuki1999} in the context of a Schr\"{o}dinger equation with polynomial potential (2D conformal case); in the same context more recently in \cite{MAR1,MAR2} they appear as a limiting degenerating case (conformal limit) of our massive one. Moreover, we wish here to show how the spontaneous breaking of this symmetry can be intensively used to obtain the fully fledged quantum integrability structure (functional equations, integrable equations, etc.) of the initial classical problem (\ref{ass-lin-prob}, \ref{D}, \ref{barD}, \ref{cl-sinh}) by extending the works \cite{ODE-IM1, ODE-IM2,ODE-IM3,LZ} about integrable models without moduli. Incidentally, it is interesting to note how our treatment could be repeated also for excited states \cite{BLZ-excited-ode-im}, \cite{DF-excited-ode-im}, which would present new kind of moduli, {\it i.e.} the (apparent) singularities of the potential (not of the equation/solution).

Let us, in fact, analyse the solutions $\Psi$. As in \cite {LZ}, we can fix {\it the Jost solution basis}, $\Psi _{\pm} (z;\theta,\vec{c})$, around $z=\rho e^{i\varphi}=0$, $\bar z=\rho e^{-i\varphi}=0$ univocally by the leading expansion
\be
\Psi _+ (z;\theta,\vec{c})\simeq \frac{1}{\sqrt{\cos \pi l}}
\left ( \begin{array}{c} 0 \\ e^{(i\varphi +\theta )l}\end{array} \right) \, , \quad
\Psi _- (z;\theta,\vec{c})\simeq \frac{1}{\sqrt{\cos \pi l}}
\left ( \begin{array}{c} e^{-(i\varphi +\theta )l} \\ 0 \end{array} \right), \,\, |z|\rightarrow 0 \,\, ,
\label {Psizero}
\ee
with the symbol $\simeq$ meaning equality at the leading order.

Furthermore, we introduce another Jost basis. The first element of the basis is the solution $\Xi (z; \theta ,\vec{c})$ uniquely defined by the asymptotic leading expansion 
\be
\Xi (z; \theta ,\vec{c}) \simeq \left ( \begin{array}{c} e^{-\frac{i N \varphi}{2}} \\ -e^{\frac{iN \varphi}{2}}\end{array} \right) \exp \left [ -e^{\theta} \int ^z dz \sqrt{p(z,\vec{c})}-e^{-\theta} \int ^{\bar z} d\bar z
\sqrt{\bar p(z,\vec{c})} \right ] \, , \label {Csi}
\ee
when $|z| \rightarrow +\infty$ and $\left |\textrm {Arg}z +\frac {\textrm {Im}\theta}{N+1} \right |<\frac {\pi}{2(N+1)}$. This is the Stokes sector $S_0$, where the Stokes sector $S_n$, with $n$ integer, is defined by 
$\left |\textrm {Arg}z +\frac {\textrm {Im}\theta+n\pi}{N+1} \right |<\frac {\pi}{2(N+1)}$.
Actually, the asymptotic behaviour (\ref  {Csi}) is valid in the wedge $\left |\textrm {Arg}z +\frac {\textrm {Im}\theta}{N+1} \right |<\frac {3\pi}{2(N+1)}$, {\it i.e.} in the Stokes sectors $S_1,S_0,S_{-1}$. The solution $\Xi$ is subdominant (i.e. going to zero) in $S_0$, whilst it is dominant (i.e. diverging) in $S_1,S_{-1}$.

The other function of the basis is $\Xi _1(z; \theta ,\vec{c})=\hat \Omega \Xi (z; \theta ,\vec{c})$. Solution $\Xi_1$ is subdominant in $S_1$ and dominant in $S_0,S_2$. Its asymptotic leading term the region $-\frac {5\pi}{2(N+1)}<\textrm {Arg}z +\frac {\textrm {Im}\theta}{N+1}<\frac {\pi}{2(N+1)}$ is
\be
\Xi _1(z; \theta ,\vec{c}) \simeq -i e^{i\Phi (\theta , \vec{c})} \sigma ^3 \left ( \begin{array}{c} e^{-\frac{i N \varphi}{2}} \\ -e^{\frac{i N \varphi}{2}}\end{array} \right) \exp \left [ e^{\theta} \int ^z dz \sqrt{p(z,\vec{c})}+ e^{-\theta} \int ^{\bar z} d\bar z
\sqrt{\bar p(z,\vec{c})} \right ] \, .
\ee

In this last formula we had to introduce the function
\be
\label{phase}
\Phi (\theta , \vec{c})=\frac{\pi}{N} B_{-1}(\vec{c}) e^{\theta}-\frac{\pi}{N} \bar B_{-1}(\vec{c}) e^{-\theta} \, ,
\ee
which depends on the coefficient $B_{-1}(\vec{c})=-e^{-\frac{i\pi}{N}}B_{-1}(\vec{c}^{R})$ appearing in the large $|z|$ expansion of $\sqrt{p(z,\vec{c})}$ together with $z^{-1}$:
\be
\sqrt{p(z,\vec{c})}=z^{N}+...+B_{-1}(\vec{c})z^{-1}+ O(z^{-2}) \, .
\ee
Therefore, the function $\Phi (\theta , \vec{c})=0$ if $N$ is semi-integer. It appears in the solution $\Xi_1$ as a consequence of the $\hat \Omega$-symmetry transformation only if $N$ is integer. Importantly, it enjoys the properties
\be
\Phi (\theta , \vec{c})=-\Phi \left ( \theta -\frac{i\pi}{N}, \vec{c}^{R}\right )=-\Phi (\theta +i\pi, \vec{c})  \label {phi-prop} \, ,
\ee
that we will broadly use in the following of our analysis to simplify relations.

Now, we can state in a complete form how the various solutions we introduced transform under the symmetry $\hat \Omega$: 
\be
\hat \Omega \Psi _{\pm} (z;\theta,\vec{c}) \equiv \Psi _{\pm }\left (ze^{\frac{i\pi}{N}} ;\theta-\frac{i\pi}{N},\vec{c}^R \right ) = \Psi _{\pm} (z;\theta,c) \label {Omegapsi}
\ee
\be
\hat \Omega \Xi (z;\theta,\vec{c})\equiv \Xi \left ( ze^{\frac{i\pi}{N}}; \theta - \frac{i\pi}{N}, \vec{c}^R \right )\equiv \Xi _1(z;\theta,\vec{c}) \, ,  \label {Omegachi}
\ee

Besides, it is very natural (at least having in mind the classical inverse scattering) to define the {\it connexion coefficients} of the Jost solutions, $Q_{\pm}(\theta, \vec{c})$, as
\be
\Xi (z;\theta, \vec{c})=Q_+(\theta ,\vec{c}) \Psi _- (z; \theta , \vec{c}) + Q_- (\theta , \vec{c}) \Psi _+ (z; \theta , \vec{c}) \label {maineq} \, .
\ee
The fundamental relation, {\it i.e.} the quantum Wronskian relation (or $QQ$ system) is obtained by the following simple steps: applying $\hat \Omega$ to the left hand side of previous equation, one gets
\be
\Xi _1(z;\theta, \vec{c})=\Xi \left ( ze^{\frac{i\pi}{N}}; \theta - \frac{i\pi}{N}, \vec{c}^R \right )=Q_+\left (\theta -\frac{i\pi}{N},\vec{c}^R \right ) \Psi _- (z; \theta , \vec{c}) + Q_- \left (\theta - \frac{i\pi}{N} , \vec{c}^R \right) \Psi _+ (z; \theta , \vec{c}) \, ,
\ee
after using (\ref {Omegapsi});
then, taking the determinant\footnote {Given two two-dimensional vectors $A^{\alpha},B^{\beta}$, with $\alpha,\beta =1,2$, the expression $\det (A,B)$ means $A^1B^2-A^2B^1$. If two such vectors $A(z),B(z)$ are solutions of the linear problem $ (\partial _z +M)\Psi =0$, with Tr$M=0$, then $\det (A(z),B(z))$ does not depend on $z$. Two solutions $A(z),B(z)$ are linearly dependent iff $\det (A(z),B(z))=0$. }  of both equations upon using $\det (\Xi, \Xi _1)=-2ie^{i\Phi (\theta , \vec{c})}$ and $\det (\Psi _+,\Psi _-)=-\frac {1}{\cos \pi l}$, we end up with the functional relation that may be thought of as the definition of the quantum integrable theory, the $QQ$-system\footnote{We discuss below the suitable asymptotic conditions on $Q_{\pm}(\theta ,\vec{c})$ for the ground and excited states.}
\be
Q_+ \left (\theta + \frac{i\pi}{2N}, \vec{c} \right ) Q_- \left (\theta - \frac{i\pi}{2N}, \vec{c}^R \right )- Q_+ \left (\theta - \frac{i\pi}{2N}, \vec{c}^R\right )Q_- \left (\theta + \frac{i\pi}{2N}, \vec{c} \right )=-2i e^{i\Phi \left (\theta + \frac{i\pi}{2N} , \vec{c}\right )}\cos \pi l \label {Qw} \, .
\ee
Its form is a suitable extension of the usual one as it involves the rotation of the moduli, a phenomenon that will take place in the following, too, and must be taken care of. In the case of 4D and 3D theories ${\cal N}=4$ SYM and ${\cal N}=6$ SCS the analogous relation is called {\it quantum spectral curve} \cite{QSC1,QSC2}. Before going on, we remark that, by virtue of results here discussed, the functions $Q_\pm$ enjoy also the alternative definition
\be
Q_\pm(\theta ,\vec{c})= \pm \cos \pi l \det (\Xi , \Psi _\pm ) \, , \quad \label  {Qdet}
\ee
in terms of determinant of vector solutions of (\ref {ass-lin-prob}).

\medskip

Besides $Q_{\pm}$ (eigenvalues) one usually defines also the transfer matrices (eigenvalues). Considering as a basis of independent solutions $\Xi$ and $\hat \Omega \Xi$, the transfer matrices can be seen as Stokes coefficients \cite{ODE-IM1,ODE-IM2,ODE-IM3} (namely connexion coefficients of $z=\infty$ Jost solutions), {\it i.e.}  
\be
\hat \Omega ^n \Xi (z; \theta , \vec{c})=e^{-i\Phi (\theta , \vec{c})}T_{\frac{n-1}{2}} \left (\theta - i\pi \frac{n}{2N}, \vec{c} \right ) \hat \Omega \Xi (z; \theta , \vec{c}) - e^{-i\Phi (\theta , \vec{c})}T_{\frac{n-2}{2}} \left (\theta - i\pi \frac{n+1}{2N}, \vec{c}^R \right ) \Xi (z; \theta , \vec{c}) \, , \quad n \geq 1
\label {as-omega}
\ee
where $\hat \Omega ^n \Xi$ is subdominant in $S_n$ and dominant in $S_{n\pm 1}$.
%or using the fact that $\hat \Omega \circ \hat \Pi$ is unbroken on $\Xi$:
%\ba
%&&  \hat \Pi ^{-n} \Xi (z;\theta, \vec{c})=i^{n-1} e^{i(-1)^n n \Phi (\theta , \vec{c})} T_{\frac{n-1}{2}}\left (\theta - \frac{i\pi}{2N}n, \vec{c}\right ) \hat \Pi ^{-1} \Xi (z;\theta, \vec{c}) - \nonumber \\
%&&- i^n  e^{i(-1)^n(n-1) \Phi (\theta , \vec{c})}
% T_{\frac{n-2}{2}}\left (\theta - \frac{i\pi}{2N}(n+1), \vec{c} \right )
%\Xi (z;\theta, \vec{c}) \, .
%\ea
In a completely equivalent way, using the fact that $\det (\hat \Omega \Xi, \Xi)=2ie^{i\Phi (\theta , \vec{c})}$, we have that
\be 
T_{\frac{n-1}{2}}\left (\theta - \frac{i\pi}{2N}n, \vec{c}\right )=\frac {1}{2i}\det ( \hat \Omega^{n} \Xi , \Xi)  \, .
\ee 
On the other hand, if we express\footnote {The expression we use is 
\be
\hat \Omega ^n \Xi (z;\theta ,  \vec{c})=\Xi \left ( ze^{\frac{i\pi n}{N}}; \theta - \frac{i\pi n}{N}, \vec{c}^{R^{n}} \right )=Q_+\left (\theta -\frac{i\pi n}{N},\vec{c}^{R^{n}} \right ) \Psi _- (z; \theta , \vec{c}) + Q_- \left (\theta - \frac{i\pi n}{N} , \vec{c}^{R^{n}} \right) \Psi _+ (z; \theta , \vec{c}) \, .
\ee}  $\Xi$, $\hat \Omega \Xi$ and $\hat \Omega ^n \Xi$ in terms of $\Psi_{\pm} $ through (\ref  {maineq}) we find $T_j$ with $j$ semi-integer in terms of $Q_{\pm}$ as generalisations (quadratic constructs) of the quantum Wronskian (\ref{Qw}):
\ba
&& T_{j}(\theta , \vec{c})=\frac{i}{2\cos \pi l}\Bigl [ Q_+ \left (\theta + \frac{i\pi}{2N}(2j+1) , \vec{c}) \right ) Q_- \left (\theta - \frac{i\pi }{2N}(2j+1) , \vec{c}^{R^{2j+1}} \right ) - \nonumber \\
&& - Q_+ \left (\theta - \frac{i\pi }{2N}(2j+1) , \vec{c}^{R^{2j+1}} \right )Q_-\left (\theta + \frac{i\pi }{2N}(2j+1) , \vec{c}) \right ) \Bigr ] \, . \label {trmat}
\ea
We remark that  $T_{-1/2}(\theta , \vec{c})=0$ and, because of  (\ref{Qw}), $T_0(\theta , \vec{c})=e^{i\Phi \left ( \theta + \frac{i\pi}{2N}, \vec{c}\right )}$. 
%which reduces to (\ref{Qw}) for the smallest representation $j=0$ along with the identities $T_{-1/2}(\theta , \vec{c})=0$ and $T_0(\theta , \vec{c})=e^{i\Phi \left ( \theta + \frac{i\pi}{2N}, \vec{c}\right )}$. 

Furthermore, as a simple consequence of only the quantum Wronskian relation (\ref {Qw}) and the expression (\ref {trmat}), we obtain the functional relation 
\ba
&& T_j(\theta,  \vec{c})Q_{\pm }\left (\theta + \frac {i\pi}{2N}(2j-1), \vec{c}^R \right )- T_{j-\frac{1}{2}}\left (\theta -\frac {i\pi}{2N}, \vec{c}^R \right )
Q_{\pm }\left (\theta + \frac {i\pi}{2N}(2j+1), \vec{c} \right )= \nonumber \\
&&=e^{i \Phi \left (\theta +\frac{i\pi}{2N}(2j+1), \vec{c}\right )}  
Q_{\pm }\left (\theta - \frac {i\pi}{2N}(2j+1), \vec{c}^{R^{2j+1}} \right ) \, . \label  {TjQ}
\ea
This expression in the case $j=1/2$ realises an extension of the Baxter $TQ$-system with the novelty of the rotation of the moduli
\be
T_{\frac{1}{2}}(\theta ,  \vec{c}) Q_{\pm }(\theta ,  \vec{c}^R)= e^{-i\Phi \left (\theta +\frac{i\pi}{N}, \vec{c}\right )} Q_{\pm }\left (\theta + \frac{i\pi}{N},  \vec{c} \right )+ e^{i\Phi \left (\theta +\frac{i\pi}{N}, \vec{c}\right )}Q_{\pm }\left (\theta - \frac{i\pi}{N},  \vec{c}^{R^2} \right ) \, . \label {TQ}
\ee 
Now, we can use the real-analyticity of the $Q$-functions 
\be 
\overline { Q_{\pm}(\theta , \vec{c})}=Q_{\pm}(\bar \theta , \bar {\vec{c}}) \, , \quad 
\label {real-an}
\ee
which implies the property $\overline { T_j(\theta , \vec{c})}=T_j(\bar \theta , \bar {\vec{c}}^{R^{-2j-1}})$ and of the function $\Phi$, $\bar \Phi (\theta, \vec{c})=\Phi (\bar \theta ,  \bar {\vec{c}})$
and obtain, after doing the complex conjugation of (\ref  {TjQ})
\ba
&&T_j(\theta , {\vec{c}}^{R^{-2j-1}}) Q_{\pm }\left (\theta - \frac {i\pi}{2N}(2j-1), \vec{c}^{R^{-1}} \right ) - T_{j-\frac{1}{2}}\left (\theta +\frac {i\pi}{2N}, \vec{c}^{R^{-2j-1}} \right )
Q_{\pm }\left (\theta - \frac {i\pi}{2N}(2j+1), \vec{c} \right )= \nonumber \\
&& =e^{-i \Phi \left (\theta -\frac{i\pi}{2N}(2j+1), \vec{c}\right )}  
Q_{\pm }\left (\theta + \frac {i\pi}{2N}(2j+1), \vec{c}^{R^{-2j-1}} \right ) \, . 
\label  {TjQbar} 
\ea
Together, relations (\ref {TjQ}) and (\ref {TjQbar}) can be considered as discrete Lax pairs for the wave functions $Q_\pm$. Their compatibility condition gives the $T$-system (notice again the rotation of the moduli $\vec{c}$), through a procedure we explain in Appendix \ref {app1}
\be
T_j \left ( \theta - \frac{i\pi}{2N}, \vec{c}^R \right ) T_j \left ( \theta + \frac{i\pi}{2N}, \vec{c} \right )= e^{i[1+(-1)^{2j+1}]\Phi \left (\theta +\frac{i\pi}{2N}(2j+2), \vec{c}\right )}  + T_{j+\frac{1}{2}}(\theta , \vec{c} ) T _{j-\frac{1}{2}}(\theta , \vec{c}^R) \, . \label {fusion2}
\ee
This relation can also be proven starting from the fact that $\det (\hat \Omega ^{n-1}\Xi , \hat \Omega ^n \Xi)=-2i e^{-i(-1)^n \Phi (\theta , \vec {c})}$ and then computing again this determinant after expanding $\hat \Omega ^{n-1}\Xi , \hat \Omega ^n \Xi$ using (\ref {as-omega}). Finally, a more direct derivation comes again by the use of only the quantum Wronskian relation (\ref {Qw}) and the formula (\ref {trmat}). 

Another useful relation, which however involves only the 'first' transfer matrix $T_{\frac{1}{2}}$ is 
\be
\hat \Omega ^{n-1} \Xi (z;\theta , \vec {c})=e^{-i(-1)^n\Phi (\theta , \vec{c})}T_{\frac{1}{2}} \left (\theta - i\pi \frac{n}{N}, \vec{c}^{R^{n-1}} \right ) \hat \Omega ^n \Xi  (z; \theta , \vec{c}) + e^{-2i(-1)^n\Phi (\theta , \vec{c})} \hat \Omega ^{n+1} \Xi (z; \theta , \vec{c}) \label {relDT} \, . 
\ee 
This relation explains the terminology 'Stokes coefficient for $\hat \Omega ^{n-1} \Xi $ with respect to $\hat \Omega ^n \Xi $' which is usually \cite {ODE-IM2} addressed to the transfer matrix $T_{\frac{1}{2}} \left (\theta - i\pi \frac{n}{N}, \vec{c}^{R^{n-1}} \right )$. 

If in (\ref {relDT}) we express each $\hat \Omega ^n \Xi$ by means of (\ref {as-omega}), we find  
for $T$-functions also the fusion identities 
\ba
T_{\frac{1}{2}}(\theta ,  \vec{c})T_j \left ( \theta + \frac{i\pi}{2N} (2j+1) , \vec{c}^{R^{-2j}} \right )&=&
e^{-i\Phi \left (\theta +\frac{i\pi}{N}, \vec{c}\right )}T_{j-\frac{1}{2}} \left ( \theta + \frac{i\pi}{2N} (2j+2) , \vec{c}^{R^{-2j}} \right )+ \nonumber \\
&+& e^{i\Phi \left (\theta +\frac{i\pi}{N}, \vec{c}\right )}T_{j+\frac{1}{2}} \left ( \theta + \frac{i\pi}{2N} 2j  , \vec{c}^{R^{-2j}} \right ) \, . \label {fusion1}
\ea
Again, relations (\ref {fusion1}) can be alternatively obtained by virtue of the formul{\ae}  (\ref {Qw}, \ref {trmat}).
It is important to say that (\ref {fusion1}) can be used for building higher $T$-functions $T_j$ from $T_0=1$ and the fundamental one $T_{\frac {1}{2}}$. For this reason $T_j$ with $j\geq 1$ can be called 'fused' Stokes coefficients.

\medskip
 
Another important quadratic functional equation, usually derived from the Thermodynamic Bethe Ansatz (TBA) integral equation, is the $Y$-system. In our construction, it enters the stage, instead, not via a physical thermodynamic procedure, but in a different way upon defining the product of two next neighbour $T$ functions:
\be
Y_j(\theta , \vec{c} )=e^{-i[1+(-1)^{2j+1}]\Phi \left (\theta +\frac{i\pi}{2N}(2j+2), \vec{c}\right )} T_{j-\frac{1}{2}}(\theta , \vec{c}^R) T_{j+\frac{1}{2}}(\theta , \vec{c}).
\label {Ydef0}
\ee
In fact, as a consequence of the $T$-system (\ref {fusion2}), they must satisfy the recursive functional relations (again with moduli rotation, too)
\be
Y_j\left ( \theta - \frac{i\pi}{2N} , \vec{c}^R\right ) Y_j\left ( \theta + \frac{i\pi}{2N} , \vec{c}\right )=
\left [ 1+ Y_{j-\frac{1}{2}}(\theta , \vec{c}^R) \right ]  \left [ 1+ Y_{j+\frac{1}{2}}(\theta , \vec{c}) \right ] \label {Y-sys} \, ,
\ee
in the $Y$-system form, from which TBA equations can be derived (see below); because of this peculiarity, we call this 'rotating' $Y$-system (and 'rotating' TBA).

In summary, we have found functional relations, (\ref {Qw}, \ref {TjQ}, \ref {fusion1}, \ref {fusion2}, \ref {Y-sys}), in the case of general polynomials $p(z,\vec{c})$. The major novelty of these relations is that they are not closed (similarly to the ones in \cite {MAS}), in the sense that they involve functions with different (rotated) moduli. The only exception is the sine-Gordon case, where only one modulus, $c_0$ survives as $c_n=0$, $n\geq 1$ and then $ \vec{c}^R= \vec{c}$. This is a crucial and relevant difference for the explicitly evaluation of the functions $Q$, $T$ and $Y$. In fact, the $Y$-system is of particular relevance as it is usually convertible into (non-linear) integral equations, the TBA ones ({\it cf.} \cite{ZamPLB} for the inverse procedure), and these can be solved rather efficiently in some regimes and numerically. But here the usual inversion procedure is obstructed on $Y$-system (\ref {Y-sys}) by the presence of rotated moduli $Y$ functions in the left hand side. We are going to illustrate later how to transform this apparent difficulty into an advantage, and, instead, modify the form of (\ref{Y-sys}) so to make it manageable with the standard procedure.   

In fact, we can exploit a {\it universal} symmetry, always present in the eingenvector equations like the ones involving the (linear) operators $ { \mathcal{D}}_z$ and $ { \mathcal{D}}_{\bar z}$

\be
\hat \Pi : { \mathcal{D}}_z (\theta ) \rightarrow \sigma ^3  { \mathcal{D}}_z (\theta -i\pi) \sigma ^3 ={ \mathcal{D}}_z (\theta )\, , \ \  { \mathcal{D}}_{\bar z}(\theta) \rightarrow \sigma ^3 { \mathcal{D}}_{\bar z} (\theta -i\pi) \sigma ^3={ \mathcal{D}}_{\bar z} (\theta ) \, .
\label{Pi-symm}
\ee
which similarly is broken (changes the {\it vacuum} solutions) in that $\hat \Pi \Xi (\theta )=\sigma _3 \Xi (\theta -i\pi)$ and $\hat \Pi \Psi _{\pm}(\theta )=\sigma_3 \Psi _\pm (\theta -i\pi)$. Explicit action on them\footnote {Combining for instance (\ref {Omegapsi}) with (\ref {omega-pi-psi}) we find 
\be
\hat \Pi \Psi _{\pm} (z;\theta,\vec{c})\equiv \sigma ^3 \Psi _{\pm} (z;\theta-i\pi,\vec{c})=\mp e^{\mp i\pi l} 
\Psi _{\pm} (z;\theta,\vec{c})
\ee
Analogous formul\ae for $\hat \Pi \Xi$ can be obtained by combining  
(\ref {Omegachi}) with (\ref {omega-pi-xi}). 
}
can be derived by (\ref {Omegapsi}, \ref {Omegachi}) and by the 
remarking property that the composed symmetry $\left ( \hat \Omega \circ \hat \Pi \right )=\left ( \hat \Pi \circ \hat \Omega \right )$ is {\it unbroken} when acting on $\Xi$, $\Psi _{\pm}$:
\be
\left ( \hat \Omega \circ \hat \Pi \right ) \Xi (z;\theta,\vec{c})\equiv \sigma ^3 \Xi \left ( ze^{\frac{i\pi}{N}}; \theta -i\pi - \frac{i\pi}{N}, \vec{c}^R \right )=-i e^{-i\Phi (\theta , \vec{c})} \Xi (z;\theta,\vec{c}) \, ,
\label {omega-pi-xi}
\ee
\be
\left ( \hat \Omega \circ \hat \Pi \right )  \Psi _{\pm} (z;\theta,c)\equiv \sigma ^3 \Psi _{\pm}\left ( ze^{\frac{i\pi}{N}}; \theta -i\pi - \frac{i\pi}{N}, \vec{c}^R \right )=\mp e^{\mp i\pi l} 
\Psi _{\pm} (z;\theta,c) \, .
\label {omega-pi-psi}
\ee
In quite a general fashion, an unbroken symmetry implies (quasi)periodicity properties. In our particular case
the use of $\hat \Omega \circ \hat \Pi$ on equation (\ref {maineq}) brings us to a new form of {\it quasi-periodicity} involving the rotation of the moduli $\vec{c}$
\be
Q_{\pm } \left (\theta -i\pi - \frac{i\pi}{N} , \vec{c} ^R \right )=e^{\mp i\pi \left (l+\frac{1}{2}\right )} e^{-i\Phi (\theta , \vec{c})}Q_{\pm} (\theta ,  \vec{c} ) \label {qper} \, .
\ee
Through (\ref {trmat}) the quasi-periodicity of $Q_{\pm }$ translates into an analogous relations for the $T$-functions 
\be
T_{j} \left (\theta -i\pi - \frac{i\pi}{N} , \vec{c} ^R \right )=
e^{-i[1+(-1)^{2j+1}]\Phi \left (\theta +\frac{i\pi}{2N}(2j+1), \vec{c}\right )}
T_{j}(\theta ,  \vec{c} ) 
\label{Tjper} \, .
\ee
Finally, the $Y$-functions (\ref {Ydef0}) show a property of quasi-periodicity without extra phase factor:
\be
Y_{j} \left (\theta -i\pi - \frac{i\pi}{N} , \vec{c} ^R \right )=
Y_{j}(\theta ,  \vec{c} )
\label{Yjper} \, .
\ee

\medskip

Now, combining (\ref {qper}) with (\ref {Qw}), we arrive at
\be
e^{i\pi l} Q_+(\theta ,\vec{c})Q_-(\theta +i\pi, \vec{c}) + e^{-i\pi l}Q_-(\theta ,\vec{c} )Q_+(\theta +i\pi, \vec{c})=-2\cos \pi l
\label{QQalt}
\,
\ee
which is an alternative form for the $QQ$-system, with no rotations of moduli and no extra phase $\Phi$, but with decorative factors $e^{\pm i \pi l}$ and  \textit{universal} shifts $\pm i \pi$ of the spectral parameter. Because of the absence of rotations of the moduli and the absence of $N$ in the shifts, we call (\ref {QQalt}) \textit{universal} $QQ$-system.

Similarly, a careful use of quasi-periodicity properties permits to write functional relations for $T_j$ and $Y_j$ without rotations of the moduli. Indeed, the new definitions
\ba
&& T^{new}_j\left (\theta +i\pi \frac{N+1}{N}\left (j+\frac{1}{2}\right ) , \vec{c} \right ) =
e^{-i(-1)^{2j}(2j+1)\Phi \left (\theta + \frac{i\pi}{2
N}(2j+1), \vec{c}\right )}T_{j}(\theta , \vec{c})
\label{Tredef}
\, , \\
&& Y^{new}_j\left (\theta +i\pi \frac{N+1}{N}\left (j+1\right ) , \vec{c} \right )=Y_{j}(\theta , \vec{c})
\label{redef} \, , 
\ea
which imply, with the aid of the rotated moduli periodicity (\ref{qper}),
\ba
&& T_{j}^{new}(\theta , \vec{c})=\frac{i}{2\cos \pi l}\Bigl [e^{i\pi \left (l+\frac{1}{2}\right )(2j+1)} Q_+ \left (\theta - \frac{i\pi}{2}(2j+1) , \vec{c}) \right ) Q_- \left (\theta + \frac{i\pi }{2}(2j+1) , \vec{c} \right ) - \nonumber \\
&& - e^{-i\pi \left (l+\frac{1}{2}\right )(2j+1)} Q_+ \left (\theta + \frac{i\pi }{2}(2j+1) , \vec{c} \right )Q_-\left (\theta - \frac{i\pi }{2}(2j+1) , \vec{c}) \right ) \Bigr ] \, , \label{tnewQ}
\ea
and
\be
Y_j^{new}(\theta , \vec{c} )=T^{new}_{j-\frac{1}{2}}(\theta , \vec{c}) T^{new}_{j+\frac{1}{2}}(\theta , \vec{c}) 
\label{Tnew-Ynew}
\ee
bring the relation (\ref {TjQ}) into
\ba
&& T_j^{new}(\theta,  \vec{c} )Q_{\pm}\left (\theta -\frac {i\pi}{2}(2j-1),\vec{c} \right )-e^{\pm i\pi \left (l+\frac{1}{2}\right )}T_{j-\frac{1}{2}}^{new}\left (\theta + \frac  {i\pi}{2}, \vec{c} \right ) Q_{\pm}\left (\theta -\frac {i\pi}{2}(2j+1) , \vec{c} \right ) = \nonumber \\
&&  e^{\mp 2i\pi j\left (l+\frac{1}{2}\right )}Q_{\pm}\left (\theta +\frac {i\pi}{2}(2j+1), \vec{c} \right ) \, , \label {TjQnew}
\ea
which in the case $j=1/2$ reduces to the familiar form of the $TQ$-system:
\be
T_{\frac{1}{2}}^{new}(\theta , \vec{c})Q_{\pm}(\theta, \vec{c})=
e^{\pm i\pi \left (l+\frac{1}{2}\right )} Q_{\pm}(\theta -i\pi, \vec{c})+e^{\mp i\pi \left (l+\frac{1}{2}\right )}Q_{\pm}(\theta +i\pi, \vec{c}) \, .
\label{closed-TQ}
\ee
Taking the complex conjugate of (\ref {TjQnew}), we get, after noticing that the real-analyticity property (\ref {real-an}) of $Q_\pm$ implies an analogous property for the new $T$-functions
\be
\overline { T^{new}_j(\theta , \vec{c})}=T^{new}_j(\bar \theta , \bar {\vec{c}}) \, , 
\ee
the following relation
\ba
&& T_j^{new}(\theta,  \vec{c} )Q_{\pm}\left (\theta +\frac {i\pi}{2}(2j-1),\vec{c} \right )-e^{\mp i\pi \left (l+\frac{1}{2}\right )}T_{j-\frac{1}{2}}^{new}\left (\theta - \frac  {i\pi}{2}, \vec{c} \right ) Q_{\pm}\left (\theta +\frac {i\pi}{2}(2j+1) , \vec{c} \right ) = \nonumber \\
&&  e^{\pm 2i\pi j\left (l+\frac{1}{2}\right )}Q_{\pm}\left (\theta -\frac {i\pi}{2}(2j+1), \vec{c} \right ) \, , \label {TjQbarnew}
\ea
which is obviously a rewriting of (\ref {TjQbar}) for the functions $T_j^{new}$ after use of quasi-periodicity. 

We remark that (\ref {TjQnew}, \ref {TjQbarnew}), after the redefinition $Q_\pm (\theta) \rightarrow Q_\pm (\theta) e^{\pm \theta \left (l+\frac {1}{2}\right )}$ coincide with (1.4,1.3) of \cite {FN}, respectively, in the case $\phi =1$.

Moreover, the $T$-system (\ref {fusion2}) and the fusion (\ref {fusion1}) become the following closed relations without rotation of the moduli
%\footnote {We observe {\it en passant} that the definitions \be \lim _{j\rightarrow +\infty}T_j^{new}\left (\theta -i\pi \left (j+\frac {1}{2}\right ),  \vec{c}  \right )=Q_+(\theta ,  \vec{c})e^{-\left (l+\frac {1} {2}\right )\theta } \, , \quad Q_-(\theta,  \vec{c})=Q_+(-\theta,  \vec{c}) \ee transform the fusion relation (\ref {closed-T-system0}) into the $TQ$-system (\ref {closed-TQ}). } 
\be
T_j^{new} \left ( \theta - \frac{i\pi}{2}, \vec{c}\right ) T_j^{new} \left ( \theta + \frac{i\pi}{2}, \vec{c} \right )= 1 + T_{j+\frac{1}{2}}^{new}(\theta , \vec{c} ) T _{j-\frac{1}{2}}^{new}(\theta , \vec{c}) \, ,
\label{closed-T-system}
\ee
\be
T_{\frac{1}{2}}^{new}(\theta ,  \vec{c}) T^{new}_j \left (\theta -\frac{i\pi}{2}(2j+1) , \vec{c} \right )=
T^{new}_{j-\frac{1}{2}} \left (\theta -\frac{i\pi}{2}(2j+2) , \vec{c} \right )+
T^{new}_{j+\frac{1}{2}} \left (\theta -\frac{i\pi}{2}(2j) , \vec{c} \right )
\label{closed-T-system0} \, .
\ee
Similarly, our $Y$-system (\ref{Y-sys}) takes the form
\be
Y_j^{new}\left ( \theta - \frac{i\pi }{2}, \vec{c}\right ) Y_j^{new}\left ( \theta + \frac{i\pi}{2} , \vec{c}\right )=
\left [ 1+ Y_{j-\frac{1}{2}}^{new}(\theta , \vec{c}) \right ]  \left [ 1+ Y_{j+\frac{1}{2}}^{new}(\theta , \vec{c}) \right ] \, .
\label{closed-Y-system}
\ee
Moreover, we have $T^{new}_0 (\theta , \vec{c})=1$ and $T^{new}_{-\frac {1}{2}} (\theta , \vec{c})=0$ and then $Y_0^{new}(\theta, \vec{c})=0$.

As an intermediate remark, we say that functional relations (\ref {closed-TQ}, \ref {closed-T-system}, \ref {closed-T-system0}, \ref {closed-Y-system}) can also be obtained as consequences of only the universal $QQ$-relation (\ref {QQalt}) and definitions (\ref {tnewQ}, \ref {Tnew-Ynew}).
Besides the adjective {\it closed}, relations (\ref  {tnewQ}-\ref {closed-Y-system}) may be denoted as {\it universal} in the specific sense that their shifts in $\theta$ do not depend on $N$ and moduli do not rotate.
If one wants to connect functions with different rotated moduli one has to use quasi-periodicity: using (\ref {redef}, \ref {Yjper}) one finds for functions $Y_j^{new}$  
\be
Y_{j}^{new} \left (\theta -i\pi - \frac{i\pi}{N} , \vec{c} ^R \right )=
Y_{j}^{new}(\theta ,  \vec{c} )
\label{Yjnewper} \, ;
\ee
using (\ref {Tredef}, \ref {Tjper}) one gets for $T_{j}^{new}$
\be
T_{j}^{new} \left (\theta -i\pi - \frac{i\pi}{N} , \vec{c} ^R \right )=
e^{-i[1+(-1)^{2j+1}]\Phi \left (\theta -\frac{i\pi}{2}(2j+1), \vec{c}\right )}
T_{j}^{new}(\theta ,  \vec{c} ) \, . \label{Tnewper}
\ee

Interestingly, these functional relations not involving rotations of the moduli can be derived by use of the $\hat \Pi$-symmetry. For instance, $QQ$-system (\ref {QQalt}) can also be found starting from (\ref {maineq}) and from the relation obtained applying the $\hat \Pi$-symmetry on it; then, taking the determinant between $\Xi $ and $\hat \Pi \Xi$ which equals $-1$, one finds (\ref {QQalt}).  In an analogous fashion the functions $T_j^{new}$ are monodromy coefficients between Jost solutions of the linear problem obtained by using the $\hat \Pi$ symmetry. One has
\be
\hat \Pi ^{-n} \Xi (z;\theta, \vec{c})=i^{n-1}  T_{\frac{n-1}{2}}^{new}\left (\theta +\frac{i\pi n}{2}, \vec{c}\right )  \hat \Pi ^{-1} \Xi (z;\theta, \vec{c}) - i^n  T_{\frac{n-2}{2}}^{new}\left (\theta +\frac{i\pi(n+1)}{2}, \vec{c} \right )
\Xi (z;\theta, \vec{c})
\label{as-pi} \, , 
\ee
from which using the fact that $\det \left ( \hat \Pi ^{-1} \Xi , \Xi \right )=-2$, one gets
\be
T^{new}_{\frac{n-1}{2}}\left (\theta + i\pi \frac{n}{2}, \vec{c} \right )=-\frac {1}{2} i^{1-n} \det
(\hat \Pi ^{-n} \Xi (z; \theta , \vec{c}) , \Xi (z; \theta , \vec{c}) ) \, . \label {Twronsk}
\ee
Relations (\ref {as-pi}) are obviously related to (\ref {as-omega}), through the use of the unbroken symmetry $\hat \Omega \circ \hat \Pi$ (\ref {omega-pi-xi}), the definition (\ref {Tredef}) of $T_j^{new}$ and the quasi-periodicity (\ref {Tjper}). In particular, $\hat \Pi ^{-n} \Xi$ is subdominant in $S_n$ and dominant in $S_{n\pm 1}$.

It is useful for practical applications to remove the shift in the left hand side of relation (\ref {Twronsk}): we find
\ba
T^{new}_{n-\frac {1}{2}}(\theta)&=&\frac {1}{2i} \det (\hat \Pi ^{-n} \Xi (\theta) , \hat \Pi ^n \Xi (\theta))  \, , \label {Twronskbis} \\
T^{new}_{n}\left (\theta +\frac {i\pi}{2}\right )&=&-\frac {1}{2} \det (\hat \Pi ^{-n-1} \Xi (\theta) , \hat \Pi ^n \Xi (\theta))  
\label {Twronskter} \, .
\ea
The closed $T$- and $Y$-systems (\ref{closed-T-system}, \ref{closed-Y-system}) have come to light in \cite {YSA} as relations (16,18), where the moduli are not explicit as they are always the same. In  \cite {YSA}, and also in \cite{GMN}, only the $\hat \Pi$ symmetry (\ref{Pi-symm}) is used, without any mention to further ($\hat \Omega$) symmetries: among other things we wish to stress here the importance of the $\hat \Omega$ symmetry (\ref{Omega-symm}), which is also more ostensive when there are more ({\it e.g.} two) irregular singularities of the opers or partial differential operators (\ref{D}, \ref{barD}) \cite {F-G-pure-Liouville}. Nevertheless we wish to stress also the fact that the universal relations (also in the following) can be derived also where there is no further ($\hat \Omega$) symmetries.

\medskip

Finally, we want to mention another property of $Q_{\pm }$ concerning a particular behaviour under complex conjugation:
\be
\overline {Q_{\pm}(\theta , \vec{c} )}=-Q_{\mp} (-\bar \theta , \vec{c} ) \label {qcompl} \, .
\ee
This property derives from using relations
\be
\sigma ^1 \overline { \Xi (z;-\theta , \vec{c})}=-\Xi (z;\bar \theta , \vec{c}) \, , \quad
\sigma ^1 \overline { \Psi _{\pm}(z;-\theta , \vec{c})}=\Psi _{\mp}(z;\bar \theta , \vec{c}) \, 
\ee
on (\ref  {maineq}) and complements usual real-analyticity (\ref {real-an}).
Property (\ref {qcompl}) immediately extends to the functions $T_j$:
\be
\overline {T_j(\theta , \vec{c})}= T _j(-\bar \theta , \vec{c})
\label{Tjcompl}
\ee
under complex conjugation. 

%However, functions defined in (\ref {trmat}) are in general not real-analytic, i.e. $\overline { T_j(\theta , \vec{c})} \not= T_j (\bar \theta , \bar \vec{c})$. This is caused by the presence of different rotated moduli in the $Q$ functions contained in definition (\ref {trmat}). This flaw is not shared by the new $T$ functions, which satisfy not only property (\ref {Tjcompl}), i.e. $T_j^{new}(\theta , \vec{c})=\bar T _j^{new}(-\bar \theta , \vec{c})$, but, differently from $T_j$, also real-analyticity: $\bar T^{new}_j (\theta , \vec{c})=T^{new}_j (\bar \theta , \bar \vec{c})$. 

\medskip

In general, the $T$-systems (\ref{closed-T-system}, \ref {closed-T-system0}) are functional relations which can involve infinite functions. However, in the particular case under analysis, the functional relations  (\ref {closed-T-system}, \ref{closed-T-system0}) can be restricted to the first $2N+2$ functions $T_j$, with $j=0,\frac {1}{2},1,...,N+\frac {1}{2}$, due to the periodicity property\footnote{Property (\ref {Qperiod}) implies the (quasi)periodicities
\be
T_j^{new}(\theta, \vec{c})=T_j^{new}(\theta -2i\pi(N+1), \vec{c}) e^{2iN\Phi \left (\theta +i\pi\left (j+\frac {1}{2}\right ),\vec{c}\right )+ 2iN\Phi \left (\theta -i\pi\left (j+\frac {1}{2}\right ),\vec{c}\right )} \, , \quad Y_j^{new}(\theta, \vec{c})=Y_j^{new}(\theta -2i\pi(N+1), \vec{c}) \, .
\ee}
\be
Q_{\pm} (\theta -2i\pi (N+1), \vec{c})=e^{\mp i\pi N (2l+1) } e^{-2iN \Phi (\theta , \vec{c})}Q_{\pm}(\theta , \vec{c}) \, , \label{Qperiod}
\ee
which implies
\be
e^{i(2N-1+(-1)^{2N})\Phi \left (\theta - \frac{i\pi}{2}(2N+3), \vec{c}\right) }T_{N+1}^{new}(\theta , \vec{c})-e^{i(2N-1+(-1)^{2N})\Phi \left (\theta - \frac{i\pi}{2}(2N+1),\vec{c}\right) }T_N^{new} (\theta , \vec{c})=2 \cos \pi (2l+1) \, .
\ee
In other words, the $T$-systems (\ref {closed-T-system}, \ref{closed-T-system0}) involving $T_j$, with $j=0,\frac {1}{2},1,...,N+\frac {1}{2}$, are closed functional relations. 
And, for the same reason, the $2N+1$ functions $Y_j^{new}$ with $j=\frac {1}{2},1,..., N$ and $\hat Y(\theta , \vec{c})=e^{i(2N-1+(-1)^{2N})\Phi \left (\theta - \frac{i\pi}{2}(2N+1),\vec{c}\right) } T_N^{new}(\theta , \vec{c})$ close\footnote {Functional relations (\ref {Ydef}) look similar to the $Y$-system of sine-Gordon model at rational values of $\beta ^2$. However, there are important differences. The shifts in (\ref {Ydef}), {\it i.e.} $\pm i\pi/2$, are universal, in the sense that they do not depend on the coupling constant ($N$ in this case), from which they do depend in the sine-Gordon case. %In addition, the number of $Y$ functions of (\ref {Ydef}) do not coincide with $Y_j$ with $j\geq N+1/2$ are non zero.
This reflects in different periodicity for the $Y$-functions with respect to sine-Gordon.
}
 %but they are all functions of $Y$ and $Y_j$, with $j\leq N$. Analogously, for what concerns the $T$ functions, $T_j$, with $j\geq N+1$ are all expressible in terms of $T_j$, with $j\leq N+1/2$.}
the $Y$-system
\ba
Y_j^{new}\left ( \theta - \frac{i\pi }{2}, \vec{c}\right ) Y_j^{new}\left ( \theta + \frac{i\pi}{2} , \vec{c}\right )&=&
\left [ 1+ Y_{j-\frac{1}{2}}^{new}(\theta , \vec{c}) \right ]  \left [ 1+ Y_{j+\frac{1}{2}}^{new}(\theta , \vec{c}) \right ] \, , \quad j=\frac {1}{2} ,...,N-\frac{1}{2} \nonumber \\
Y_N^{new}\left ( \theta - \frac{i\pi }{2}, \vec{c}\right ) Y_N^{new}\left ( \theta + \frac{i\pi}{2} , \vec{c}\right )&=&\left [ 1+ Y_{N-\frac{1}{2}}^{new}(\theta , \vec{c}) \right ] \left [ 1+ e^{2\pi i \left (l+\frac{1}{2}\right )}\hat Y (\theta , \vec{c}) \right ]
 \left [ 1+ e^{-2\pi i \left (l+\frac{1}{2}\right )}\hat Y (\theta , \vec{c}) \right ]  \nonumber \\
\hat Y \left ( \theta - \frac{i\pi }{2}, \vec{c}\right ) \hat Y \left ( \theta + \frac{i\pi }{2}, \vec{c}\right )&=& 1+Y_N^{new}(\theta ,  \vec{c} ) \label {Ydef} \, ,
\ea
which resembles in form a $D_{2N+2}$-type $Y$-system. By construction the functions $Y_j^{new}$ and $\hat Y$ have the periodicity property
\be
Y_j^{new}(\theta +2i\pi (N+1), \vec {c})=Y_j^{new}(\theta, \vec {c}) \, ,\quad
\hat Y (\theta +2i\pi (N+1), \vec {c})=\hat Y(\theta, \vec {c}) \, .
\ee

\medskip

A more drastic phenomenon appear if $l=0$. In this case there is the extra property
\be
T_j^{new}(\theta ,  \vec{c})=-e^{2iN\Phi \left (\theta +i\pi \left (j+\frac {1}{2}\right ),\vec{c}\right )- 2iN\Phi \left (\theta -i\pi \left (j+\frac {1}{2}\right ),\vec{c}\right )}
T_{2N-j+1}^{new}(\theta ,  \vec{c}) \, , 
\ee
which, for $j=N+\frac {1}{2}$, means $T_{N+\frac {1}{2}}^{new}(\theta ,  \vec{c})=0$: this last relation implies $Y_N^{new} (\theta ,  \vec{c})=0$. Therefore, we are left with $2N-1$ functions $Y^{new}_j$ which close a $Y$-system of the type $A_{2N-1}$, given by relations
(\ref {closed-Y-system}) for $j=1/2,...,N-\frac {1}{2}$ with the extra condition $Y_N^{new} (\theta ,  \vec{c})=0$.
This $Y$-system appears also in the study of minimal area surfaces in $AdS_3$ or, equivalently, in the computations of null polygonal Wilson loops in $\mathcal{N} = 4$ SYM at strong coupling \cite {YSA,HISS}. The number $2N-1$ of independent $Y$-functions equals the number of complex moduli $\vec{c}=(c_0,c_1,...,c_{2N-2})$ of the problem, in terms of which one expresses the $4N-2$ real cross ratios characterising scattering states dual to $AdS_3$.

%Particularly simple when $l=0$ is the case $N=\frac {3}{2}$, since $T_{\frac {3}{2}}^{new}(\theta ,  \vec{c})=1$ and 
%$T_{1}^{new}\left (\theta \pm \frac {5i\pi}{2} ,  \vec{c}\right )=T_{\frac {1}{2}}^{new} ( \theta ,  \vec{c})$. Then, the $T$-system (\ref {closed-T-system}) can be written in terms of the single function $T_{\frac {1}{2}}^{new}$, 
%\be
%T_{\frac {1}{2}}^{new} \left ( \theta - \frac{i\pi}{2}, \vec{c}\right ) T_{\frac {1}{2}}^{new} \left ( \theta + \frac{i\pi}{2}, \vec{c} \right )= 1 + T _{\frac{1}{2}}^{new}\left (\theta + \frac {5i\pi}{2}, \vec{c}\right ) \, .
%\label{closed-T-system-3}
%\ee
%Similarly, the $Y$-system (\ref {closed-Y-system}) can be expressed in terms of only $Y_{\frac {1}{2}}^{new}$, since 
%\be
%Y_{\frac {1}{2}}^{new}\left (\theta,  \vec{c}\right )=T_{1}^{new}\left (\theta,  \vec{c}\right )=T_{\frac {1}{2}}^{new} \left ( \theta \pm \frac {5i\pi}{2},  \vec{c}\right )=Y_1^{new} \left ( \theta \pm \frac {5i\pi}{2},  \vec{c}\right ) \, . 
%\label {YT-funct-rel}
%\ee
%Actually, because of (\ref {YT-funct-rel}), the form of the $Y$-system is identical to the form of the $T$-system (\ref {closed-T-system-3}):
%\be
%Y_{\frac {1}{2}}^{new}\left ( \theta - \frac{i\pi }{2}, \vec{c}\right ) Y_{\frac {1}{2}}^{new}\left ( \theta + \frac{i\pi}{2} , \vec{c}\right )=
%1 + Y_{\frac{1}{2}}^{new}\left (\theta + \frac {5i\pi}{2}, \vec{c}\right ) \, .
%\label{closed-Y-system-3}
%\ee 

\medskip

Eventually, from the $Y$-system (\ref {Y-sys}, \ref {Ydef}) TBA equations for the functions  $\ln Y_j(\theta ,  \vec{c})$, $\ln  Y_j^{new}(\theta ,  \vec{c})$ can be derived, if one knows the limits at Re$\theta \rightarrow \pm \infty$ of these functions. These asymptotic behaviours can be derived by expressing the $Y$-functions in terms of $Q_\pm$ by means of (\ref {Tnew-Ynew}, \ref {tnewQ}) and then by using the limits of the $Q$-functions at Re$\theta \rightarrow \pm \infty$.
These limits will be derived in next Section, together with other properties of $Q_\pm$.

%{\bf $Y$-systems new, TBA, relation with NLIEs of \cite {FRS}}

\section{$Q$-functions and solutions of the linear problem}
\setcounter{equation}{0}
\label{QT}

The basic objects in our construction and in this paper are the $Q$-functions and then they must be better investigated by exploiting their connection with the solutions of the associated linear problem. In this respect,
when $|z|\rightarrow 0$ with $\varphi$ fixed, the two component of the solution $\Xi$ become proportional to $Q_{\pm}$, respectively:
\be
\Xi (|z|=0, \arg z=\varphi;\theta , \vec{c})=\frac{1}{\sqrt{\cos \pi l}}
\left ( \begin{array}{c}
e^{-\left (\theta +i\varphi\right) l}Q_+(\theta, \vec{c}) \\
e^{\left (\theta +i\varphi \right) l}Q_-(\theta, \vec{c})
\end{array} \right ) \, . \label{QXi}
\ee
This relation is a simple consequence of (\ref {maineq}) and the form (\ref {Psizero}) of the solutions $\Psi_{\pm}$ near the origin.

For what concerns the polynomial $p(z, \vec{c})$, which defines the quantum integrable model along with $\eta(z,\bar z, \vec{c})$, in principle it can be very general. Yet, for performing specific calculation we must know to be in a domain where it is different from zero as we need to define in terms of it an invertible (analytic) map $z\rightarrow w(z)$ such that $\frac {dw}{dz}= \sqrt{p(z, \vec{c})}\neq 0$. For instance a rather technical assumption might be to move from the case $p_{lz}(z)=z^{2N} + c_0$, $c_0<0$ of \cite{LZ} by adding moduli $c_n$, $n=1,...,2N-2$ and then
%In this way, integration on positive real axis is free of problems in the the rotated coordinate $z'=ze^{\frac{i\pi}{2N}}$.
to change variable and sign of the polynomial
\be
P(z', \vec{c})\equiv -p(z, \vec{c}) ={z'}^{2N}+\sum _{n=0}^{2N-2} c_n e^{i\pi\frac{2N-n}{2N}} z'^n \, , \quad z=z' e^{-\frac{i\pi}{2N}} \, .
\ee
The 'rotation' $z\rightarrow z'$ makes $P(z', \vec{c})$ free of zeroes for $z'>0$ and this property allows to define properly the map $z'\rightarrow w$ around the positive real axis if the added moduli $c_n$ are not very large. 
%The 'amount' of the rotation has been chosen to be $\pi/2N$, since for this value it is easier to write an integral equation for $Q_{\pm}$ (see next Section for details).

As announced, after these manipulations we pass from $z'$ to $w$ by means of the map
\be
\frac{dw}{dz'}=\sqrt{P(z', \vec{c})} \, , \quad w(z')=\mathcal{Q}_N(z')-\int _{z'}^{\infty} dx \left [ \sqrt{P(x, \vec{c})}-q_N(x, \vec{c}) \right ] \, , \quad
\frac{d\mathcal{Q}_N(z', \vec{c})}{dz'}=q_N(z', \vec{c}) \, ,
\ee
where integration is made on a suitable contour (possibly the real axis) avoiding cuts and
$q_N(x, \vec{c})$ is a function chosen to make convergent the integration.

The choice of both the contour and the functions $q_N(x,\vec{c})$, $\mathcal{Q}_N(x,\vec{c})$ depends on the form of $P(z,\vec{c})$. For what concerns $q_N(x,\vec{c})$, if $N=1/2$ and $P(x,\vec{c})=x-c_0$, we choose $q_{1/2}(x,\vec{c})=x^{1/2}-\frac{c_0}{2}x^{-1/2}$ (then, $\mathcal{Q}_N(x,\vec{c})=\frac{2}{3}x^{\frac{3}{2}}-c_0 x^{\frac{1}{2}}$). In general, if $N$ is semi-integer, $q_N(x,\vec{c})=x^N +\tilde B_{N-2} x^{N-2}+...+\tilde B_{-1/2}x^{-1/2}$, $ \mathcal{Q}_N(x,\vec{c})=\frac{x^{N+1}}{N+1}+\tilde B_{N-2} \frac{x^{N-1}}{N-1}+...+\tilde B _{-\frac{1}{2}}x^{\frac{1}{2}}$;
if $N$ is integer, but $N\not=1$, $q_N(x,\vec{c})=x^N +\tilde B_{N-2} x^{N-2}+...+\tilde B_{-1}/(x+a)$ with $a>0$ and $\mathcal{Q}_N(x,\vec{c})=\frac{x^{N+1}}{N+1}+\tilde B_{N-2} \frac{x^{N-1}}{N-1}+...+\tilde B _{0}x +\tilde B_{-1} \ln (x+a)$; in both cases, $\tilde B_i$ are suitable coefficients which are real analytic functions of $\vec{c}$.

The utility of the changes of variables $z\rightarrow z'\rightarrow w$ is that the function
\be
\hat \Xi (w; \theta ,\vec{c})= e^{- \left (\frac{i\pi}{4} + \frac{1}{8} \ln \frac{\bar P(z', \vec{c})}{P(z', \vec{c})} \right) \sigma^3}\Xi \left (z; \theta + \frac{i\pi}{2}+\frac{i\pi}{2N} , \vec{c}\right ) \label {newXi}
\ee
is a solution of the linear problem
\ba
&& \left [ \partial _w + \frac{\sigma ^3}{2}\partial _w \hat \eta -e^{\theta} \left ( \sigma ^+ e^{\hat \eta}+\sigma ^- e^{-\hat \eta} \right ) \right ] \hat \Xi (w; \theta , \vec{c} ) \equiv { \mathcal{D}}_{w}  \hat \Xi (w; \theta , \vec{c} )=0 \nonumber \\
&& \left [ \partial _{\bar w} - \frac{\sigma^3}{2}\partial _{\bar w} \hat \eta -e^{-\theta} \left ( \sigma ^- e^{\hat \eta}+\sigma ^+ e^{-\hat \eta} \right ) \right ] \hat \Xi (w; \theta , \vec{c} ) \equiv { \mathcal{D}}_{\bar w}\hat \Xi (w; \theta , \vec{c} )=0  \label {linhat}
\ea
with (real) 'potential'
\be
\hat \eta = \eta -\frac{1}{4}\ln (P\bar P)
\ee
satisfying the ordinary sinh-Gordon equation $\partial _w \partial _{\bar w} \hat \eta -2\sinh 2\hat \eta =0$.

In this context an important quantity is the apex of the cone in the $w, \bar w$ plane, {\it i.e.} the image of the point $z'=\bar z'=0$ under the map $(z', \bar z')\rightarrow (w, \bar w)$: it  has coordinates
$w_0(\vec{c}), \bar w_0(\vec{c})$, where
\be
w_0(\vec{c})=-\int _{0}^{\infty} dx \left [ \sqrt{P(x, \vec{c})}-q_N(x, \vec{c}) \right ] \, ,
\label {apex1}
\ee
for $N$ semi-integer and
\be
w_0(\vec{c})=-\int _{0}^{\infty} dx \left [ \sqrt{P(x, \vec{c})}-q_N(x, \vec{c}) \right ] + \tilde B _{-1}\ln a \, ,
\label {apex2}
\ee
for $N$ integer different from $1$. Added moduli $c_n$, $n=1,...,2N-2$ are chosen in such a way that integrations can be safely performed on the real axis: this is true if $|c_n|$ are sufficiently small, but in principle this is not the only case we can consider.

Explicit expression of $w_0$ in the simplest case on only one modulus $c_0<0$ (quantum sine-Gordon model) is
\be
w_0=\frac {2\pi ^{\frac {3}{2}}N}{(N+1)\Gamma \left (-\frac {1}{2N} \right ) \Gamma \left (\frac {N+1}{2N} \right ) \sin \frac {\pi}{N}}(-c_0)^{\frac {1+N}{2N}} \, . \label {w0sg}
\ee

\medskip

The importance of the apex is that, as a simple consequence of relations (\ref {newXi}) and (\ref {QXi}), there is a direct a connection between the solution $\hat \Xi$ of the (\ref {linhat}) at the apex and the $Q$ functions:
\be
\hat \Xi (w_0; \theta ,\vec{c})= e^{-i\pi \frac{\sigma ^3}{4}}
\frac{1}{\sqrt{\cos \pi l}}
\left ( \begin{array}{c}
e^{-\left (\theta +i\varphi+\frac{i\pi}{2}+\frac{i\pi}{2N}\right) l}Q_+\left(\theta +\frac{i\pi}{2}+\frac{i\pi}{2N}, \vec{c} \right ) \\
e^{\left (\theta +i\varphi+\frac{i\pi}{2}+\frac{i\pi}{2N}\right) l}Q_-\left(\theta +\frac{i\pi}{2}+\frac{i\pi}{2N}, \vec{c} \right )
\end{array} \right ) \, . \label{hatXi-Q}
\ee
In other words we can say that (\ref {hatXi-Q}) is the 'initial condition' which fixes the solution $\hat \Xi$ of the linear problem (\ref{linhat}). The value $w_0$ is reminiscent of the polynomial $p(z,\vec{c})$ we started from.

This connection and the simple behaviour of $\hat \Xi$ at large $\theta$,
\be
\ln \hat \Xi (w; \theta ,\vec{c})\simeq -(w+a(\vec{c}))e^{\theta} -(\bar w+\bar a (\vec{c}))e^{-\theta} \, ,
\ee
with $a(\vec{c})$ an arbitrary constant, imply that
the leading behaviour of $Q_{\pm}$ when Re$\theta \rightarrow \pm \infty$ and $|\textrm {Im}\theta | <\pi \frac{N+1}{2N}$, 
\be
\ln Q_{\pm} \left ( \theta +i\pi \frac{N+1}{2N};\vec{c}\right )
 \simeq  -(w_0(\vec{c})+a(\vec{c}) )e^{\theta}-(\bar w_0(\vec{c})+\bar a (\vec{c}))e^{-\theta}
%= r e^\theta- \bar r e^{-\theta}
\label {largetheta} \, ,
\ee
with $w_0(\vec{c})+a(\vec{c})$ and $\bar w_0(\vec{c})+\bar a(\vec{c})$ Renormalisation Group (RG) parameters.
$a (\vec{c})$ is fixed by imposing the asymptotic behaviour (\ref {largetheta}) on quasi-periodicity (\ref{qper}) and $QQ$-system (\ref {Qw}). Then, 
\be
a (\vec{c})=\frac{\pi}{2N} B_{-1}(\vec{c})e^{\frac{i\pi}{2N}} \, , \label{alfac}
\ee
which means that the large $\theta $ behaviour of $\ln \hat \Xi$ can be expressed alternatively as
\be
\ln \hat \Xi (w; \theta ,\vec{c})\simeq -we^{\theta}-\bar w e^{-\theta}-\frac{1}{2}\Phi \left (\theta + \frac{i\pi}{2N},\vec{c} \right ) \, ,
\ee
with $\Phi$ defined in (\ref{phase}). 

Formul{\ae} (\ref {largetheta}, \ref {alfac}) were announced in \cite {FRS} as formul{\ae} (2.28, 2.29). Here, they have been proved.

\subsection{Zeroes of $Q_{\pm}$ and bound states of the radial problem}

The zeroes of the functions $Q_{\pm} (\theta ,\vec{c})$ have a particular meaning in the sense that, for these particular values of $\theta$, the solutions $\Psi _{\pm}$ are identified with the solution $\Xi$, which vanishes exponentially in a particular sector of the complex plane. The starting point is the relation (see equation \ref {maineq})
\be
\Xi (z;\theta, \vec{c})=Q_+(\theta ,\vec{c}) \Psi _- (z; \theta , \vec{c}) + Q_- (\theta , \vec{c}) \Psi _+ (z; \theta , \vec{c}) \label {maineq2} \, .
\ee
Then, for the particular values $\theta _n^{+}$ ($\theta _n^{-}$) for which $Q_+$ ($Q_-$) vanishes,
the solution $\Psi _{+}$ ($\Psi _{-}$) coincides with the solution $\Xi $, which vanishes exponentially in the Stokes sector $S_0$ of the complex $z$-plane. If $\theta _n^{+}$ ($\theta _n^{-}$) is real, this sector contains the positive real axis and then the solution to the linear problem we are considering identifies a bound state of a radial problem, {\it i.e.} a solution that is regular at the origin and vanishes as $z \rightarrow +\infty$. 

\section{Leading orders and TBA equations for rotating $Y$-system}
\setcounter{equation}{0}
\label {l0N32case}

As written, the novelty of the rotating $Y$-system (\ref  {Y-sys}) is the rotation of the new moduli and this represents also the main difficulty in deriving now the TBA equations from it. Let us show how to overcome this problem. We consider the case $l=0$ and $N$ semi-integer, but extensions to the general case do not have obstructions. Since $Y_0=Y_N=0$, there are $2N-1$ functions $Y_j$, with semi-integer $j$, $\frac {1}{2}\leq j \leq N-\frac {1}{2}$, which satisfy
 (\ref  {Y-sys}). 
First step is to derive the asymptotic limits of 
\be
Y_j (\theta , \vec{c})=T_{j-\frac {1}{2}} (\theta , \vec {c}^R ) T_{j+\frac {1}{2}} (\theta , \vec {c}) \, ,
\ee
using (\ref {largetheta}) and (\ref {trmat}). The final result is that, if Re$\theta  \rightarrow \pm \infty$ and 
\be
\frac {\pi}{N} \textrm {max} \{ -j, j-N \}\leq \textrm {Im}\theta \leq \frac {\pi}{N}  \textrm {min} \{ j, N-j \}
\ee
\ba
&& \ln Y_j (\theta,  \vec {c}) \simeq -2 \cos \frac {\pi}{2N} w_0 ( \vec {c}) e^{\theta +\frac {i\pi}{2N}(2j-N)}-
2 \cos \frac {\pi}{2N} \overline  {w_0 ( \vec {c})} e^{-\theta -\frac {i\pi}{2N}(2j-N)} - \nonumber \\
&& - w_0 (c^{R^{2j+1}}) e^{\theta +\frac {i\pi}{2N}(N-2j-1)}- w_0 (c^{R^{2j-1}}) e^{\theta +\frac {i\pi}{2N}(N-2j+1)}
- \nonumber \\
&& -\overline  { w_0 (c^{R^{2j+1}})} e^{-\theta -\frac {i\pi}{2N}(N-2j-1)}- \overline  { w_0 (c^{R^{2j-1}})} e^{-\theta -\frac {i\pi}{2N}(N-2j+1)} \equiv {\cal A}_{j}\left ( \theta  ,  \vec{c} \right ) \label {Y-asy-general} \, .
\ea
It is important to remark that, in the case in which $c_n=0$ for $n\geq 1$, 
\be
{\cal A}_{j} ( \theta  , c_0)=-2 w_0(c_0) \cosh \theta \left [ 2 \cos  \frac {\pi (2j+1-N)}{2N} + 
 2 \cos  \frac {\pi (2j-1-N)}{2N} \right ]  \label {Ajc_0}
\ee
with 
\be
w_0(c_0)=-|c_0|^{\frac {N+1}{2N}} \int _0^{+\infty} dx [ \sqrt {x^{2N}+1} - x^N]=  |c_0|^{\frac {N+1}{2N}} \frac {\Gamma \left (-\frac {1}{2}-\frac {1}{2N} \right ) \Gamma \left (\frac {1}{2N} \right )}{4N\sqrt {\pi}} \label {w0c0}
\ee
Then, since $w_0(c_0)<0$, Re${\cal A}_{j} ( \theta  , c_0)>0$, when $|\textrm {Im}\theta |<\pi/2$. By continuity, the condition  Re${\cal A}_{j} ( \theta  , \vec {c})>0$ holds true in a strip in $\theta$ around the real axis also in a neighbourhood of $c_n=0$ (small $c_n$, with $n \not=0$). And also in general the leading order ${\cal A}_{j}( \theta  , \vec {c})$  satisfies an extension (by rotation of the moduli) of the functional relations that were discussed in \cite {ZamPLB, DTBA}, in specific
\be
{\cal A}_{j}\left ( \theta  -\frac {i\pi}{2N},  \vec{c}^R \right )+{\cal A}_{j}\left ( \theta  +\frac {i\pi}{2N},  \vec{c} \right )
-{\cal A}_{j-\frac {1}{2}}\left ( \theta  ,  \vec{c}^R \right )- {\cal A}_{j+\frac {1}{2}}\left ( \theta  ,  \vec{c} \right )=0 \, . \label {Aj-funct-rel}
\ee
Then, defining 
\be
\varepsilon _{j}(\theta , \vec{c})=\ln Y_{j}  \left ( \theta  ,  \vec{c} \right ) \, , 
\ee
as a consequence of (\ref  {Y-sys}, \ref {Aj-funct-rel}), one has the functional relations
\ba
&& \left [\varepsilon_{j}\left ( \theta  -\frac {i\pi}{2N},  \vec{c}^R \right )-{\cal A}_{j}\left ( \theta  -\frac {i\pi}{2N},  \vec{c}^R \right )\right ] + \left [\varepsilon_{j}\left ( \theta  +\frac {i\pi}{2N},  \vec{c} \right )-{\cal A}_{j}\left ( \theta  +\frac {i\pi}{2N},  \vec{c} \right )\right ]- \nonumber \\
&& - \left [ \varepsilon _{j-\frac {1}{2}}\left ( \theta  ,  \vec{c}^R \right )-{\cal A}_{j-\frac {1}{2}}\left ( \theta  ,  \vec{c}^R \right ) \right ]- \left [ \varepsilon _{j+\frac {1}{2}}\left ( \theta  ,  \vec{c} \right )-{\cal A}_{j+\frac {1}{2}}\left ( \theta  ,  \vec{c} \right ) \right ]= \\
&&= \ln \left ( 1+e^{-\varepsilon _{j-\frac {1}{2}}\left ( \theta  ,  \vec{c}^R \right )} \right ) +
\ln \left ( 1+e^{-\varepsilon _{j+\frac {1}{2}}\left ( \theta  ,  \vec{c} \right )} \right ) \nonumber \, , 
\ea
which are conveniently rewritten as
\ba
&& \left [\varepsilon_{j}\left ( \theta  -\frac {i\pi}{2N},  \vec{c}^R \right )-{\cal A}_{j}\left ( \theta  -\frac {i\pi}{2N},  \vec{c}^R \right )\right ] + \left [\varepsilon_{j}\left ( \theta  +\frac {i\pi}{2N},  \vec{c} \right )-{\cal A}_{j}\left ( \theta  +\frac {i\pi}{2N},  \vec{c} \right )\right ] = \nonumber  \\
&& = \ln \left ( e^{-{\cal A}_{j-\frac {1}{2}}\left ( \theta  ,  \vec{c}^R \right )} +e^{\varepsilon _{j-\frac {1}{2}}\left ( \theta  ,  \vec{c}^R \right )-{\cal A}_{j-\frac {1}{2}}\left ( \theta  ,  \vec{c}^R \right )} \right ) +
\ln \left (e^{-{\cal A}_{j+\frac {1}{2}}\left ( \theta  ,  \vec{c} \right )} +e^{\varepsilon _{j+\frac {1}{2}}\left ( \theta  ,  \vec{c} \right )-{\cal A}_{j+\frac {1}{2}}\left ( \theta  ,  \vec{c} \right )} \right ) \, . \label {funct-rel-rot}
\ea
In the particular case $c_n=0, n \geq 1$, the associated TBA equation read \cite {DTBA,ZamPLB}
\ba
\varepsilon _j (\theta)&=&{\cal A}_{j}(\theta , c_0)+\int _{-\infty}^{+\infty} \frac {d\theta '}{2\pi}\frac {N}{\cosh [N(\theta -\theta ')]}\Bigl [ \ln \left (e^{-{\cal A}_{j-\frac {1}{2}}\left ( \theta  ' \right )} +e^{\varepsilon _{j-\frac {1}{2}}\left ( \theta '  \right )-{\cal A}_{j-\frac {1}{2}}\left ( \theta '  \right )} \right )+ \nonumber \\
&+& \ln \left (e^{-{\cal A}_{j+\frac {1}{2}}\left ( \theta  ' \right )} +e^{\varepsilon _{j+\frac {1}{2}}\left ( \theta '  \right )-{\cal A}_{j+\frac {1}{2}}\left ( \theta '  \right )} \right ) \Bigr ] \, , \quad \frac {1}{2}\leq j \leq N-\frac {1}{2} \, , \quad |\textrm {Im}\theta |<\pi/2N \, , 
\label {TBA-eq-nomoduli}
\ea
where ${\cal A}_{j}(\theta , c_0)$ is given by (\ref {Ajc_0}, \ref {w0c0}). Here, we have used the important positivity of Re${\cal A}_{j(\theta , c_0)}$. The kernel in (\ref  {TBA-eq-nomoduli}) is the 'universal' kernel.

%A particular case is $N=3/2$

When $c_n \not=0$, $n \geq 1$, generalising a procedure discussed in \cite {MAS}, we perform the discrete Fourier sums (the integer $l$ is defined mod $2N$) 
\ba
\chi _{j,l}  (\theta , \vec{c} )&=& \sum _{k=1}^{2N} e^{\frac {2i\pi kl}{2N}} \varepsilon _j \left (\theta ,  \vec{c}^{R^k} \right ) \, , \quad \hat {\cal A}_{j,l}(\theta , \vec{c} )=  \sum _{k=1}^{2N} e^{\frac {2i\pi kl}{2N}} {\cal A}_{j} \left (\theta ,  \vec{c}^{R^k} \right )  \nonumber \\
\Lambda _{j,l} (\theta , \vec{c})  &=& \sum _{k=1}^{2N} e^{\frac {2i\pi kl}{2N}} 
\ln \left (e^{-{\cal A}_{j}\left ( \theta , \vec{c}^{R^k}   \right )} +e^{\varepsilon _{j}\left ( \theta , \vec{c}^{R^k}   \right )-{\cal A}_{j}\left ( \theta , \vec{c}^{R^k}   \right )} \right ) 
\, , \label {disc-fou-tra}
\ea
whose inversion is
\ba
&&\varepsilon _j \left (\theta ,  \vec{c}^{R^k} \right ) = \frac {1}{2N}\sum _{l=1/2-N}^{N-1/2} e^{-\frac {2i\pi kl}{2N}}  \chi_{j,l}\left ( \theta  ,  \vec{c}\right ) \, , \quad 
{\cal A}_{j} \left (\theta ,  \vec{c}^{R^k} \right ) = \frac {1}{2N}\sum _{l=1/2-N}^{N-1/2} e^{-\frac {2i\pi kl}{2N}}  \hat {\cal A}_{j,l}(\theta , \vec{c} ) \nonumber \\
&& \ln \left (e^{-{\cal A}_{j}\left ( \theta , \vec{c}^{R^k}   \right )} +e^{\varepsilon _{j}\left ( \theta , \vec{c}^{R^k}   \right )-{\cal A}_{j}\left ( \theta , \vec{c}^{R^k}   \right )} \right ) =\frac {1}{2N}\sum _{l=1/2-N}^{N-1/2} e^{-\frac {2i\pi kl}{2N}} \Lambda _{j,l} (\theta , \vec{c}) \, . \label {inv-disc-fou-tra} 
\ea
As a consequence of (\ref {funct-rel-rot}), the quantities (\ref {disc-fou-tra}) satisfy 
\ba
&&  \left [\chi_{j,l}\left ( \theta  -\frac {i\pi}{2N},  \vec{c}\right )-\hat {\cal A}_{j,l}\left ( \theta  -\frac {i\pi}{2N},  \vec{c} \right )\right ] e^{-\frac {2i\pi l}{2N}} +  \left [\chi_{j,l}\left ( \theta  +\frac {i\pi}{2N},  \vec{c}\right )-\hat {\cal A}_{j,l}\left ( \theta  +\frac {i\pi}{2N},  \vec{c} \right )\right ]= \nonumber \\
&& e^{-\frac {2i\pi l}{2N}}\Lambda _{j-\frac {1}{2},l}(\theta , \vec{c} )+
\Lambda _{j+\frac {1}{2},l}(\theta , \vec{c} ) \label {funct-rel-four}
\ea
With $\frac {1}{2}-N \leq l \leq N-\frac {1}{2}$, these functional equations can be inverted by crucially using the  positivity of Re${\cal A}_{j}(\theta , \vec{c})$
\be
\chi_{j,l}(\theta , \vec{c})=\hat {\cal A}_{j,l}(\theta , \vec{c})+\int _{-\infty}^{+\infty} \frac {d\theta '}{2\pi}\frac {Ne^{-l (\theta -\theta ') }}{\cosh [N(\theta -\theta ')]} \left [ e^{-\frac {i\pi l}{2N}} \Lambda _{j-\frac {1}{2},l}(\theta ', \vec{c} ) + e^{\frac {i\pi l}{2N}}\Lambda _{j+\frac {1}{2},l}(\theta ', \vec{c} ) \right ] \, , \label {TBA-final-general-chi}
\ee
which are the TBA equations in the $\chi $ variables. Instead, by using 
the inverse discrete Fourier sum (\ref {inv-disc-fou-tra}), we obtain those in the $\varepsilon _j $ variables (Fock-Gancharov coordinates)
\ba
&&\varepsilon _j \left (\theta ,  \vec{c}^{R^k} \right ) ={\cal A}_{j} \left (\theta ,  \vec{c}^{R^k} \right )+ \nonumber \\
&+& \frac {i}{2}\sum _{k'=1}^{2N}\left [ \int _{-\infty}^{+\infty}\frac {d\theta '}{2\pi} \frac {(-1)^{k-k'}}{\sinh \left ( \frac {\theta -\theta '}{2}-\frac {i\pi}{2N}\left (k'-k-\frac {1}{2}\right ) \right )}\ln \left (e^{-{\cal A}_{j-\frac {1}{2}}\left ( \theta , \vec{c}^{R^{k'}}   \right )} +e^{\varepsilon _{j-\frac {1}{2}}\left ( \theta , \vec{c}^{R^{k'}}   \right )-{\cal A}_{j-\frac {1}{2}}\left ( \theta , \vec{c}^{R^{k'}}   \right )} \right ) + \right.\nonumber \\
&+&  \left. \int _{-\infty}^{+\infty}\frac {d\theta '}{2\pi} \frac {(-1)^{k-k'}}{\sinh \left ( \frac {\theta -\theta '}{2}-\frac {i\pi}{2N}\left (k'-k+\frac {1}{2}\right ) \right )}\ln \left (e^{-{\cal A}_{j+\frac {1}{2}}\left ( \theta , \vec{c}^{R^{k'}}   \right )} +e^{\varepsilon _{j+\frac {1}{2}}\left ( \theta , \vec{c}^{R^{k'}}   \right )-{\cal A}_{j+\frac {1}{2}}\left ( \theta , \vec{c}^{R^{k'}}   \right )} \right ) \right] \label {TBA-final-general}
\ea
%xxx

%Using in the expression of $\Lambda _{j\pm \frac {1}{2},l}$ the relation $\varepsilon _j \left (\theta ,  \vec{c}^{R^k} \right ) = \frac {1}{2N}\sum _{l=1/2-N}^{N-1/2} e^{-\frac {2i\pi kl}{2N}}  \chi_{j,l}\left ( \theta  ,  \vec{c}\right )$, they become equations for $\chi _{j,l}$. 

%xxx

Equations (\ref  {TBA-final-general}) hold for $\theta $ belonging to a strip around the real axis in the $\theta $ plane in which Re${\cal A}_{j}\left ( \theta  , \vec{c}^{R^k}\right )>0$. As we discussed before, by continuity with the case $c_n =0, n \geq 1$, this strip exists by continuity also when $c_n\not=0$, but $|c_n|\ll 1$ for $n\geq 1$.

%Again, in the particular case $N=3/2$,.....

\subsection {The case $l=0, N=3/2$}

The simplest case in which the general discussion of last Section holds is $l=0, N=\frac {3}{2}$. However, this case benefits also of further simplifications which eventually halves the number of TBA equations. Indeed, there are two $Y$-functions,  $Y_{\frac {1}{2}} \left ( \theta  , \vec{c} \right )$ and $Y_{1} \left ( \theta  , \vec{c} \right )$, which satisfy the functional relations (see (\ref {Y-sys}))
\ba
&& Y_{\frac {1}{2}}\left ( \theta - \frac{i\pi }{3}, \vec{c}^R\right ) Y_{\frac {1}{2}}\left ( \theta + \frac{i\pi}{3} , \vec{c}\right )=
1 + Y_{1}\left (\theta , \vec{c}\right ) \label {Y12-funct-rel}\\
&& Y_{1}\left ( \theta - \frac{i\pi }{3}, \vec{c}^R\right ) Y_{1}\left ( \theta + \frac{i\pi}{3} , \vec{c}\right )=
1 + Y_{\frac {1}{2}} \left (\theta , \vec{c}\right ) \label {Y1-funct-rel}
\ea
However, the relation $Y_{1}\left (\theta , \vec{c}\right )=Y_{\frac {1}{2}} \left ( \theta  , \vec{c}^{R^{-1}} \right )$ holds. 
Then, there is only one relevant functional relation,
\be
Y_{\frac {1}{2}}\left ( \theta - \frac{i\pi }{3}, \vec{c}^R\right ) Y_{\frac {1}{2}}\left ( \theta + \frac{i\pi}{3} , \vec{c}\right )=
1 + Y_{\frac {1}{2}} \left ( \theta  , \vec{c}^{R^{-1}} \right ) \, , \label {Y-only}
\ee
which involves, for given $\vec{c}$, three different $Y$-functions, $ Y_{\frac {1}{2}} \left (\theta , \vec{c}\right )$, 
$Y_{\frac {1}{2}} \left ( \theta  , \vec{c}^{R} \right )$, $Y_{\frac {1}{2}} \left ( \theta  , \vec{c}^{R^{2}} \right )$.
The asymptotic leading term when Re$\theta \rightarrow \pm \infty$ and $|\textrm {Im}\theta | <\frac {2\pi}{3}$ can be read from (\ref {Y-asy-general}) and equals
\be
{\cal A}_{\frac {1}{2}}\left ( \theta  ,  \vec{c} \right )=-w_0\left (\vec{c}^{R^{-1}} \right)e^{\theta -\frac {i\pi}{6}}-\overline  {w_0\left (\vec{c}^{R^{-1}} \right)} e^{-\theta +\frac {i\pi}{6}}-w_0(\vec{c})e^{\theta +\frac {i\pi}{6}}- \overline  {w_0(\vec{c})}e^{-\theta - \frac {i\pi}{6}} \label  {Y1-asy-offcrit} \, , 
\ee
%The expression of $Y_{\frac {1}{2}} \left ( \theta  , \vec{c} \right )$ in terms of $Q$-functions is
%\be
%Y_{\frac {1}{2}} \left ( \theta  , \vec{c} \right )=-\frac {1}{2}\left [Q_+\left (\theta +  \frac {2i\pi}{3}, \vec{c}^{R^{-1}} \right )
%Q_-(\theta +i\pi, \vec{c} ) + Q_-\left (\theta +  \frac {2i\pi}{3}, \vec{c}^{R^{-1}} \right )
%Q_+(\theta +i\pi, \vec{c} ) \right ] \, 
%\ee
where the quantity $w_0( \vec{c})$ for this case has the form (remembering that $c_0<0$)
\be
w_0( \vec{c})\simeq -\int _{0}^{+\infty} dx \left [ \sqrt {x^3+c_1 e^{\frac{2i\pi}{3}}x-c_0}-x^{\frac {3}{2}}-\frac {c_1 e^{\frac {2i\pi}{3}}}{2\sqrt {x}} \right ] \, .
\ee
%It is convenient, from now on, to define $\varepsilon _{\frac {1}{2}}(\theta , \vec{c})=\ln Y_{\frac {1}{2}}  \left ( \theta  ,  \vec{c} \right )$. 
%The use of (\ref  {largetheta}) for the asymptotics of $Q_\pm$ brings to the conclusion that, when Re$\theta  \rightarrow \pm \infty$ and $|\textrm {Im}\theta | <\frac {2\pi}{3}$, 
%\be
%\varepsilon _{\frac {1}{2}}(\theta , \vec{c}) \simeq  -w_0\left (\vec{c}^{R^{-1}} \right)e^{\theta -\frac {i\pi}{6}}-\overline  {w_0\left (\vec{c}^{R^{-1}} \right)} e^{-\theta +\frac {i\pi}{6}}-w_0(\vec{c})e^{\theta +\frac {i\pi}{6}}- \overline  {w_0(\vec{c})}e^{-\theta - \frac {i\pi}{6}} \equiv {\cal A}_{\frac {1}{2}}\left ( \theta  ,  \vec{c} \right )
%\label {Y1-asy-offcrit}
%\ee
When $c_1=0$ explicit computations are possible and 
\be
{\cal A}_{\frac {1}{2}}\left ( \theta  ,  \vec{c} \right )=2\sqrt {3\pi}\frac {\Gamma (1/3)}{\Gamma (11/6)} |c_0|^{\frac {5}{6}}\cosh \theta \, .
\ee
In this case, there is only one TBA equation, which reads
\be
\varepsilon _{\frac {1}{2}} (\theta)=2\sqrt {3\pi}\frac {\Gamma (1/3)}{\Gamma (11/6)} |c_0|^{\frac {5}{6}}\cosh \theta +\frac {2\sqrt {3}}{\pi}\int _{-\infty}^{+\infty} d\theta ' \frac {\cosh (\theta -\theta ')}{1+2\cosh (2\theta -2\theta ')} \ln  \left (1+e^{-\varepsilon _{\frac {1}{2}} \left ( \theta ' \right ) } \right )  \label {TBA-eq-nomoduli-12}
\ee
and which holds for $|\textrm {Im}\theta |<\pi/3$: in this strip, Re${\cal A}_{\frac {1}{2}}\left ( \theta  ,  \vec{c} \right )>0$. 
This equation can be also obtained by specialising the general case (\ref {TBA-eq-nomoduli}) to $j=1/2$ and by re-arranging  the non-linear term in the integral, which eventually produces a modification of the integration kernel: this computation is shown in Appendix  \ref {app4}.

For $c_1\not=0$, the $Y$-system (\ref {Y-only}) translates into the relations 
\be
\varepsilon _{\frac {1}{2}}  \left ( \theta - \frac{i\pi }{3}, \vec{c}^R\right ) +\varepsilon _{\frac {1}{2}} \left ( \theta + \frac{i\pi }{3}, \vec{c}\right )- \varepsilon _{\frac {1}{2}} \left ( \theta , \vec{c}^{R^{-1}} \right ) = \ln \left (1+e^{-\varepsilon _{\frac {1}{2}} \left ( \theta , \vec{c}^{R^{-1}} \right ) } \right )  \label {epsilon-only} \, , 
\ee
while the asymptotic behaviour (\ref {Y1-asy-offcrit}) is a zero mode of the shift operators appearing in the l.h.s of (\ref {epsilon-only}):
\be
{\cal A}_{\frac {1}{2}} \left ( \theta - \frac{i\pi }{3}, \vec{c}^R\right ) + {\cal A}_{\frac {1}{2}} \left ( \theta + \frac{i\pi }{3}, \vec{c}\right )-  {\cal A}_{\frac {1}{2}} \left ( \theta , \vec{c}^{R^{-1}} \right )=0 \, . 
\ee
Then, the quantities ($l$ is defined mod $3$)
\ba
&& \chi _{\frac {1}{2},l}(\theta )=\sum _{k=1}^3 e^{\frac {2i\pi kl}{3}} \varepsilon _{\frac {1}{2}}\left (\theta , \vec {c}^{R^k}\right ) \, , \quad \tilde \Lambda _{\frac {1}{2},l}(\theta )= \sum _{k=1}^3 e^{\frac {2i\pi kl}{3}} \ln \left (1+e^{-\varepsilon _{\frac {1}{2}} \left ( \theta , \vec{c}^{R^{k}} \right ) } \right ) \, , \\
&& \hat {\cal A}_{\frac {1}{2},l}(\theta)= \sum _{k=1}^3 e^{\frac {2i\pi kl}{3}} 
{\cal A}_{\frac {1}{2}} \left ( \theta , \vec{c}^{R^{k}} \right ) \, , \label {disc-four-sum-32}
\ea
satisfy the functional equations
\ba
&& e^{\frac {2i\pi l}{3}}\left [\chi _{\frac {1}{2},l} \left (\theta -\frac {i\pi}{3} \right ) - \hat {\cal A}_{\frac {1}{2},l} \left (\theta -\frac {i\pi}{3} \right ) \right ]+
e^{-\frac {2i\pi l}{3}}\left [\chi _{\frac {1}{2},l} \left (\theta +\frac {i\pi}{3} \right ) - \hat {\cal A}_{\frac {1}{2},l} \left (\theta +\frac {i\pi}{3} \right ) \right ] - \nonumber \\
&& -  [\chi _{\frac {1}{2},l} \left (\theta \right ) - \hat {\cal A}_{\frac {1}{2},l}  (\theta )]=\tilde \Lambda _{\frac {1}{2},l}(\theta ) \, , 
\ea
which imply the TBA equations
\be
\chi _{\frac {1}{2},l} (\theta)= \hat {\cal A}_{\frac {1}{2},l}  (\theta )+\int _{-\infty}^{+\infty} d\theta '  {\cal K}_l^{\frac {3}{2}}(\theta -\theta ') \tilde \Lambda _{\frac {1}{2},l} (\theta ')
\, ,  \label  {TBA-final}
\ee
where
\be
{\cal K}_0^{\frac {3}{2}}(\theta)=\frac {\sqrt {3}}{\pi} \frac {2\cosh \theta}{1+2\cosh 2\theta} \, , \quad {\cal K}_1^{\frac {3}{2}}(\theta)=-  
\frac {\sqrt {3}}{\pi} \frac {e^{-\theta}}{1+2\cosh 2\theta} \, , \quad \quad {\cal K}_2^{\frac {3}{2}}(\theta)=-  
\frac {\sqrt {3}}{\pi} \frac {e^{\theta}}{1+2\cosh 2\theta} \, . 
\ee
Inverting (\ref {disc-four-sum-32}), one eventually gets
\ba
&& \varepsilon _{\frac {1}{2}}\left (\theta , \vec {c}^{R^k}\right ) = {\cal A}_{\frac {1}{2}} \left ( \theta , \vec{c}^{R^{k}} \right ) +\frac {1}{\sqrt {3}\pi }\int d\theta ' \frac {3\cosh (\theta -\theta ')+i\sqrt {3}\sinh (\theta -\theta ')}{1+2\cosh (2(\theta -\theta '))}\ln \left (1+e^{- \varepsilon _{\frac {1}{2}}\left (\theta ', \vec {c}^{R^{k+1}}\right )}\right ) + \nonumber \\
&&+ \frac {1}{\sqrt {3}\pi }\int d\theta ' \frac {3\cosh (\theta -\theta ')-i\sqrt {3}\sinh (\theta -\theta ')}{1+2\cosh (2(\theta -\theta '))}\ln \left (1+e^{- \varepsilon _{\frac {1}{2}}\left (\theta ', \vec {c}^{R^{k-1}}\right )}\right ) \label {TBA-final-j}
\ea
As in the general case of last subsection, when $c_1\not=0$, but $|c_1|\ll 1$, by continuity there is a strip around the real axis in the $\theta $ plane in which Re$ {\cal A}_{\frac {1}{2}} \left ( \theta , \vec{c}^{R^{k}} \right )>0$. Equations (\ref  {TBA-final-j}) hold for $\theta $ belonging to this strip. Obviously, they can be also obtained by specialising general equations (\ref {TBA-final-general}) to the case $N=3/2$. Actually, it is simpler to prove the equivalent statement that (\ref {TBA-final}) can also be obtained by specialising the general equations (\ref {TBA-final-general-chi}) to the case $N=3/2$: this is reported in Appendix \ref {app4}.

\section {The leading order and TBA for the universal $Y$-system }
\setcounter{equation}{0}

The 'universal' $Y$-system does not contain rotations of the moduli and for this reason it seems amenable for inversion in TBA equations for $\varepsilon _j =\ln Y_j$. 
The possibility of inversion, however, depends also on the fundamental property of negativity of the driving terms, ${\cal A}_j (\theta , \vec{c})$, in a strip around the real axis. This property holds for some regions of the space of $c_n$, but, unfortunately, not for the region in which we are interested in, which is the region in which $c_0$ is arbitrary negative and $c_n, n\geq 1$, are very small.

This is already evident in the simplest example, $N=3/2$. 
In this case (see (\ref {closed-Y-system}))
\ba
&& Y_{\frac {1}{2}}^{new}\left ( \theta - \frac{i\pi }{2}, \vec{c}\right ) Y_{\frac {1}{2}}^{new}\left ( \theta + \frac{i\pi}{2} , \vec{c}\right )=
1 + Y_{1}^{new}\left (\theta , \vec{c}\right ) \label {Y12-funct-rel-new}\\
&& Y_{1}^{new}\left ( \theta - \frac{i\pi }{2}, \vec{c}\right ) Y_{1}^{new}\left ( \theta + \frac{i\pi}{2} , \vec{c}\right )=
1 + Y_{\frac {1}{2}}^{new} \left (\theta , \vec{c}\right ) \label {Y1-funct-rel-new}
\ea
%The relevant relation is now  $Y_{1}^{new}\left (\theta \pm \frac {5i\pi}{2}, \vec{c}\right )=Y_{\frac {1}{2}} ^{new}\left ( \theta  , \vec{c}\right )$, which gives, as unique functional relation, 
%\be
%Y_{\frac {1}{2}}^{new}\left ( \theta - \frac{i\pi }{2}, \vec{c}\right ) Y_{\frac {1}{2}}^{new}\left ( \theta + \frac{i\pi}{2} , \vec{c}\right )=
%1 + Y_{\frac {1}{2}} ^{new}\left ( \theta  \pm \frac {5i\pi}{2}, \vec{c}\right )
%\ee
%The shift in $\theta $ in the r.h.s. creates technical problems if one wants to write TBA equations.
The asymptotic behaviour of $\varepsilon _{\frac {1}{2}}^{new}\left (\theta , \vec{c}\right )=\ln Y_{\frac {1}{2}}^{new} \left (\theta , \vec{c}\right )$,  when Re$\theta  \rightarrow \pm \infty$ and $|\textrm {Im}\theta | <\frac {\pi}{6}$, is
\be
\varepsilon _{\frac {1}{2}}^{new} (\theta , \vec{c}) \simeq  -w_0\left (\vec{c}^{R^{-1}} \right)e^{\theta -\frac {2i\pi}{3}}-\overline  {w_0\left (\vec{c}^{R^{-1}} \right)} e^{-\theta +\frac {2i\pi}{3}}-w_0(\vec{c})e^{\theta +\frac {2i\pi}{3}}- \overline  {w_0(\vec{c})}e^{-\theta - \frac {2i\pi}{3}} \equiv {\cal A}_{\frac {1}{2}}^{new}\left ( \theta  ,  \vec{c} \right )
\, , 
\ee
while the one of $\varepsilon _{1}^{new}\left (\theta , \vec{c}\right )=\ln Y_{1}^{new} \left (\theta , \vec{c}\right )$,  when Re$\theta  \rightarrow \pm \infty$ and $|\textrm {Im}\theta | <\frac {2\pi}{3}$, is
\be
\varepsilon _{1}^{new}(\theta , \vec{c})  \simeq  -w_0\left (\vec{c}^{R^{-1}} \right)e^{\theta -\frac {i\pi}{6}}-\overline  {w_0\left (\vec{c}^{R^{-1}} \right)} e^{-\theta +\frac {i\pi}{6}}-w_0(\vec{c})e^{\theta +\frac {i\pi}{6}}- \overline  {w_0(\vec{c})}e^{-\theta - \frac {i\pi}{6}} \equiv {\cal A}_{\frac {1}{2}}^{new}\left ( \theta  ,  \vec{c} \right ) \, .
\ee
Then, the TBA equations read
\be
\varepsilon _{\frac {1}{2}}^{new} (\theta , \vec{c})= {\cal A}_{\frac {1}{2}}^{new}\left ( \theta  ,  \vec{c} \right )+\int _{-\infty}^{+\infty} \frac {d\theta '}{2\pi} \frac {1}{\cosh (\theta -\theta ')} \ln \left (1+e^{\varepsilon _{1}^{new}(\theta ', \vec{c})} \right )
\ee
\be
\varepsilon _{1}^{new} (\theta , \vec{c})= {\cal A}_{1}^{new}\left ( \theta  ,  \vec{c} \right )+\int _{-\infty}^{+\infty} \frac {d\theta '}{2\pi} \frac {1}{\cosh (\theta -\theta ')} \ln \left (1+e^{\varepsilon _{\frac {1}{2}}^{new}(\theta ', \vec{c})} \right )
\ee
When $c_1=0$, ${\cal A}_{\frac {1}{2}}^{new}\left ( \theta  ,  \vec{c} \right ) \rightarrow -\infty$, whilst ${\cal A}_{1}^{new}\left ( \theta  ,  \vec{c} \right ) \rightarrow +\infty$. 
Then, for small $c_1$ we are in the so-called maximal chamber, where there are three $\varepsilon ^{new}$ functions which satisfy three TBA equations. To reach the 'minimal chamber', where there are two TBA equations for the functions $\varepsilon _{1}^{new}\left (\theta , \vec{c}\right ), \varepsilon _{\frac {1}{2}}^{new}\left (\theta , \vec{c}\right )$ we need a wall-crossing in the space of the parameter $c_1$, which therefore becomes 'big'. This phenomenon is discussed in \cite {MAR1}, where the authors derived TBA equations from 'new' $Y$-systems in the conformal case. 

Eventually, for our aim, which is to write TBA equations when $c_n, n\geq 1$ are small, the $Y$-system (\ref {Y-sys}) seems more convenient.  

%xxx

%Indeed, for integer $j$, $\frac {1}{2}\leq j \leq N-\frac {1}{2}$, when Re$\theta \rightarrow \pm \infty$, 
%\ba
%&& \ln Y_j^{new}(\theta ,  \vec{c}) \simeq -w_0 \left ( \vec{c}^{R^{-j}} \right ) e^{\theta +i\pi \frac {2j-N-1}{2N}}
% -w_0 \left ( \vec{c}^{R^{j-1}} \right ) e^{\theta - i\pi \frac {2j-N-1}{2N}} - \nonumber \\
%&& - w_0 \left ( \vec{c}^{R^{-j-1}} \right ) e^{\theta + i\pi \frac {2j-N+1}{2N}} - 
%w_0 \left ( \vec{c}^{R^{j}} \right ) e^{\theta - i\pi \frac {2j-N+1}{2N}} - \nonumber \\
%&& - \overline { w_0 \left ( \vec{c}^{R^{-j}} \right ) }e^{-\theta -i\pi \frac {2j-N-1}{2N}}
% -\overline { w_0 \left ( \vec{c}^{R^{j-1}} \right ) } e^{-\theta + i\pi \frac {2j-N-1}{2N}} - \nonumber \\
%&& - \overline {w_0 \left ( \vec{c}^{R^{-j-1}} \right ) } e^{-\theta - i\pi \frac {2j-N+1}{2N}} - 
%\overline { w_0 \left ( \vec{c}^{R^{j}} \right ) } e^{-\theta +  i\pi \frac {2j-N+1}{2N}}\equiv {\cal A}_j ^{new}(\theta ,  \vec{c})
%\ea
%Then, if $c_n=0, n\geq 1$, 
%\be
%{\cal A}_j ^{new}(\theta ,  c_0)= |w_0(c_0)| \left [ \cos  \left (\pi \frac {2j-N-1}{2N} \right ) +   \cos  \left (\pi \frac {2j-N+1}{2N} \right )   \right ] \cosh \theta >0 
%\ee
%On the other hand, for semi-integer $j$.... 

\section{Conformal limit}
\setcounter{equation}{0}
\label{conf-lim}

The conformal limit was proposed and discussed in \cite  {FRS} as realised by
\be
\label{clim}
|c_0| \rightarrow 0 \, , \quad c_m =|c_0|^{\frac{2N-m}{2N}} c_m^{c} \rightarrow 0 \, , \quad \theta =\theta ^{c}-
\frac{1+N}{2N}\ln |c_0| \rightarrow +\infty \, ,
\ee
upon keeping $c_m^{c}$, $\theta ^{c}$ finite. The linear problem (\ref {D}) for the vector $\Psi=\left ( \begin{array}{c} e^{\frac {\theta +\eta}{2}}\psi (z) \\ e^{-\frac{\theta +\eta}{2}}\bar \psi (z) \end{array}  \right)$, after the scaling
\be
z=xe^{-\frac{\theta}{1+N}} \, ,
\ee
becomes
\be
\left [ -\frac{d^2}{dx^2}+\frac{l(l+1)}{x^2}+\sum _{n=1}^{2N} c_n^{c}e^{\frac{2N-n}{1+N}\theta ^{c}}x^n \right ]
\psi ^c(x)= e^{\frac{2N}{1+N}\theta ^{c}}\psi ^c (x) \, , \label {ODE-cft}
\ee
which is a Schr\"{o}dinger equation with energy $E=e^{\frac{2N}{1+N}\theta ^{c}}$ ($c_0^c=-1$) and potential
\be
\label{potential}
V(x)=\sum _{m=1}^{2N} c_m^{c}e^{\frac{2N-m}{1+N}\theta ^{c}}x^m \, 
\ee
for the function $\psi ^c (x)=\psi (z)$.
Interestingly, the limiting $\hat \Omega$ symmetry acts as ($m\geq 1$)
\be
\hat \Omega _{c}: \quad x\rightarrow x e^{\frac{i\pi}{1+N}} \, , \quad
c_m^{c} e^{\frac{2N-n}{1+N}\theta ^{c}}\rightarrow c_m^{c} e^{\frac{2N-m}{1+N}\theta ^{c}} e^{i\pi\frac{2N-n}{1+N}} \, , \quad e^{\frac{2N}{1+N}\theta ^{c}} \rightarrow e^{\frac{2N}{1+N}\theta ^{c}}
e^{-\frac{2i\pi}{1+N}}
\label{Omegac}
\ee
and the limiting $\hat \Pi$ symmetry as
\be
\hat \Pi _{c}: \quad x\rightarrow x e^{-\frac{i\pi}{1+N}} \, , \quad
c_m^{c} e^{\frac{2N-m}{1+N}\theta ^{c}}\rightarrow c_m^{c} e^{\frac{2N-n}{1+N}\theta ^{c}} e^{-i\pi\frac{2N-n}{1+N}} \, , \quad e^{\frac{2N}{1+N}\theta ^{c}} \rightarrow e^{\frac{2N}{1+N}\theta ^{c}}
e^{\frac{2i\pi}{1+N} }\, .
\ee
Therefore in the ODE limit $\hat \Omega _{c}=(\hat \Pi _{c})^{-1}$ and then one can use only one type of symmetry.

With an abuse of notation, the symbols $Q_\pm, T_j, Y_j,T_j^{new}, Y_j^{new}$ will be used to denote also the conformal limit of the corresponding functions.
Functional relations of the conformal case are obtained by the 'off-critical' ones by replacing $\theta$ with $\theta ^c$ and $c_n $ with $ c_n^{c}$.  First instance is that the $T$-system comes from the conformal limit of (\ref {fusion2}):
\be
T_j \left ( \theta ^c - \frac{i\pi}{2N}, \vec{c^c}^R \right ) T_j \left ( \theta ^c+ \frac{i\pi}{2N}, \vec{c^c} \right )= e^{i[1+(-1)^{2j+1}]\Phi \left (\theta ^c+\frac{i\pi}{2N}(2j+2), \vec{c^c}\right )}  + T_{j+\frac{1}{2}}(\theta ^c, \vec{c^c} ) T _{j-\frac{1}{2}}(\theta , \vec{c^c}^R) \, . \label {fusion2cft}
\ee
Second instance comes from the conformal limit of (\ref {Y-sys}), which gives the $Y$-system
\be
Y_j\left ( \theta ^c- \frac{i\pi}{2N} , \vec{c^c}^R\right ) Y_j\left ( \theta ^c+ \frac{i\pi}{2N} , \vec{c^c}\right )=
\left [ 1+ Y_{j-\frac{1}{2}}(\theta ^c, \vec{c^c}^R) \right ]  \left [ 1+ Y_{j+\frac{1}{2}}(\theta ^c, \vec{c^c}) \right ] \label {Y-syscft} \, ,
\ee
Analogously, the conformal limit of (\ref{closed-T-system}) gives
\be
T_{j}^{new} \left ( \theta ^c- \frac{i\pi}{2}, \vec{c^c}\right ) T_{j}^{new} \left ( \theta ^c+ \frac{i\pi}{2}, \vec{c^c} \right )= 1 + T _{j+\frac{1}{2}}^{new}\left (\theta ^c , \vec{c^c}\right )T _{j-\frac{1}{2}}^{new}\left (\theta ^c , \vec{c^c}\right ) .
\label{closed-T-system-cft-prov}
\ee
and that of (\ref{closed-Y-system}),
\be
Y_j^{new}\left ( \theta ^c- \frac{i\pi }{2}, \vec{c^c}\right ) Y_j^{new}\left ( \theta ^c+ \frac{i\pi}{2} , \vec{c^c}\right )=
\left [ 1+ Y_{j-\frac{1}{2}}^{new}(\theta ^c, ) \right ]  \left [ 1+ Y_{j+\frac{1}{2}}^{new}(\theta ^c, \vec{c^c}) \right ] \, ,\label{closed-Y-system-cft-prov}
\ee
that, in the case $l=0$, matches the $Y$-system (3.37) of \cite {MAR1}.
Obviously, all the expressions of $T$ (and then of $Y$) functions in terms of $Q_\pm$ (e.g. (\ref {trmat}) and (\ref {tnewQ})) have the same form as the ones written before in the off-critical case, with $\theta ^c$ and $\vec{c^c}$ in the place of $\theta, \vec{c}$. 

Finally, the leading behaviour of $Q_{\pm}$ when Re$\theta ^c \rightarrow + \infty$ and $|\textrm {Im}\theta ^c| <\pi \frac{N+1}{2N}$, is
\be
\ln Q_{\pm} \left ( \theta ^c +i\pi \frac{N+1}{2N};\vec{c^c}\right )
 \simeq  -(w_0(\vec{c^c})+a(\vec{c^c}) )e^{\theta ^c}
\label {largethetacft} \, ,
\ee
with $w_0(\vec{c^c}), a(\vec{c^c})$ the same functions (\ref {apex1}, \ref {apex2}, \ref {alfac}) of the off-critical case, but now with arguments $c_n^c$, instead of $c_n$. 
On the other hand, if Re$\theta ^c \rightarrow -\infty$ and $|\textrm {Im}\theta ^c| < \pi \frac{N+1}{2N}$,
\be
Q_\pm \left (\theta ^c + i\pi \frac{N+1}{2N};\vec{c^c}\right )  \simeq \textrm {constant} \, .
\label {largethetacftbis}
\ee
The use of (\ref {largethetacft}, \ref {largethetacftbis}) in the expressions of $T_j ( \theta ^c, \vec{c^c} )$, $Y_j  ( \theta ^c, \vec{c^c} ) $, $T_{j}^{new} ( \theta ^c, \vec{c^c} ), Y_{j}^{new} ( \theta ^c, \vec{c^c} )$ in terms of $Q_\pm$ provides, eventually, the asymptotic behaviours of the $T$- and $Y$-functions. These can be used in order to derive from (\ref {closed-Y-system-cft-prov}) a system of TBA equations. 

\medskip

It can be of interest to reinterpret all the above functional relations and asymptotic behaviours in terms of more natural variables. Then, 
with a little abuse of notation, let us define 
\be
c_m = c_m^{c} e^{\frac{2N-m}{1+N}\theta ^{c}} \, , \quad 1 \leq m \leq 2N \, .
\ee
Neglecting also the superscript $c$ ($\psi^c\rightarrow \psi$ and $\theta^c \rightarrow \theta$) the ODE (\ref {ODE-cft}) becomes
\be
\left [ -\frac{d^2}{dx^2}+\frac{l(l+1)}{x^2}+\sum _{m=1}^{2N} c_m x^m \right ]
\psi (x)= e^{\frac{2N}{1+N}\theta}\psi (x) \, , \label {ODE-cft-2}
\ee
where the rapidity $\theta$ enters only the energy $E\equiv -c_0\equiv e^{\frac{2N}{1+N}\theta}$ and potential
\be
\label{potential2}
V(x)=\sum _{m=1}^{2N} c_m x^m \, .
\ee
Now, the key-point is to rewrite all the above relations in these variables: in fact, these would be the actual relations derived by applying the above off-critical procedure directly in the conformal case with ODE (\ref {ODE-cft-2}). As usual, we start by the fundamental conformal symmetry (\ref{Omegac}), which would serve to derive the functional equations and takes the form, deleting the subscript $c$: 
\be
\hat \Omega: \quad x\rightarrow x e^{\frac{i\pi}{1+N}} \, ; \quad
c_m \rightarrow c_m e^{i\pi\frac{2N-m}{1+N}} \, , \, 1 \leq m \leq 2N \, ; \quad c_0=-e^{\frac{2N}{1+N}\theta } \rightarrow c_0 e^{-\frac{2i\pi}{1+N}} \, .
\ee
As noticed above, since $\Pi=(\Omega)^{-1}$ these two symmetries are not independent and only one can be used effectively. Moreover, the periodicity transformation $\theta \rightarrow  \theta + i\pi \frac {1+N}{N}, c_n^c \rightarrow c_n^c e^{\frac {i\pi n}{N}}$, $0\leq n \leq 2N$, leaves invariant the variables $c_n$. It is then natural in the conformal case to change independent variables from $c_n^c$ to $c_n$ and then to define the functions  
\be
\tilde T_j (c_0, c_m)=T_j (\theta , c_m^c) \, , \quad \tilde Y_j (c_0, c_m)=Y_j (\theta , c_m^c) \, , \quad c_0=-e^{\frac {2N\theta}{1+N}2N} \, , \quad c_m=c_m^c e^{\frac {2N-m}{1+N}\theta}
\ee
Having in mind the idea of writing TBA equations, we consider the $Y$-system, which, as a consequence of (\ref  {Y-syscft}), in terms  of the functions $\tilde Y_j$ reads
\be
\tilde Y_j \left ( c_0e^ {-\frac {i\pi}{1+N}}, c_m e^{-\frac {i\pi m}{N}-\frac {i\pi(2N-m)}{2N(1+N)}} \right )
\tilde Y_j \left ( c_0e^ {\frac {i\pi}{1+N}}, c_m e^{\frac {i\pi(2N-m)}{2N(1+N)}} \right )=
\left [1+\tilde Y_{j-\frac {1}{2}}\left (c_0, c_m e^{-\frac {i\pi m}{N}} \right ) \right ] 
\left [1+\tilde Y_{j+\frac {1}{2}}\left (c_0, c_m  \right ) \right ]  \, .
\ee 
In terms of new parameters
\be
\tilde c_m =c_m e^{\frac {i\pi (2N-m)}{2N}}
\ee
the $Y$-system can be put in a more symmetric form
\ba
&& \tilde Y_j \left ( c_0e^ {-\frac {i\pi}{1+N}}, \tilde c_m e^{\frac {i\pi(2N-m)}{2(1+N)}} \right )
\tilde Y_j \left ( c_0e^ {\frac {i\pi}{1+N}}, \tilde c_m e^{-\frac {i\pi(2N-m)}{2(1+N)}} \right )= \nonumber \\
&&=
\left [1+\tilde Y_{j-\frac {1}{2}}\left (c_0, \tilde c_m e^{\frac {i\pi (2N-m)}{2N}} \right ) \right ] 
\left [1+\tilde Y_{j+\frac {1}{2}}\left (c_0, \tilde c_m e^{-\frac {i\pi (2N-m)}{2N}} \right )  \right ] \, .
\ea 
With the aim to write TBA equations, we introduce
\be
{\cal Y}_{j,k}(\theta , \tilde c_m)=\tilde Y_j \left (c_0, \tilde c_m e^{\frac {i\pi (m-2N)j}{N}}e^{-\frac {i\pi(2N-m)k}{2(1+N)}} \right ) \, , \quad c_0=-e^{\frac {2N\theta}{1+N}} \, , 
\ee
which satisfy
\be
{\cal Y}_{j,k-1}\left (\theta -\frac {i\pi}{2N}  , \tilde c_m \right ) {\cal Y}_{j,k+1}\left (\theta +\frac {i\pi}{2N}  , \tilde c_m \right ) = \left [1+{\cal Y}_{j-\frac {1}{2},k}(\theta , \tilde c_m ) \right ]  \left [1+{\cal Y}_{j+\frac {1}{2},k}(\theta , \tilde c_m ) \right ] 
\label {Y-syst-conf}
\ee
and have the following invariance properties 
\be
{\cal Y}_{j,k}\left (\theta +i\pi \frac {N+1}{N}  , \tilde c_m \right )= {\cal Y}_{j,k}(\theta , \tilde c_m) 
\, , \quad {\cal Y}_{j,k}(\theta, \tilde c_ {m})={\cal Y}_{j,k+4(N+1)}\left (\theta, \tilde c_ {m}\right ) \, . \label {inv-Y-system-cft}
\ee
The first invariance descends from quasiperiodicity of $Y_j$ since the combined transformation $\theta \rightarrow i\pi \frac {N+1}{N}, c_m^c \rightarrow c_m^c e^{\frac {i\pi m}{N}}$ leaves invariant $\tilde c_m$. The second property is $Z_{4N+4}$ symmetry. 
This symmetry increases to $Z_{2N+2}$ either if $N$ is semi-integer and, for $m$ even, $c_m=0$ or
if $N$ is integer and, for $m$ odd, $c_m=0$.

It is worth also mentioning that one can express also the 'new' $T$ and $Y$-functions
in terms of the 'natural' parameters $c_n$:
%Defining ($c_0=-e^{\frac {2N\theta}{1+N}}$)
\be
\tilde T_j^{new} (c_0, c_m)=T_j^{new} (\theta , c_m^c) \, , \quad \tilde Y_j^{new} (c_0, c_m)=Y_j^{new} (\theta , c_m^c) \, , \quad c_0=-e^{\frac {2N\theta}{1+N}2N} \, , \quad c_m=c_m^c e^{\frac {2N-m}{1+N}\theta}
\ee
%\tilde T_{j}^{new} (c_0,c_1,...,c_{2N})=\tilde T_{j}^{new} \left ( \{ c_n \} \right )=T_{j}^{new} \left (\theta ,  \{ c_m^c = c_m e^{-\frac{2N-m}{1+N}\theta }, 1\leq m\leq 2N \}\right )
%\ee
%and 
%\be
%\tilde Y_{j}^{new} (c_0,c_1,...,c_{2N})=\tilde Y_{j}^{new} \left ( \{ c_n \} \right )=Y_{j}^{new} \left (\theta ,  \{ c_m^c = c_m e^{-\frac{2N-m}{1+N}\theta }, 1\leq m\leq 2N \}\right ) \, ,
%\ee
%in first instance, the $T$-system (\ref {closed-T-system-cft-prov}) can be written in terms of $\tilde T_{j}^{new} $ as
%\be
%\tilde T_{j}^{new} \left ( \{ c_n e^{-\frac {(2N-n)i\pi}{2(1+N)}} \} \right ) \tilde T_{j}^{new} \left (  \{ c_n e^{\frac {(2N-n)i\pi}{2(1+N)}}\} \right )= 1 +\tilde T _{j+\frac{1}{2}}^{new}\left ( \{ c_n \}\right )\tilde T _{j-\frac{1}{2}}^{new}\left ( \{ c_n \}\right ) \, .
%\label{closed-T-system-cft}
%\ee
%and, in second instance, 
Then, in particular, the $Y$-system (\ref{closed-Y-system-cft-prov}) can be written in terms of $\tilde Y_{j}^{new} $ as 
\be
\tilde Y_j^{new}\left (   c_n e^{-\frac {(2N-n)i\pi}{2(1+N)}}  \right ) \tilde Y_j^{new}\left (   c_n e^{\frac {(2N-n)i\pi}{2(1+N)}} \right )=
\left [ 1+ \tilde Y_{j-\frac{1}{2}}^{new}( c_n ) \right ]  \left [ 1+ \tilde Y_{j+\frac{1}{2}}^{new}( c_n ) \right ] 
\label{closed-Y-system-cft} \, .
\ee
Defining 
\be
{\cal Y}_{j,k} ^{new}(\theta , c_m)=\tilde Y_j ^{new}\left (c_0, c_m e^{-\frac {i\pi(2N-m)k}{2(1+N)}} \right ) \, , 
\ee
these functions satisfy the functional relations
\be
{\cal Y}_{j,k-1}^{new}\left (\theta -\frac {i\pi}{2}  , c_m \right ) {\cal Y}_{j,k+1}^{new}\left (\theta +\frac {i\pi}{2}  , c_m \right ) = \left [1+{\cal Y}_{j-\frac {1}{2},k}^{new}(\theta , c_m ) \right ]  \left [1+{\cal Y}_{j+\frac {1}{2},k}^{new}(\theta , c_m ) \right ] \, . \label {Y-syst-new-conf}
\ee
Moreover, they have the invariance properties 
\be
{\cal Y}_{j,k}^{new}\left (\theta +i\pi \frac {N+1}{N}  , \tilde c_m \right )= {\cal Y}_{j,k}^{new} (\theta , \tilde c_m) 
\, , \quad {\cal Y}_{j,k}^{new}(\theta, \tilde c_ {m})={\cal Y}_{j,k+4(N+1)}^{new}\left (\theta, \tilde c_ {m}\right ) \, . \label {inv-Y-system-new-cft}
\ee
The two relations (\ref {Y-syst-new-conf}, \ref {inv-Y-system-new-cft}) coincide in form with (\ref {Y-syst-conf}, \ref {inv-Y-system-cft}).

\subsection {Leading orders and TBA}

Now, if we want to convert the functional relations (\ref {Y-syst-conf}) into TBA equations, we need to compute the asymptotic limit at large $\theta $ of  ${\cal Y}_{j,k}$. This in general is laborious, however possible, since the quantity $e^\theta w_0 \left (c_m ^c= c_m e^{-\frac{2N-m}{1+N}\theta }\right )$ as a functions of $c_m$ will contain many different fractional powers of $e^\theta$. With the aim of just giving an instructive example, we then restrict to the 
case $N=\frac  {4p+3} {2}$, with $p$ non-negative integer and with the only non-zero module $c_ {N-\frac  {1} {2}}$.

The asymptotic limit of ${\cal Y}_{j,k}$ when Re$\theta \rightarrow +\infty$  derives from the conformal limit of (\ref {Y-asy-general}) in which, however, the quantity $w_0 e^\theta$ is expressed in terms of $c_{N-\frac  {1} {2}}$ as follows. We start from 
\be
w_0\left (c_{N-\frac {1}{2}} ^c \right )=-\int _0^{+\infty}dx \left [ \sqrt {x^{2N}+c_{N-\frac {1}{2}}^c e^{i\pi \frac {2N+1}{4N}}x^{N-\frac {1}{2}}+1} -x^N - \frac {c_{N-\frac {1}{2}}^c e^{i\pi \frac {2N+1}{4N}}}{2\sqrt {x}}\right ]
\ee
and insert in it $c_{N-\frac  {1} {2}}^c =c_{N-\frac  {1} {2}} e^{-\frac {(2N+1)\theta}{2(N+1)}}$. Then, 
in the limit Re$\theta \rightarrow +\infty$ and $|\textrm {Im}\theta | <\pi \frac{N+1}{2N}$, 
\ba
&& -w_0\left (c_{N-\frac {1}{2}}^c \right ) e^\theta  \simeq e^\theta \int _{0}^{+\infty} dx \left [\sqrt {x^{2N}+1}-x^{N}+ \frac {e^{-\frac {(2N+1)\theta}{2(N+1)}} c_{N-\frac  {1} {2}} e^{i\pi \frac {2N+1}{4N}}x^{N-\frac  {1} {2}}}{2\sqrt {x^{2N}+1}}-\frac {e^{-\frac {(2N+1)\theta}{2(N+1)}} c_{N-\frac {1}{2}} e^{i\pi \frac {2N+1}{4N}}}{2\sqrt {x}}\right ] = \nonumber \\
&&= -e^\theta \frac {\Gamma \left (-\frac {1}{2}-\frac {1}{2N} \right ) \Gamma \left (\frac {1}{2N} \right )}{4N\sqrt {\pi}}
-e^{\frac {\theta}{2(N+1)}}c_{N-\frac {1}{2}}e^{i\pi \frac {2N+1}{4N}}\frac {\Gamma \left (1-\frac {1}{4N}\right )\Gamma \left (\frac {1}{2}+\frac {1}{4N}\right )}{\sqrt {\pi}} \label {Qpmlimit} \, ,  = \\
&&= -e^\theta A_N - e^{\frac {\theta}{2(N+1)}}c_{N-\frac {1}{2}}e^{i\pi \frac {2N+1}{4N}} B_N \, . \nonumber 
\ea
%whilst, if Re$\theta  \rightarrow -\infty$ and $|\textrm {Im}\theta | < \pi \frac{N+1}{2N}$, it 
%\be
%\ln Q_\pm \left (\theta +i\pi \frac{N+1}{2N}; c_{N-\frac {1}{2}}^c \right ) \simeq \textrm {constant} \, . 
%\label {Qpmlimitbis}
%\ee
The quantity $\ln {\cal Y}_{j,k}(\theta , \tilde c_{N-\frac {1}{2}})=\varepsilon _{j,k}(\theta , \tilde c_{N-\frac {1}{2}}) $ when Re$\theta \rightarrow +\infty$ 
\ba
&& \varepsilon _{j,k} (\theta , \tilde c_{N-\frac {1}{2}}) \simeq -4 e^\theta A_N \cos \frac {\pi}{2N}\sin \frac {\pi j}{N}-2e^{\frac {\theta}{2(N+1)}}B_N  \tilde c_{N-\frac {1}{2}}e^{\frac {i\pi}{4N}-i\pi \frac {(2N+1)k}{4(N+1)}-i\pi j }\left (\cos \frac {\pi}{2N}e^{\frac {i\pi j}{2N}}+e^{-2i\pi j}e^{-\frac {i\pi j}{2N}} \right ) = \nonumber \\
&& \equiv {\cal A}_{j,k}^c(\theta , \tilde c_{N-\frac {1}{2}})
\ea
Removing in the notation the dependence on $\tilde c_{N-\frac {1}{2}}$, this function satisfies the functional equation
\be
{\cal A}_{j,k-1}^c\left (\theta -\frac {i\pi}{2N}\right )+
{\cal A}_{j,k+1}^c\left (\theta +\frac {i\pi}{2N}\right )-
{\cal A}_{j-\frac {1}{2} ,k}^c\left (\theta \right )-
{\cal A}_{j+\frac {1}{2} ,k}^c\left (\theta \right )=0
\ee
and together with $\varepsilon$
\ba
&& \left [\varepsilon_{j,k-1}^c\left (\theta -\frac {i\pi}{2N}\right )-{\cal A}_{j,k-1}^c\left (\theta -\frac {i\pi}{2N}\right )\right ] 
+\left [\varepsilon_{j,k+1}^c\left (\theta +\frac {i\pi}{2N}\right )-{\cal A}_{j,k+1}^c\left (\theta +\frac {i\pi}{2N}\right )\right ] = \nonumber \\
&& = \ln \left ( e^{-{\cal A}_{j-\frac {1}{2},k}\left ( \theta  \right )} +e^{\varepsilon _{j-\frac {1}{2},k}\left ( \theta  \right )-{\cal A}_{j-\frac {1}{2},k}\left ( \theta  \right )} \right ) +
\ln \left (e^{-{\cal A}_{j+\frac {1}{2},k}\left ( \theta  \right )} +e^{\varepsilon _{j+\frac {1}{2},k}\left ( \theta  \right )-{\cal A}_{j+\frac {1}{2},k}\left ( \theta  \right )} \right )
\ea
We perform the discrete Fourier sums (the integer $l$ is defined mod $2N+2$) 
\ba
\chi _{j,l}  (\theta , \vec{c} )&=& \sum _{k=1}^{2N+2} e^{\frac {2i\pi kl}{2N+2}} \varepsilon _{j,k}\left (\theta \right ) \, , \quad \hat {\cal A}_{j,l}(\theta , \vec{c} )=  \sum _{k=1}^{2N+2} e^{\frac {2i\pi kl}{2N+2}} {\cal A}_{j,k} \left (\theta \right )  \nonumber \\
\Lambda _{j,l} (\theta , \vec{c})  &=& \sum _{k=1}^{2N+2} e^{\frac {2i\pi kl}{2N+2}} 
\ln \left (e^{-{\cal A}_{j,k}\left ( \theta  \right )} +e^{\varepsilon _{j,k}\left ( \theta  \right )-{\cal A}_{j,k}\left ( \theta  \right )} \right ) 
\, , \label {disc-fou-tra-cft}
\ea
and find the equations
\ba
&&  \left [\chi_{j,l}\left ( \theta  -\frac {i\pi}{2N},  \vec{c}\right )-\hat {\cal A}_{j,l}\left ( \theta  -\frac {i\pi}{2N},  \vec{c} \right )\right ] e^{\frac {2i\pi l}{2N+2}} +  \left [\chi_{j,l}\left ( \theta  +\frac {i\pi}{2N},  \vec{c}\right )-\hat {\cal A}_{j,l}\left ( \theta  +\frac {i\pi}{2N},  \vec{c} \right )\right ]e^{-\frac {2i\pi l}{2N+2}}= \nonumber \\
&& \Lambda _{j-\frac {1}{2},l}(\theta , \vec{c} )+
\Lambda _{j+\frac {1}{2},l}(\theta , \vec{c} ) \label {funct-rel-four-cft} \, . 
\ea
The associated TBA equations are
\be
\chi_{j,l}(\theta , \vec{c})=\hat {\cal A}_{j,l}(\theta , \vec{c})+\int _{-\infty}^{+\infty} \frac {d\theta '}{2\pi} 
{\cal K}^{c}_l(\theta -\theta ') 
\left [  \Lambda _{j-\frac {1}{2},l}(\theta ', \vec{c} ) + \Lambda _{j+\frac {1}{2},l}(\theta ', \vec{c} ) \right ] \, , \label {TBA-final-general-chi-cft}
\ee
where
\be
{\cal K}^{c}_l(\theta)=\frac {N}{2\pi}\frac {e^{\frac {2lN\theta}{N+1}}}{\cosh N\theta} \, , \quad |l| \leq \frac {N}{2}+\frac {1}{4} \, \quad ; \quad {\cal K}^{c}_l(\theta)=-\frac {N}{2\pi}\frac {e^{\frac {2lN\theta}{N+1}-2N\theta}}{\cosh N\theta}\, ,   \quad \frac {N}{2}+\frac {5}{4}\leq l \leq \frac {3N}{2}+\frac {3}{4} 
\ee
%xxxx

%Asintotici difficili, one module, N=3/2.

%alla fine il caso universale.

\medskip

Again, the particular case $l=0, N=3/2$ is of interest and can be treated in a simpler way. The only non-zero module is $c_1$ and the symmetry is $Z_5$.
Since   $Y_{1}\left (\theta , \vec{c^c}\right )=Y_{\frac {1}{2}} \left ( \theta  , \vec{c^c}^{R^{-1}} \right )$ there is only one functional equation which reads
\be
{\cal Y}_{\frac {1}{2},k-1}\left (\theta -\frac {i\pi}{3},\tilde c_1\right ) {\cal Y}_{\frac {1}{2},k+1} \left (\theta +\frac {i\pi}{3},\tilde c_1\right )=1+{\cal Y}_{\frac {1}{2},k}(\theta , \tilde c_1)
\label {cal Y-system} \, 
\ee
Since  ${\cal Y}_{\frac {1}{2},k}(\theta , \tilde c_1)=\tilde Y_{\frac {1}{2}}\left (c_0 , \tilde c_1 e^{-\frac {2i\pi}{3}} 
e^{-\frac {2i\pi k}{5}}\right )=\tilde Y_{\frac {1}{2}}\left (c_0 , c_1  e^{-\frac {2i\pi k}{5}}\right )=Y_{\frac {1}{2}}\left (\theta , c_1^c e^{-\frac {2i\pi k}{5}}=c_1 e^{-\frac {4\theta }{5}}e^{-\frac {2i\pi k}{5}}\right )$,
when Re$\theta  \rightarrow \pm \infty$ and $|\textrm {Im}\theta | <\frac {2\pi}{3}$, 
\be
\ln {\cal Y}_{\frac {1}{2},k}(\theta , \tilde c_1) \simeq  -w_0\left (c_1^c e^{\frac{2i\pi}{3}}e^{-\frac {2i\pi k}{5}}=c_1 e^{-\frac {4\theta }{5}}
e^{\frac{2i\pi}{3}}e^{-\frac {2i\pi k}{5}} \right)e^{\theta -\frac {i\pi}{6}}-w_0\left (c_1^ce^{-\frac {2i\pi k}{5}}=c_1 e^{-\frac {4\theta }{5}}e^{-\frac {2i\pi k}{5}}\right)e^{\theta +\frac {i\pi}{6}}
\label {Y1-asy-offcrit2}
\ee
where
\be
w_0(c_1^c)= -\int _{0}^{+\infty} dx \left [ \sqrt {x^3+c_1^c e^{\frac{2i\pi}{3}}x+1}-x^{\frac {3}{2}}-\frac {c_1^c e^{\frac {2i\pi}{3}}}{2\sqrt {x}} \right ] \, , 
\ee
In terms of $c_1$ one has
\ba
w_0\left (c_1^c=c_1e^{-\frac {4\theta }{5}}\right ) e^\theta &=& -e^\theta \int _0^{+\infty} dx \left ( \sqrt {x^3+1}-x^{\frac {3}{2}} \right ) -e^{\frac {\theta }{5}}c_1 e^{\frac {2i\pi}{3}}\int _0 ^{+\infty} dx \left ( \frac {x}{2\sqrt {x^3+1}}-\frac {1}{2\sqrt {x}} \right )  \nonumber \\
&=& -e^\theta \frac {\sqrt {\pi}}{3}\frac {\Gamma (1/3)}{\Gamma (11/6)}+\frac {e^{\frac {\theta}{5} } e^{\frac{2i\pi}{3}}c_1}{\sqrt {\pi}}\Gamma (2/3) \Gamma (5/6) \, , 
\label {w0fract}
\ea
Then, when Re$\theta  \rightarrow \pm \infty$ and $|\textrm {Im}\theta | <\frac {2\pi}{3}$,
\be
\ln {\cal Y}_{\frac {1}{2},k}(\theta , \tilde c_1) \simeq \sqrt {\pi/3}\frac {\Gamma (1/3)}{\Gamma (11/6)}e^\theta - 
\frac  {\tilde c_1 e^ {-\frac  {2i\pi k} {5}}\Gamma (2/3) \Gamma (5/6)} {\sqrt  {\pi}}\left (\frac  {3i} {2}+\frac  {\sqrt  {3}} {2} \right ) e^ {\frac  {\theta} {5}}\equiv {\cal A}_{\frac {1}{2},k}^c(\theta , \tilde c_1) \label {Y1-asy-crit}
\ee
Functional relations (\ref  {cal Y-system}) and asymptotic behaviour (\ref {Y1-asy-crit}) match the ones obtained by Masoero in \cite {MAS}. 
The TBA equations obtained in that paper are shown to be equivalent to (\ref {TBA-final-general-chi-cft}) specialised to $N=3/2$ and $j=1/2$ by using a strategy similar to the one outlined in Appendix \ref {app4}.

\section{Beyond the polymers: small $l$}
\label{lzero}
\setcounter{equation}{0}

The $Y$-systems (\ref {Ydef}) for 'new' $Y$-functions has a relevant application in the limit in which $N$ is generic, but $l$ is small: the realisation of the solutions of the TBA-like equations appearing in \cite{FS,CFIV} when computing Witten index in integrable supersymmetric ${\cal N}=2$ theories and which Al. Zamolodchikov used to describe polymers in \cite{Zam-poly}, in terms of $Q$ functions, derived (in the spirit of this paper) from the solution of the linear problem (\ref {linhat}). Results of this Section extend findings of J. Suzuki \cite{JSUZ}, related to second order ODEs with potentials $(x-E)^k$ and to third order ODEs.

\medskip

We go back to constructions done in Section \ref  {funrel} in the context of scattering amplitudes in $AdS_3$ and move from the $Y$-system (\ref {Ydef})
\ba
Y_j^{new}\left ( \theta - \frac{i\pi }{2}, \vec{c}\right ) Y_j^{new}\left ( \theta + \frac{i\pi}{2} , \vec{c}\right )&=&
\left [ 1+ Y_{j-\frac{1}{2}}^{new}(\theta , \vec{c}) \right ]  \left [ 1+ Y_{j+\frac{1}{2}}^{new}(\theta , \vec{c}) \right ] \, , \quad j=1,...,N-\frac{1}{2} \nonumber \\
Y_N^{new}\left ( \theta - \frac{i\pi }{2}, \vec{c}\right ) Y_N^{new}\left ( \theta + \frac{i\pi}{2} , \vec{c}\right )&=&\left [ 1+ Y_{N-\frac{1}{2}}^{new}(\theta , \vec{c}) \right ] \left [ 1+ e^{2\pi i \left (l+\frac{1}{2}\right )}\hat Y (\theta , \vec{c}) \right ]
 \left [ 1+ e^{-2\pi i \left (l+\frac{1}{2}\right )}\hat Y (\theta , \vec{c}) \right ]  \nonumber \\
\hat Y \left ( \theta - \frac{i\pi }{2}, \vec{c}\right ) \hat Y \left ( \theta + \frac{i\pi }{2}, \vec{c}\right )&=& 1+Y_N^{new}(\theta ,  \vec{c} ) \label {Ydef2} \, ,
\ea
in the small $l$ limit.
As claimed at the end of Section \ref  {funrel}, when $l=0$, $Y_N^{new} (\theta ,  \vec{c})=0$. In addition, by using the $QQ$-system (\ref {QQalt}), periodicity (\ref {Qperiod}) and definition (\ref {tnewQ}), one gets also the condition $\hat Y(\theta ,  \vec{c} )=1$ when $l=0$.
This entails that if $l$ is small we can parametrise
\be
Y_N^{new} (\theta )=-2\pi l y_N (\theta) + O(l^2) \, , \quad \hat Y(\theta)=1-2\pi l \hat y(\theta)+ O(l^2) \, .
\label {yNhaty}
\ee
Then, the last two equations of (\ref {Ydef2}) become
\ba
&& y_N \left (\theta + \frac{i\pi}{2}\right ) y_N \left (\theta - \frac{i\pi}{2}\right )=f_N(\theta) [1+\hat y ^2(\theta)] \, , \\
&& \hat y\left (\theta + \frac{i\pi}{2}\right )+ \hat y\left (\theta - \frac{i\pi}{2}\right )=y_N(\theta) \, , 
\ea
where the only reminiscence of the rest of (\ref {Ydef2}) is the function $f_N(\theta)=1+Y_{N-\frac{1}{2}}^{new}(\theta)$ computed at $l=0$, which is supposed to be known. A simple redefinition $y_N(\theta)=e^{-\epsilon (\theta)}$ and 
$\hat y(\theta)=\eta (\theta)$ yields the familiar functional relations
\ba
&& \epsilon \left (\theta + \frac{i\pi}{2}\right )+ \epsilon \left (\theta - \frac{i\pi}{2}\right )=-\ln f_N(\theta) -\ln [1+\eta ^2 (\theta ) ] \, ,\\
&&\eta \left (\theta + \frac{i\pi}{2}\right )+ \eta \left (\theta - \frac{i\pi}{2}\right )=e^{-\epsilon (\theta)} \, , 
\ea
which, using the property
\be 
\frac {1}{\cosh \left (x+\frac {i\pi}{2}-i\varepsilon \right )}+
\frac {1}{\cosh \left (x-\frac {i\pi}{2}+i\varepsilon \right )}=2\pi \delta (x)
\label  {shift-cosh} \, , 
\ee
can be translated into the TBA-like equations  \cite{FS,CFIV}
\ba
\epsilon (\theta)&=&2u(\theta)-\int _{-\infty}^{+\infty} \frac{d\theta '}{2\pi} \frac{\ln \left [ 1+\eta (\theta ')^2\right ]}{\cosh (\theta -\theta ')} \, , \nonumber \\
\eta (\theta)&=&g(\theta )+ \int _{-\infty}^{+\infty} \frac{d\theta '}{2\pi} \frac{e^{-\epsilon (\theta ')}}{\cosh (\theta -\theta ')} \label {TBA-off-2} \, ,
\ea
where we introduced two functions $u(\theta)$ and $g(\theta)$, which have to satisfy the functional relations $2u\left (\theta +\frac {i\pi}{2}\right )+2u\left (\theta -\frac {i\pi}{2}\right )=-\ln [1+Y_{N-\frac{1}{2}}^{new}(\theta)]$ and $g\left (\theta +\frac {i\pi}{2}\right )+g\left (\theta -\frac {i\pi}{2}\right )=0$. The first relation gives
\be
2u\left (\theta +\frac {i\pi}{2}\right )+2u\left (\theta -\frac {i\pi}{2}\right )=-\ln [1+T^{new}_{N-1}(\theta) T^{new}_N(\theta)]=-\ln T_{N-\frac{1}{2}}^{new} \left (\theta +\frac {i\pi}{2}\right) -\ln T_{N-\frac{1}{2}}^{new}\left (\theta -\frac {i\pi}{2}\right ) \, , 
\ee
which provides the solution
\be
2u(\theta)=-\ln T_{N-\frac{1}{2}}^{new} (\theta ) \, , 
\ee
up to zero modes of the shift operator: $f(\theta) \rightarrow f\left (\theta +\frac {i\pi}{2}\right)+f\left (\theta -\frac {i\pi}{2}\right )$. 
Alternative expression of $u$ in terms of $T_{\frac{1}{2}}^{new}$, obtained by using periodicity properties of $Q$ functions contained in the definition of $T_j^{new}$ is 
\be
2u(\theta)=-\ln \left [ e^{-2iN \Phi (\theta +i\pi N +2i\pi, \vec{c})}T_{\frac{1}{2}}^{new}(\theta +i\pi N+i\pi, \vec{c}) \right ] + \textrm{zero} \ \ \textrm{modes}\,  .
\ee
On the other hand, $g(\theta)$ is given only by zero modes of the shift operator $f(\theta) \rightarrow f\left (\theta +\frac {i\pi}{2}\right)+f\left (\theta -\frac {i\pi}{2}\right )$. Zero modes are fixed by analysis of the asymptotic behaviour of $\epsilon $ and $\eta$ and  this gives the complete form of $u$ and $g$. 

To summarise, the $l\rightarrow 0$ limit of $Y$-system (\ref {Ydef2}) gives (\ref {TBA-off-2}), which is the TBA of \cite{FS,CFIV}. Now, expressing the $Y$-functions in terms of $Q$-functions and going to the limit $l\rightarrow 0$ provides formul{\ae} for the functions $\epsilon $, $\eta$, $u$ and $g$ appearing in the TBA equations (\ref {TBA-off-2}). For instance, by simply using periodicity of $Q$ functions, $Y_N^{new}$ enjoys the expression
\ba
Y_N^{new}(\theta , \vec{c})&=&i \tan \pi l \Bigl [Q_+(\theta +i\pi N+2i\pi,\vec{c})Q_{-}(\theta +i\pi N,\vec{c})- \nonumber \\
&-& Q_-(\theta +i\pi N+2i\pi,\vec{c})Q_{+}(\theta +i\pi N,\vec{c})\Bigr ] \, Q_+(\theta +i\pi N + i\pi,\vec{c}) Q_-(\theta +i\pi N + i\pi,\vec{c}) \nonumber \, .
\ea
The limit $l\rightarrow 0$ identifies $y_N$ according to the first of (\ref {yNhaty}):
\ba
y_N(\theta , \vec{c})=e^{-\epsilon (\theta)}&=&\frac{1}{2i} \Bigl [Q_+(\theta +i\pi N+2i\pi,\vec{c})Q_{-}(\theta +i\pi N,\vec{c})- \nonumber \\
&-& Q_-(\theta +i\pi N+2i\pi,\vec{c})Q_{+}(\theta +i\pi N,\vec{c})\Bigr ] \, Q_+(\theta +i\pi N + i\pi,\vec{c}) Q_-(\theta +i\pi N + i\pi,\vec{c})  \nonumber \\
&=& - T_{\frac{1}{2}}^{new}(\theta +i\pi N+i\pi, \vec{c}) \, Q_+(\theta +i\pi N + i\pi,\vec{c}) Q_-(\theta +i\pi N + i\pi,\vec{c}) \label {yNexpr} \, , 
\ea
where all the functions $Q_{\pm}, T_{\frac{1}{2}}^{new}$ are evaluated at $l=0$. On the other hand for $\hat Y (\theta ,\vec{c} )$ one has the exact expression
\ba
\hat Y (\theta ,\vec{c} )&=&-\frac {1}{2\cos \pi l} \Bigl [ e^{i\pi l} Q_+\left ( \theta +i\pi N + \frac{3i\pi}{2},\vec{c} \right )
Q_-\left ( \theta +i\pi N + \frac{i\pi}{2},\vec{c} \right ) + \nonumber \\
&+&  e^{-i\pi l} Q_+\left ( \theta +i\pi N + \frac{i\pi}{2},\vec{c} \right )
Q_-\left ( \theta +i\pi N + \frac{3i\pi}{2},\vec{c} \right ) \Bigr ] \, , 
\ea
which gives when $l \rightarrow 0$ the function $\hat y$ in terms of $Q_{\pm}$ evaluated at $l=0$:
\ba
\hat y(\theta )=\eta (\theta)&=&\frac{i}{2}\Bigl [ Q_+\left ( \theta +i\pi N + \frac{3i\pi}{2},\vec{c} \right )
 Q_-\left ( \theta +i\pi N + \frac{i\pi}{2},\vec{c} \right )- \nonumber \\
 &-& Q_-\left ( \theta +i\pi N + \frac{3i\pi}{2},\vec{c} \right ) Q_+\left ( \theta +i\pi N + \frac{i\pi}{2},\vec{c} \right ) \Bigr ] \, .
\label {yexpr}
\ea
%Finally, the analysis of the asymptotic behaviour of $\epsilon, \eta$ when $\theta \rightarrow \pm \infty$ fixes the zero modes appearing in $u$ and $g$ and then the form of these functions:
%\be
%2u(\theta)=-\ln \left [ e^{-2iN \Phi (\theta +i\pi N +2i\pi, \vec{c})}T_{\frac{1}{2}}^{new}(\theta +i\pi N+i\pi, \vec{c}) \right ] \, , \quad g(\theta )=0 \, .
%\ee

\subsection {Realisation of TBA functions in terms of integral kernels}
\label{real}

The aim of this part is to provide a realisation for the TBA functions $e^{-\epsilon (\theta)}$ and $\eta (\theta)$ in terms, finally, of the function $T_{\frac{1}{2}}^{new}$ and a peculiar integral kernel. This fact is a direct consequence of the classical/quantum integrable theories correspondence in its massive form, the off-critical ODE/IM correspondence. Actually, it is crucial to use the inverse procedure, from quantum to classical, as illustrated in \cite {FRletter}, by going through three steps. The first step is to recall  relation (2.44) of \cite {FRletter}, which expresses the $Q$-functions in terms of wave functions $\psi _\pm$:
\be
 \lim _{w\rightarrow w_0(\vec{c})} \bigl (w-w_0(\vec{c})\bigr )^{\pm l}\psi _{\pm}(w',\bar w'|\theta)= D_{\pm}\bigl (w_0(\vec{c})\bigr )e^{\mp \theta l}Q_{\pm} \left (\theta + \frac{i\pi}{2}+\frac {i\pi}{2N}, \vec{c} \right )  \, , \quad w'=-iw \, , \ \bar w'=i\bar w \, .
 \label{waveQdef}
\ee
In (\ref {waveQdef}) $D_{\pm}(w_0(\vec{c}))$ are quantities not depending on $\theta$ which can be fixed for instance by matching the asymptotic behaviour of the left hand side with that of  $Q_{\pm} \left (\theta + \frac{i\pi}{2}+\frac {i\pi}{2N}, \vec{c} \right )$.

The second step is to elaborate expression (2.46) in \cite {FRletter}, to find the wave functions as a series 
\be
\psi_{\pm}(w',\bar w'|\theta )=-e^{-iw'e^{\theta}+i\bar w' e^{-\theta}}\sqrt{2} e^{v(\theta)-\frac{\theta}{2}}\left ( \frac{1}{\hat I\pm \hat K_{v}} E_{v} \right ) (\theta) \, , \label{psiexact}
\ee
where
\be
e^{-2 v(\theta)}=e^{-2iw'e^{\theta}+2i\bar w' e^{-\theta}}
T_{\frac{1}{2}}^{new}\left ( \theta + \frac{i\pi}{2}+\frac {i\pi}{2N}, \vec{c}\right ) \, , \quad \frac{ E_{v}(\theta)}{\sqrt{2}}=e^{-v(\theta)}e^{\frac{\theta}{2}} \label{ushort}
\ee
and the kernel $\hat K_v$ acts as 
\be
(\hat K_{v}f)(\theta)=\int_{-\infty}^{+\infty} \frac{d\theta '}{4\pi} \frac{e^{-v(\theta)-v(\theta ')}}{\cosh \frac{\theta -\theta'}{2}}f(\theta ') \, . \label{Kshort}
\ee
The last step is to use (\ref {psiexact}, \ref {waveQdef}) to find an expression for $Q$ functions. Since the interest is in finding a realisation for $e^{-\epsilon (\theta)}$ and $\eta (\theta)$, written as (\ref {yNexpr}, \ref {yexpr}), the expressions for the $Q$ functions are needed for $l=0$. In this case from (\ref {psiexact}, \ref {waveQdef}) one finds
\be
l=0 \quad \Rightarrow \quad Q_{\pm} \left (\theta + \frac{i\pi}{2}+\frac {i\pi}{2N}, \vec{c} \right ) =-\frac{\sqrt{2}}{D_{\pm}\bigl (w_0(\vec{c})\bigr )e^{\frac{\theta}{2}}}\frac{1}{\sqrt{T_{\frac{1}{2}}^{new}\left (\theta + \frac{i\pi}{2}+\frac {i\pi}{2N}, \vec{c} \right )}}
\left ( \frac{1}{\hat I\pm \hat K_{v_0}} E_{v_0}\right ) (\theta)
\label{Qlzero}
\ee
In this case 
\be
D_{\pm}\bigl (w_0(\vec{c}))=-2 e^{\mp \frac {\hat \eta (w_0,w_0)}{2}}e^{\mp \frac {i\pi}{4}} \label {Dpmform} \, .
\ee

To make contact with notations of this paper, we observe that the wave functions $\psi _\pm$ of \cite {FRletter} appear in the 
two components of $\hat \Xi$ as
\be
\hat \Xi (w; \theta ,\vec{c})= -
\frac{1}{2}
\left ( \begin{array}{c}
e^{\frac {\eta}{2}}\psi _+(w,\bar w) \\
e^{-\frac {\eta}{2}}\psi _-(w,\bar w)
\end{array} \right ) \, . \label{Xipsi}
\ee

Now, in order to obtain realisations for $\epsilon $ and $\eta$ in terms of actions of integral operators 
we distinguish the case $N$ semi-integer from the case $N$ integer.

\medskip

If $N$ is semi-integer, we have that $\Phi (\theta, \vec{c})=0$. Then, by using $N-\frac{1}{2}$ times the quasiperiodicity of $Q_{\pm}$ (\ref {qper}) we get
\be
 Q_{\pm}(\theta +i\pi N + i\pi,\vec{c})=e^{\pm \frac{i\pi}{2}\left(N-\frac{1}{2}\right)}Q_{\pm} \left ( \theta + \frac{i\pi}{2}+\frac{i\pi}{2N}, \vec{c}^{R^{N-\frac{1}{2}}} \right ) \, .
\ee
Using (\ref {Qlzero}), we arrive at the expression
\be
 Q_{\pm}(\theta +i\pi N + i\pi,\vec{c})=-\frac{\sqrt{2}}{D_{\pm}}e^{\pm \frac{i\pi}{2}\left(N-\frac{1}{2}\right)}\frac{e^{-\frac{\theta}{2}}}{\sqrt{T _{\frac{1}{2}}^{new}\left ( \theta + \frac{i\pi}{2}+\frac{i\pi}{2N}, \vec{c}^{R^{N-\frac{1}{2}}} \right ) }} \left ( \frac{1}{\hat I \pm \hat K _{0,N-\frac{1}{2}}} E_{0,N-\frac{1}{2}} \right ) (\theta) \, ,
\ee
where $\hat K_{0,N-\frac{1}{2}}$, $E_{0,N-\frac{1}{2}}$ stand for quantities (\ref{ushort}, \ref{Kshort}) evaluated when $w'=-iw_0( \vec{c}^{R^{N-\frac{1}{2}}} )$ and with the vector $\vec{c}$ in $T_{\frac{1}{2}}^{new}$ replaced by $\vec{c}^{R^{N-\frac{1}{2}}}$.
Since periodicity (\ref {qper}) implies
\be
T_{\frac{1}{2}}^{new}\left ( \theta +i\pi + i\pi N, \vec{c} \right )=
T_{\frac{1}{2}}^{new}\left ( \theta + \frac{i\pi}{2}+\frac{i\pi}{2N}, \vec{c}^{R^{N-\frac{1}{2}}} \right )  \, ,
\ee
and $D_+D_-=-4$, as a consequence of (\ref {Dpmform}), 
the final expressions for $y_N(\theta)=e^{-\epsilon (\theta)}$ and $\eta (\theta)$ simplify as follows
\ba
e^{-\epsilon (\theta)}&=&\frac{1}{2e^{\theta}}\left ( \frac{1}{\hat I + \hat K _{0,N-\frac{1}{2}}} E_{0,N-\frac{1}{2}} \right ) (\theta)\left ( \frac{1}{\hat I - \hat K _{0,N-\frac{1}{2}}} E_{0,N-\frac{1}{2}} \right ) (\theta)
\label{yN1} \, , \\
\eta (\theta)&=&\frac{1}{4ie^{\theta}}\frac {1}{\sqrt {T_{\frac{1}{2}}^{new}\left ( \theta + i\pi+\frac{i\pi}{2N}, \vec{c}^{R^{N-\frac{1}{2}}} \right )T_{\frac{1}{2}}^{new}\left ( \theta +\frac{i\pi}{2N}, \vec{c}^{R^{N-\frac{1}{2}}} \right )}} \cdot  \nonumber \\
&\cdot&  \left [  \left ( \frac{1}{\hat I + \hat K _{0,N-\frac{1}{2}}}E_{0,N-\frac{1}{2}} \right ) \left (\theta+\frac{i\pi}{2}\right )\left ( \frac{1}{\hat I - \hat K _{0,N-\frac{1}{2}}} E_{0,N-\frac{1}{2}} \right ) \left (\theta-\frac{i\pi}{2} \right ) - \right. \nonumber \\
&-& \left.   \left ( \frac{1}{\hat I - \hat K _{0,N-\frac{1}{2}}} E_{0,N-\frac{1}{2}} \right ) \left (\theta+\frac{i\pi}{2} \right )\left ( \frac{1}{\hat I + \hat K _{0,N-\frac{1}{2}}} E_{0,N-\frac{1}{2}} \right ) \left (\theta-\frac{i\pi}{2}\right ) \right ] \, .
\label{eta11}
\ea
The behaviour of $y_N(\theta)=e^{-\epsilon (\theta)}$ and $\eta (\theta)$ at large $\theta$ is
\be
y_N(\theta ) \simeq e^{-2w_0( \vec{c}^{R^{N-\frac{1}{2}}} )e^{\theta}-2\bar w_0( \vec{c}^{R^{N-\frac{1}{2}}} )e^{-\theta}} T_{\frac{1}{2}}^{new} \left ( \theta +i\pi + i\pi N, \vec{c} \right ) \, , \quad \eta (\theta) \simeq 0 \, .
\ee
Then $g(\theta)=0$ and the function $2u(\theta )$ equals
\ba
2u(\theta)&=&2w_0( \vec{c}^{R^{N-\frac{1}{2}}} )e^{\theta}+2\bar w_0( \vec{c}^{R^{N-\frac{1}{2}}} )e^{-\theta}-\ln T_{\frac{1}{2}}^{new}\left ( \theta +i\pi + i\pi N, \vec{c}) \right ) = \, , \nonumber \\
&=& 2w_0( \vec{c}^{R^{N-\frac{1}{2}}} )e^{\theta}+2\bar w_0( \vec{c}^{R^{N-\frac{1}{2}}} )e^{-\theta}-\ln T _{\frac{1}{2}}^{new}\left ( \theta +i\pi \frac{N+1}{2N}, \vec{c}^{R^{N-\frac{1}{2}}} \right ) \, ,
\label{2u1}
\ea
where the first two terms in the right hand side are the sought zero modes of the shift operator of $\pm i\pi/2$.
Eventually, $\hat K _{0,N-\frac{1}{2}}$ and $E_{0,N-\frac{1}{2}}$ are given by definitions
(\ref{Kshort}) and the second of  (\ref{ushort}), with $v$ replaced by $u$ (\ref {2u1}). Then, in agreement with notations used in (\ref{ushort},\ref {Kshort}), they can be called $\hat K_u$ and $ E_u$, respectively. In conclusion, for $N$ semi-integer we have found the representations of the TBA solutions  in terms of an integral kernel by using the classical/quantum integrable theories correspondence in the inverse direction \cite {FRletter}. 

This proves the two Zamolodchikov's conjectures \cite{Zam-poly} in that the realisations (\ref {yN1}, \ref {eta11}) coincide with those given by the first of (2.4) and (3.14) of Tracy and Widom \cite{TW2}, respectively (with the function $u$ given by (\ref {2u1})). 

\medskip

The case of $N$ integer is more delicate, since the function $\Phi $ is not zero. Now, by using $N$ times the quasiperiodicity of $Q_{\pm}$ (\ref{qper})  we get
\ba
 Q_{\pm}(\theta +i\pi N + i\pi,\vec{c})&=&e^{\pm \frac{i\pi N}{2}}e^{iN\Phi (\theta +i\pi +i\pi N, \vec{c})}Q_{\pm} \left ( \theta , \vec{c}^{R^{N}} \right ) \,  \\
 T_{\frac{1}{2}}^{new}(\theta +i\pi N + i\pi,\vec{c})&=&e^{-2iN\Phi (\theta +i\pi +i\pi N, \vec{c})} T_{\frac{1}{2}}^{new} \left ( \theta , \vec{c}^{R^{N}} \right )
\ea
Using (\ref {Qlzero}), we get
\be
Q_{\pm} \left ( \theta , \vec{c}^{R^{N}} \right )=-\frac{\sqrt{2}}{D_{\pm}}\frac{e^{-\frac{\theta}{2}+i\pi \frac {N+1}{4N} }}{\sqrt{T _{\frac{1}{2}}^{new}\left ( \theta, \vec{c}^{R^{N} }\right ) }} \left ( \frac{1}{\hat I \pm \hat K _{0,N}} E_{0,N} \right ) \left (\theta-i\pi \frac {N+1}{2N}\right )  \, , \label {QNlzero}
\ee
where $\hat K _{0,N}$, $E_{0,N}$ stand for quantities (\ref{ushort}, \ref{Kshort}) evaluated when $w'=-iw_0( \vec{c}^{R^{N}} )$ and with the vector $\vec{c}$ in $T_{\frac{1}{2}}^{new}$ replaced by $\vec{c}^{R^{N}}$.

Restricting for the moment to the function $y_N(\theta)=e^{-\epsilon (\theta)}$ we find
\be
y_N(\theta)=e^{-\epsilon (\theta)}=\frac{1}{2e^{\theta -\frac{i\pi}{2}-\frac{i\pi}{2N}}}
\left ( \frac{1}{\hat I + \hat K _{0,N}}E_{0,N} \right ) \left (\theta -\frac{i\pi(N+1)}{2N}\right )\left ( \frac{1}{\hat I - \hat K _{0,N}} E_{0,N} \right ) \left (\theta -\frac{i\pi(N+1)}{2N} \right ) \, .
\ee
The asymptotic behaviour of $y_N(\theta)$ is
\be
y_N(\theta) \simeq e^{-2w_0( \vec{c}^{R^{N}} )e^{\theta -\frac{i\pi}{2}-\frac{i\pi}{2N} }-2\bar w_0( \vec{c}^{R^{N}} )e^{-\theta+\frac{i\pi}{2}+\frac{i\pi}{2N}}} T_{\frac{1}{2}}^{new} \left ( \theta, \vec{c}^{R^{N}}\right ) \, ,
\ee
which implies that the function $2u(\theta )$ equals
\be
2u(\theta)=-\ln T _{\frac{1}{2}}^{new}\left ( \theta, \vec{c}^{R^{N}}\right )+
2w_0( \vec{c}^{R^{N}} )e^{\theta -\frac{i\pi}{2}-\frac{i\pi}{2N} }+2\bar w_0( \vec{c}^{R^{N}} )e^{-\theta+\frac{i\pi}{2}+\frac{i\pi}{2N}}  \label{2u2} \, .
\ee
%&=& -\ln \left [ e^{-2iN \Phi (\theta +i\pi N +2i\pi, \vec{c})}T (\theta +i\pi N+i\pi, \vec{c}) \right ]+
%2w_0( \vec{c}^{R^{N}} )e^{\theta -\frac{i\pi}{2}-\frac{i\pi}{2N} }+2\bar w_0( \vec{c}^{R^{N}} )e^{-\theta+\frac{i\pi}{2}+\frac{i\pi}{2N}} \nonumber \, .
%\ea
Now we use the property
\be
\left ( \frac{1}{\hat I \pm \hat K _{0,N}}E_{0,N} \right ) \left (\theta -\frac{i\pi(N+1)}{2N}\right )=
e^{-i\pi \frac{N+1}{4N}}\left ( \frac{1}{\hat I \pm \hat K _u} E_u \right ) (\theta) \, , \label {useprop}
\ee
where the operator $\hat K_u$ and the function $E_u$ are given by definitions (\ref {Kshort}) and second of (\ref {ushort}) with the function $v$ given by $u$ (\ref {2u2}).  Property (\ref {useprop}) allows to write (\ref {QNlzero}) as
\be
Q_{\pm} \left ( \theta , \vec{c}^{R^{N}} \right )=-\frac{\sqrt{2}}{D_{\pm}}\frac{e^{-\frac{\theta}{2} }}{\sqrt{T _{\frac{1}{2}}^{new}\left ( \theta, \vec{c}^{R^{N} }\right ) }} \left ( \frac{1}{\hat I \pm \hat K _{u}} E_{u} \right ) (\theta)  \label {QNlzero2}
\ee
and then brings $y_N (\theta )=e^{-\epsilon (\theta)}$ in the final form
\be
e^{-\epsilon (\theta)}=\frac{1}{2e^{\theta}}\left ( \frac{1}{\hat I + \hat K _{u}} E_u \right ) (\theta)
\left ( \frac{1}{\hat I - \hat K _{u}} E_{u} \right ) (\theta) \, . \label {yN2}
\ee
For what concerns $y(\theta)=\eta (\theta)$ we need to use formula (\ref {QNlzero2}) with shifts $\pm i\pi/2$ in $\theta$. Then, we obtain
\ba
\eta (\theta)&=&\frac{1}{4ie^{\theta}}\frac {1}{\sqrt {T_{\frac{1}{2}}^{new}\left ( \theta +\frac{i\pi}{2}, \vec{c}^{R^{N}} \right )T_{\frac{1}{2}}^{new}\left ( \theta -\frac{i\pi}{2}, \vec{c}^{R^{N}} \right )}} \cdot  \nonumber \\
&\cdot&  \Bigl [  \left ( \frac{1}{\hat I + \hat K _{u}} E_{u} \right ) \left (\theta+\frac{i\pi}{2}\right )\left ( \frac{1}{\hat I - \hat K _{u}} E_{u} \right ) \left (\theta-\frac{i\pi}{2} \right ) - \nonumber \\
&-&   \left ( \frac{1}{\hat I - \hat K _{u}} E_{u} \right ) \left (\theta+\frac{i\pi}{2}\right )\left ( \frac{1}{\hat I + \hat K _{u}} E_{u} \right ) \left (\theta-\frac{i\pi}{2} \right ) \Bigr ] \, , \label{eta12}
\ea
from which one also gets that at large $\theta $ the function $\eta (\theta)$ approaches zero: then $g(\theta )=0$. Also for $N$ semi-integer we have found the representations of the TBA solutions  in terms of the same integral kernel by the correspondence in the inverse direction \cite {FRletter}. 

Eventually, this proves the two Zamolodchikov's conjectures \cite{Zam-poly} as the realisations (\ref {yN2}, \ref {eta12}) coincide with the those given by the first of (2.4) and (3.14) of Tracy and Widom \cite{TW2}, respectively (with the function $u$ given by (\ref {2u2})).

\section {Summary and conclusions}
\setcounter{equation}{0}
\label{concl}

We have found, analysed and sometimes solved the functional and integral equations for connexion coefficients ($Q$-functions), Voros-Stokes coefficients ($T$-functions) and $Y$-functions, which describe the monodromy space of the Lax pair (\ref {ass-lin-prob}) computed on a specific solution of the modified sinh-Gordon equation (\ref {cl-sinh}), depending on a parameter $l$, to which the Lax linear problems are associated. Besides, they are eigenvalues of proper operator extensions of the renowned Baxter's matrices. The modification is encoded in a degree $2N$ polynomial, depending on $2N-1$ complex coefficients, the moduli. This problem generalises with the parameter $l$ the minimal area problem for strings ending in $AdS_3$ which are dual to null polygonal Wilson loops at strong coupling, which is obtained when $l=0$. Moreover, it can be seen as one side of the off-critical ODE/IM correspondence, as those coefficients coincide with vacuum eigenvalues of $Q$ and $T$ operators of the quantum Homogeneous sine-Gordon field theory. 

Importantly, the Lax linear problems have two symmetries, the $\Omega$-symmetry (\ref  {Omega-symm}) and the $\Pi$-symmetry (\ref {Pi-symm}), which can be used to construct different solutions of the Lax problem. If one uses the $\Omega$-symmetry, the monodromy data, the $Q$-, $T$- and $Y$- functions, satisfy functional relations relating functions with different, rotated, moduli. In this sense, the $T$- and $Y$- systems (\ref {fusion2}) and (\ref {Y-sys}) can be seen as generalisations of the usual $T$ and $Y$-systems. The number of independent rotations is $2N$ and then the $T$- and $Y$-systems exhibit a $Z_{2N}$-symmetry.

Usual functional relations are instead obtained for connexion coefficient, Voros-Stokes coefficients and $Y$ functions which relate solutions of the Lax linear problem obtained by using the $\Pi$-symmetry. We called these new functions $T^{new}$ and $Y^{new}$: they satisfy the $T$-system (\ref {closed-T-system}) and the $Y$-system (\ref {closed-Y-system}), which involve functions depending on the same set of moduli and, differently, from  (\ref {fusion2}) and (\ref {Y-sys}), contain 'universal' shifts, i.e. shifts not depending on the modifying polynomial.  

However, when converting $Y$-systems into TBA equations, the 'generalised' $Y$-system (\ref {Y-sys}) 
was found to be more convenient, in the region of the space of the moduli of our interest, which is $c_n \ll 1, n\geq 1$. The final TBA equations are (\ref {TBA-final-general}) and involve a minimal number of $Y$-functions. On the contrary, when trying to convert for $c_n \ll 1$ the more natural $Y$-system (\ref {closed-Y-system}) into TBA equations, one has to face the problem of 'wall-crossing', which increases the number of $Y$-functions. 

Important application was the so-called conformal limit: in this limit the Lax pair reduces to a Schroedinger equation (\ref {ODE-cft}) with a polynomial potential and in this process of limit an interesting phenomenon happens: the coefficients of the potential are not the conformal limit of the (off-critical) moduli. However, it is natural to express the $T$ and $Y$ functions in terms of the coefficients of the polynomial potential. In doing this, the conformal limit of the $Y$-system (\ref {Y-sys}) takes the form (\ref  {Y-syst-conf}), different from  (\ref {Y-sys}), and with additional $Z_{4N+4}$ symmetry (\ref  {inv-Y-system-cft}). We wrote TBA equations (\ref {TBA-final-general-chi-cft}), in the simple case of a polynomial with 
three terms, depending on the arbitrary coefficient of the term $x^ {N-1/2}$, for which the symmetry of the $Y$-system/TBA enhances to $Z_ {2N+2}$. 

An interesting application of the $Y$-systems satisfied by 'new' $Y$-function was finally discussed in Section \ref {lzero}. It concerned 
the realisation of the solutions of the TBA-like equations appearing in the computation of Witten index in integrable supersymmetric ${\cal N}=2$ theories and which Al. Zamolodchikov used to describe polymers. This realisation holds in the limit $l\rightarrow 0$. 

\medskip

The main motivation for this work was to highlight all integrable structures lying below and permitting the computation of a rather general connexion and monodromy linear differential problem in Lax (Hitchin) form. In particular, $l=0$ gives the exact computation of $AdS_3$ Wilson loops in ${\cal N}=4 $ SYM at strong coupling. This can be done by using TBA equations (\ref {TBA-final-general}), which are alternative to TBA obtained from 'universal' $Y$-system \cite {GMN,YSA,HISS} and also to the set of NLIEs found in \cite {FRS}. 
In a broader perspective, this work, concerning infinite coupling, lays the foundation for the the long standing problem of a deformation/second quantisation in an instanton counting perspective (of ODE/IM) by a comparison with the Young diagrams representation of Wilson loops at all couplings in \cite{FPP}. In this context, a final perspective is the extension to functional relations describing minimal area problem for strings ending on $AdS_5$. This would lead to NLIEs, which would be an alternative method, with respect for instance to what discussed in \cite {a}, for computing scattering amplitudes/Wilson loops.

\vspace{1truecm}

%This research was supported in part by the National Science Foundation under Grant No. NSF PHY-1748958.
{\bf Acknowledgements}
We thank R. Tateo for important comments and A. Cavagli\`{a}, R. Conti, D. Gregori, S. Lukyanov, H. Shu, A. Zamolodchikov for discussions. This research was supported in part by the grants GAST (INFN), the MPNS-COST Action MP1210, the EC Network Gatis, the MIUR-PRIN contract 2017CC72MK\textunderscore 003 and the National Science Foundation under Grant No. NSF PHY-1748958.

\appendix 

\section {The $T$-system as compatibility condition}
\setcounter{equation}{0}
\label{app1}

We now explain how the validity of both relations (\ref {TjQ}) and (\ref {TjQbar}) implies the $T$-system relations (\ref {fusion2}).

We start by multiplying (\ref {TjQbar}) by $T_{j-\frac {1}{2}}\left (\theta -\frac {i\pi}{2N}, \vec {c}^{R^{-2j}}\right )$:
\ba
&&T_{j-\frac {1}{2}}\left (\theta -\frac {i\pi}{2N}, \vec {c}^{R^{-2j}}\right ) T_j(\theta , {\vec{c}}^{R^{-2j-1}}) Q_{\pm }\left (\theta - \frac {i\pi}{2N}(2j-1), \vec{c}^{R^{-1}} \right ) - \nonumber \\
&&-T_{j-\frac {1}{2}}\left (\theta -\frac {i\pi}{2N}, \vec {c}^{R^{-2j}}\right )T_{j-\frac{1}{2}}\left (\theta +\frac {i\pi}{2N}, \vec{c}^{R^{-2j-1}} \right )
Q_{\pm }\left (\theta - \frac {i\pi}{2N}(2j+1), \vec{c} \right )= \nonumber \\
&& =e^{-i \Phi \left (\theta -\frac{i\pi}{2N}(2j+1), \vec{c}\right )}  
T_{j-\frac {1}{2}}\left (\theta -\frac {i\pi}{2N}, \vec {c}^{R^{-2j}}\right )
Q_{\pm }\left (\theta + \frac {i\pi}{2N}(2j+1), \vec{c}^{R^{-2j-1}} \right ) \, . 
\label  {TjQbar2} 
\ea
On the other hand, starting again from (\ref {TjQbar}) we shift in it $j\rightarrow j-\frac {1}{2}$ and $\theta \rightarrow \theta -\frac {i\pi}{2N}$ and multiplies the resulting relation by $T_j(\theta , c^{R^{-2j-1}})$:
\ba
&& T_{j-\frac {1}{2}}\left (\theta -\frac {i\pi}{2N}, \vec {c}^{R^{-2j}}\right ) T_j(\theta , {\vec{c}}^{R^{-2j-1}}) Q_{\pm }\left (\theta - \frac {i\pi}{2N}(2j-1), \vec{c}^{R^{-1}} \right )- \nonumber \\
&& -T_{j}\left (\theta , \vec{c}^{R^{-2j-1}} \right )
 T_{j-1}\left (\theta , \vec{c}^{R^{-2j}} \right )
Q_{\pm }\left (\theta - \frac {i\pi}{2N}(2j+1), \vec{c} \right )= \nonumber \\
&& =e^{-i \Phi \left (\theta -\frac{i\pi}{2N}(2j+1), \vec{c}\right )}   T_{j}\left (\theta , \vec{c}^{R^{-2j-1}} \right )
Q_{\pm }\left (\theta + \frac {i\pi}{2N}(2j-1), \vec{c}^{R^{-2j}} \right ) \, . 
\label  {TjQbar3} 
\ea
Subtracting these two expressions we find
\ba
&&-T_{j-\frac {1}{2}}\left (\theta -\frac {i\pi}{2N}, \vec {c}^{R^{-2j}}\right )T_{j-\frac{1}{2}}\left (\theta +\frac {i\pi}{2N}, \vec{c}^{R^{-2j-1}} \right )
Q_{\pm }\left (\theta - \frac {i\pi}{2N}(2j+1), \vec{c} \right ) + \nonumber \\
&&+ T_{j}\left (\theta , \vec{c}^{R^{-2j-1}} \right )
 T_{j-1}\left (\theta , \vec{c}^{R^{-2j}} \right )
Q_{\pm }\left (\theta - \frac {i\pi}{2N}(2j+1), \vec{c} \right )= \nonumber \\
&& e^{-i \Phi \left (\theta -\frac{i\pi}{2N}(2j+1), \vec{c}\right )}   \Bigl [ 
T_{j-\frac {1}{2}}\left (\theta -\frac {i\pi}{2N}, \vec {c}^{R^{-2j}}\right )
Q_{\pm }\left (\theta + \frac {i\pi}{2N}(2j+1), \vec{c}^{R^{-2j-1}} \right ) - \nonumber \\
&&- T_{j}\left (\theta , \vec{c}^{R^{-2j-1}} \right )
Q_{\pm }\left (\theta + \frac {i\pi}{2N}(2j-1), \vec{c}^{R^{-2j}} \right ) \Bigr ] \, .
\ea
We now use (\ref {TjQ}) to simplify the right hand side of this expression: we get
\ba
&&-T_{j-\frac {1}{2}}\left (\theta -\frac {i\pi}{2N}, \vec {c}^{R^{-2j}}\right )T_{j-\frac{1}{2}}\left (\theta +\frac {i\pi}{2N}, \vec{c}^{R^{-2j-1}} \right )
Q_{\pm }\left (\theta - \frac {i\pi}{2N}(2j+1), \vec{c} \right ) + \nonumber \\
&&+ T_{j}\left (\theta , \vec{c}^{R^{-2j-1}} \right )
 T_{j-1}\left (\theta , \vec{c}^{R^{-2j}} \right )
Q_{\pm }\left (\theta - \frac {i\pi}{2N}(2j+1), \vec{c} \right )= \nonumber \\
&&=-e^{i[1+(-1)^{2j}] \Phi \left (\theta +\frac{i\pi}{2N}(2j+1), \vec{c}^{R^{-2j-1}}\right )}  
Q_{\pm }\left (\theta - \frac {i\pi}{2N}(2j+1), \vec{c} \right )
\ea
after using $2j+1$ times the property $\Phi (\theta , \vec{c})=-\Phi \left ( \theta -\frac{i\pi}{N}, \vec{c}^{R}\right )$, first of (\ref {phi-prop}). Removing the $Q_\pm$ functions and redefining $j \rightarrow j+\frac {1}{2}$ and $\vec {c} \rightarrow \vec {c}^{R^{2j+2}}$, we find the $T$-system relations (\ref {fusion2}). 

%\section {The $TQ$-system and Wronkians}
%\setcounter{equation}{0}
%\label{app2}

%We speculate on how the $TQ$-system constrains quadratic constructs of the $Q$-functions. Let us start from a generic $TQ$-relation
%\be 
%T(\alpha, \theta )Q_\pm (\alpha, \theta) = Q_\pm (\alpha , \theta +\alpha ) + Q_\pm (\alpha, \theta -\alpha )
%\ee 
%We multiply the relation for $Q_+$ by $Q_-(\alpha, \theta )$ and the relation for $Q_-$ by $Q_+(\alpha , \theta )$. Then we subtract the two equalities and obtain
%\be 
%Q_+(\alpha, \theta +\alpha )Q_-(\alpha, \theta )- Q_+(\alpha, \theta )Q_-(\alpha, \theta +\alpha)=
%Q_+(\alpha, \theta )Q_-(\alpha, \theta -\alpha)- Q_+(\alpha, \theta -\alpha)Q_-(\alpha, \theta )
%\ee
%which means periodicity for shifting $\theta $ by $\alpha$ (i.e. $F(\alpha, \theta )=F(\alpha, \theta -\alpha )$) for the quantity 
%\be 
%F(\alpha, \theta )=Q_+(\alpha, \theta +\alpha )Q_-(\alpha, \theta )- Q_+(\alpha, \theta )Q_-(\alpha, \theta +\alpha)
%\ee
%In the limit $\alpha \rightarrow 0$ the periodicity $F(\alpha, \theta )=F(\alpha, \theta -\alpha )$, with the extra condition $F(0,\theta)=0$, implies
%\be 
%\frac {\partial} {\partial \theta } \left . \left [ \frac {\partial} {\partial \alpha } F (\alpha,\theta ) \right ]\right |_{\alpha =0} =0 \, , 
%\ee
%where
%\be
% \left .\left [ \frac {\partial} {\partial \alpha } F (\alpha,\theta ) \right ] \right |_{\alpha =0}=\left [ \frac {\partial }{\partial \theta } Q_+(0,\theta ) \right ] 
%Q_-(0,\theta) - Q_+(0,\theta )\left [ \frac {\partial }{\partial \theta}Q_-(0,\theta ) \right ] \, .
%\ee

\section {$Y$-functions as cross-ratios}
\setcounter{equation}{0}
\label{app3}

We can now elaborate the expressions of $Y$ functions (\ref {Tnew-Ynew}) to express them in a different form.
The first step consists in using (\ref {Twronskbis}, \ref {Twronskter}) to find an expression for the $Y$-functions in terms of determinants of solutions. Using the fact that $\det (\hat \Pi ^{n}\Xi , \hat \Pi ^{n+1}\Xi )=1$,
we have 
\ba
Y_n^{new}(\theta , \vec {c})&=&  -\frac {\det (\hat \Pi ^{-n} \Xi (\theta) , \hat \Pi ^n \Xi (\theta)) \det (\hat \Pi ^{-n-1}\Xi (\theta), \hat \Pi ^{n+1}\Xi (\theta))}
{\det (\hat \Pi ^{-1-n} \Xi , \hat \Pi ^{-n} \Xi) \det (\hat \Pi ^{n}\Xi , \hat \Pi ^{n+1}\Xi )} \nonumber \\
Y_{n+\frac {1}{2}}^{new}\left (\theta +\frac {i\pi}{2}, \vec {c}\right)&=&\frac {\det (\hat \Pi ^{-n-1} \Xi (\theta), \hat \Pi ^n \Xi (\theta)) \det (\hat \Pi ^{-n-2}\Xi (\theta), \hat \Pi ^{n+1}\Xi (\theta) )}{\det (\hat \Pi ^{-2-n} \Xi (\theta), \hat \Pi ^{-1-n} \Xi (\theta)) \det (\hat \Pi ^{n}\Xi (\theta) , \hat \Pi ^{n+1}\Xi (\theta) )}
\nonumber
\ea
with $n$ non negative integer. Since the various $\hat \Pi ^n \Xi (\theta)$, with $n$ integer are all solutions of the same linear problem,  each of the determinants appearing in these formul{\ae} are independent on the point in which they are computed. We choose to compute all them in the same point $z_0$. Then, let
us call $f(\theta)$ ($g(\theta)$) the upper (lower) component of $\Xi (\theta)$ computed in $z_0$. Then, the upper (lower) component of 
$\hat \Pi ^n \Xi (\theta)$ computed in $z_0$ are $f(\theta -i\pi n)$ ($(-1)^n g(\theta -i\pi n)$). After simple manipulations, we find 
\be
Y_n^{new}(\theta , \vec {c})=-\frac {\left [ \frac {f(\theta -i\pi n)}{g(\theta -i\pi n)}-\frac {f(\theta +i\pi n)}{g(\theta +i\pi n)}\right ] \left [ \frac {f(\theta +i\pi (n+1))}{g(\theta +i\pi (n+1))}-\frac {f(\theta -i\pi (n+1))}{g(\theta -i\pi (n+1))}\right ]}{\left [ \frac {f(\theta +i\pi (n+1))}{g(\theta +i\pi (n+1))}+\frac {f(\theta +i\pi n)}{g(\theta +i\pi n)}\right ]\left [ \frac {f(\theta -i\pi n)}{g(\theta -i\pi n)}+\frac {f(\theta -i\pi (n+1))}{g(\theta -i\pi (n+1))}\right ]}
\ee 
\be
Y_{n+\frac {1}{2}}^{new}\left (\theta +\frac {i\pi}{2}, \vec {c}\right)=\frac {\left [ \frac {f(\theta +i\pi (n+1))}{g(\theta +i\pi (n+1))}+\frac {f(\theta -i\pi n)}{g(\theta -i\pi n)}\right ] \left [ \frac {f(\theta +i\pi (n+2))}{g(\theta +i\pi (n+2))}+\frac {f(\theta -i\pi (n+1))}{g(\theta -i\pi (n+1))}\right ]}{\left [ \frac {f(\theta +i\pi (n+2))}{g(\theta +i\pi (n+2))}+\frac {f(\theta +i\pi (n+1))}{g(\theta +i\pi (n+1))}\right ]\left [ \frac {f(\theta -i\pi n)}{g(\theta -i\pi n)}+\frac {f(\theta -i\pi (n+1))}{g(\theta -i\pi (n+1))}\right ]}
\ee 
We introduce the quantity 
\be
k(\theta)=e^{-\theta}\frac {f(\theta)}{g(\theta)}
\ee
and find for general $j$ the formula
\be
Y_j^{new}(\theta)=-\frac {[k(\theta -i\pi j)-k(\theta +i\pi j)][k(\theta +i\pi j+i\pi)-k(\theta -i\pi j -i\pi)]}{[k(\theta -i\pi j)-k(\theta -i\pi j-i\pi)][k(\theta +i\pi j+i\pi)-k(\theta +i\pi j )]}
\ee
This expression has the form of a cross ratio
\be
Y_j^{new}(\theta)=- \left ( k(\theta -i\pi j), k(\theta +i\pi +i\pi j); k(\theta +i\pi j), k(\theta -i\pi -i\pi j) \right ) 
\ee
where 
\be 
(a,b;c,d)=\frac {(a-c)(b-d)}{(a-d)(b-c)}
\ee
The choice $z_0=0$ allows to express, by virtue of (\ref {QXi}), the function $k$ in terms of $Q_\pm$: we have 
\be 
\frac {f(\theta )}{g(\theta )}=e^{-2\theta  l}e^{-2i\varphi l } \frac {Q_+(\theta )}{Q_-(\theta )} \, , 
\ee
which implies  
\be
k(\theta)=e^{-2\left (l+\frac {1}{2}\right )\theta}e^{-2i\varphi l } \frac {Q_+(\theta)}{Q_-(\theta)} \, .
\ee
Then, applying $n$ times the symmetry $\hat \Pi $ to equation (\ref {maineq}), we find
\be
\hat \Pi ^n \Xi (z ;\theta )= Q_+(\theta -i\pi n) e^{i\pi nl} \Psi _ - (z; \theta ) +e^{i\pi n}Q_-(\theta -i\pi n) e^{-i\pi nl } \Psi _+(z; \theta)
\, . 
\ee
Since $\hat \Pi ^n \Xi (z ; \theta ) \rightarrow 0$ if $z \rightarrow \infty$ while being in the Stokes sector $S_n$, we can write
\be
k(\theta -i\pi n)=-e^{-2\left (l+\frac {1}{2}\right )\theta}e^{-2i\varphi l }\lim _{\stackrel {z\rightarrow +\infty}{z\in S_n}} \frac {\Psi _+^\alpha(z; \theta)}{\Psi _-^\alpha(z; \theta)}\equiv -e^{-2\left (l+\frac {1}{2}\right )\theta}e^{-2i\varphi l } \omega _n \left ( \frac {\Psi _+^\alpha(z; \theta)}{\Psi _-^\alpha(z; \theta)} \right ) \, ,
\ee
with $n$ integer and $\alpha $ denoting the components of the two-dimensional vectors $\Psi _\pm$.

When computing $Y_n^{new}(\theta)$ the factors $-e^{-2\left (l+\frac {1}{2}\right )\theta}e^{-2i\varphi l }$ are inessential. Then we get 
\be
Y_n^{new}(\theta)=-\left ( \omega _n \left ( \frac {\Psi _+^\alpha (\theta)}{\Psi _-^\alpha(\theta)} \right ) , \omega _{-n-1}\left ( \frac {\Psi _+^\alpha(\theta)}{\Psi _-^\alpha(\theta)} \right ); \omega _{-n}\left ( \frac {\Psi _+^\alpha(\theta)}{\Psi _-^\alpha(\theta)} \right ), \omega _{n+1}\left ( \frac {\Psi _+^\alpha(\theta)}{\Psi _-^\alpha(\theta)} \right ) \right )
\ee
Actually the argument $\left ( \frac {\Psi _+^\alpha(\theta)}{\Psi _-^\alpha(\theta)} \right )$ can be omitted, since the cross-ratio which gives $Y_j^{new}$ is invariant if one replaces $\Psi _\pm$ with other two independent solutions of the linear problem (\ref {ass-lin-prob}).
Since $Y_n^{new}(\theta)=Y_n\left (\theta -i\pi \frac {N+1}{N}(j+1) \right)$ similar expressions hold also for the $Y_j$. Functions $Y_j$ are Fock-Goncharov coordinates...

\subsection {Being more general}

Let us consider a generic solution $\Psi (z,\theta )$ of the linear problem.
As a basis we consider the functions $\Phi _\pm (z,\theta)$ that at a certain point $z_0$ have the simple form
\be
\Phi _+ (z_0, \theta )=\left ( \begin{array}{c} 1 \\  0 \end{array} \right) \, , \quad \Phi _-(z_0,\theta)= \left ( \begin{array}{c} 0 \\  1 \end{array} \right) \, .
\ee
Let us denote by $f(\theta ), g(\theta )$ the coefficients with respect to the basis $\Phi _\pm$:
\be
\Psi (z,\theta)=f(\theta ) \Phi _+ (z,\theta ) + g(\theta ) \Phi _- (z, \theta ) \, .
\ee
Applying $n$ times the symmetry $\hat \Pi $, we get
\be 
\hat \Pi ^ n \Psi (z, \theta )=f(\theta -i\pi n) \Phi _+ (z, \theta )+ (-1)^n g(\theta -i \pi n) \Phi _ - (z, \theta )
\label {PinPsi} 
\ee
since $\hat \Pi \Phi _\pm =\pm \Phi _\pm$. If we make a change of basis\footnote {The prime symbol $'$ in this part does not mean derivative.}
\ba
\Phi _+(z,\theta )&=&\alpha (\theta ) \Phi _+^\prime (z,\theta )+ \beta (\theta ) \Phi _-^\prime (z, \theta ) \\
\Phi _-(z,\theta )&=&\gamma (\theta ) \Phi _+^\prime (z,\theta )+ \delta (\theta ) \Phi _-^\prime (z, \theta ) 
\ea
we have 
\be 
\hat \Pi ^ n \Psi (z, \theta )=f^\prime (\theta -i\pi n) \Phi _+^\prime (z, \theta )+ (-1)^n g^\prime (\theta -i \pi n) \Phi _ -^\prime  (z, \theta )
\ee
where
\be 
f^\prime (\theta - i\pi n)=\alpha (\theta ) f(\theta -i \pi n) +(-1)^n \gamma (\theta ) g(\theta -i \pi n)
\label {fprime}
\ee
\be
g^\prime (\theta -i \pi n)=\delta (\theta ) g(\theta -i \pi n)+(-1)^n \beta (\theta ) f(\theta -i \pi n)
\label {gprime}
\ee
We define
\be
Y_n^{new}(\theta , \vec {c})=  -\frac {\det (\hat \Pi ^{-n} \Psi (\theta) , \hat \Pi ^n \Psi (\theta)) \det (\hat \Pi ^{-n-1}\Psi (\theta), \hat \Pi ^{n+1}\Psi (\theta))}
{\det (\hat \Pi ^{-1-n} \Psi , \hat \Pi ^{-n} \Psi) \det (\hat \Pi ^{n}\Psi , \hat \Pi ^{n+1}\Psi )} 
\ee
and
\be
Y_{n+\frac {1}{2}}^{new}\left (\theta +\frac {i\pi}{2}, \vec {c}\right)=\frac {\det (\hat \Pi ^{-n-1} \Psi (\theta), \hat \Pi ^n \Psi (\theta)) \det (\hat \Pi ^{-n-2}\Psi (\theta), \hat \Pi ^{n+1}\Psi (\theta) )}{\det (\hat \Pi ^{-2-n} \Psi (\theta), \hat \Pi ^{-1-n} \Psi (\theta)) \det (\hat \Pi ^{n}\Psi (\theta) , \hat \Pi ^{n+1}\Psi (\theta) )}
\ee
Using (\ref {PinPsi}) and the fact that $\det (\Phi _+ , \Phi _-)=1$, we find 
\be
Y_n^{new}(\theta , \vec {c})=-\frac {\left [ \frac {f(\theta -i\pi n)}{g(\theta -i\pi n)}-\frac {f(\theta +i\pi n)}{g(\theta +i\pi n)}\right ] \left [ \frac {f(\theta +i\pi (n+1))}{g(\theta +i\pi (n+1))}-\frac {f(\theta -i\pi (n+1))}{g(\theta -i\pi (n+1))}\right ]}{\left [ \frac {f(\theta +i\pi (n+1))}{g(\theta +i\pi (n+1))}+\frac {f(\theta +i\pi n)}{g(\theta +i\pi n)}\right ]\left [ \frac {f(\theta -i\pi n)}{g(\theta -i\pi n)}+\frac {f(\theta -i\pi (n+1))}{g(\theta -i\pi (n+1))}\right ]} \label {Yn}
\ee 
\be
Y_{n+\frac {1}{2}}^{new}\left (\theta +\frac {i\pi}{2}, \vec {c}\right)=\frac {\left [ \frac {f(\theta +i\pi (n+1))}{g(\theta +i\pi (n+1))}+\frac {f(\theta -i\pi n)}{g(\theta -i\pi n)}\right ] \left [ \frac {f(\theta +i\pi (n+2))}{g(\theta +i\pi (n+2))}+\frac {f(\theta -i\pi (n+1))}{g(\theta -i\pi (n+1))}\right ]}{\left [ \frac {f(\theta +i\pi (n+2))}{g(\theta +i\pi (n+2))}+\frac {f(\theta +i\pi (n+1))}{g(\theta +i\pi (n+1))}\right ]\left [ \frac {f(\theta -i\pi n)}{g(\theta -i\pi n)}+\frac {f(\theta -i\pi (n+1))}{g(\theta -i\pi (n+1))}\right ]} \label {Ynhalf}
\ee 
Two relevant properties of (\ref {Yn}, \ref {Ynhalf}) are the following.

The first one is that they are invariant by change of basis: in other words, the right  hand sides of (\ref {Yn}, \ref {Ynhalf}) do not change if we replace $f$ and $g$ with $f^\prime$ and $g^\prime$ related to them by (\ref {fprime}, \ref {gprime}).

The second one is that $Y_j^{new}$ (\ref {Yn}, \ref {Ynhalf}) solve the $Y$-system relation  
\be
Y_j^{new}\left ( \theta - \frac{i\pi }{2}, \vec{c}\right ) Y_j^{new}\left ( \theta + \frac{i\pi}{2} , \vec{c}\right )=
\left [ 1+ Y_{j-\frac{1}{2}}^{new}(\theta , \vec{c}) \right ]  \left [ 1+ Y_{j+\frac{1}{2}}^{new}(\theta , \vec{c}) \right ] \, ,
\ee
for generic $f,g$, {\it i.e.} without imposing any conditions on $f,g$.

\section {The appearance of two Painlev\'e equations}
\setcounter{equation}{0}
\label{lp}

It can be useful for applications to write the linear problem (\ref {linhat}) in 'polar' coordinates. By this term we mean parametrising $w$ and $\bar w$ as $w=\frac {t}{4}e^{i\phi}$, $\bar w=\frac {t}{4} e^{-i\phi}$, respectively, and noticing that the operator $w { \mathcal{D}}_{w} -\bar w  { \mathcal{D}}_{\bar w} $ contains only derivatives with respect to $\phi$, whilst $w{ \mathcal{D}}_{w} +\bar w  { \mathcal{D}}_{\bar w}$ contains only derivatives with respect to $t$.  We can then write the two equivalent equations
\be
[w { \mathcal{D}}_{w} -\bar w  { \mathcal{D}}_{\bar w} ]\hat \Xi (w) \equiv  { \mathcal{D}}_{\phi} \hat \Xi (t,\phi) = 0 \, \quad [w  { \mathcal{D}}_{w}+\bar w  { \mathcal{D}}_{\bar w}]\hat \Xi (w)
 \equiv  { \mathcal{D}}_{t} \hat \Xi (t,\phi) = 0 \, , \label {lprb}
\ee
where
\be
 { \mathcal{D}}_{\phi}= \left [ -i \frac{\partial}{\partial \phi}+ \frac{t}{2} \frac{\partial \hat \eta}{\partial t} \sigma ^3 +\frac {t}{4} (e^{-i\phi-\theta-\hat \eta}-e^{i\phi+\theta+\hat \eta})\sigma ^+ +\frac {t}{4} (e^{-i\phi-\theta+\hat \eta}-e^{i\phi+\theta-\hat \eta})\sigma ^- \right ]
\label{Dphi} \, ,
\ee
and
\be
  { \mathcal{D}}_{t}=\left [ t\frac{\partial}{\partial t}- \frac{i}{2} \frac{\partial \hat \eta}{\partial \phi} \sigma ^3 - \frac {t}{4} (e^{i\phi+\theta+\hat \eta}+e^{-i\phi-\theta-\hat \eta})\sigma ^+ -\frac {t}{4} (e^{i\phi+\theta-\hat \eta}+e^{-i\phi-\theta+\hat \eta})\sigma ^- \right ] \, .
 \label{Dti}
\ee
The utility of this change of variables is that in some interesting situations the field $\hat \eta$ depends only on the coordinate $t$. Then, the differential operators (\ref {Dphi}, \ref {Dti}), which in general are differential operators in $\phi$ and $t$ and contain $\theta $ as parameter, can be seen as differential operators in $t$ and in the new variable $\theta +i\phi$. In other words, we are in the case in which we can write differential equations also with respect to the energy $E=e^{2\theta}$ (bispectrality). 
Moreover, if $\eta$ does not depend on $\phi$, the sinh-Gordon equation satisfied by this field reduces to the Painlev\'{e} III$_3$ equation
\be
\frac{1}{t} \frac{d}{dt} \left ( t \frac{d}{dt} \hat \eta (t) \right ) = \frac{1}{2} \sinh 2\hat \eta (t) \, .
\label{pain3} 
\ee
This situation occurs for instance in the case $N=1/2$ and $l=0$. Here we give a brief proof of this fact. Since $p(z, \vec{c})=z+c_0$, with $c_0$ real, the modified sinh-Gordon equation reads
\be
\partial _z \partial _{\bar z} \eta -e^{2\eta}+(z+c_0)(\bar z+c_0)e^{-2\eta}=0 \, .
\label {cl-sinh-2}
\ee
with boundary condition $\eta \simeq $ constant as $z\rightarrow 0$ and $\eta \simeq \frac {1}{2}  \ln |z| $ as $z\rightarrow \infty$.
We can then argue that it is possible to find a solution $\eta$ of (\ref {cl-sinh-2}) depending only on the variable $|z+c_0|$ and not containing any parameters (since $l=0$ and $c_0$ is incorporated in the independent variable $|z+c_0|$).
Since $w=\frac {2}{3}(-z-c_0)^\frac {3}{2}$, the field $\eta $ depends only on $|w|$. Then, from the fact that $P\bar P=|z+c_0|^2=\left (\frac {3|w|}{2}\right )^{\frac {2}{3}}$, it eventually follows that also $\hat \eta =\eta -\frac {1}{4}\ln P\bar P $ depends only on $|w|=\frac {t}{4}$. We conclude that we can write $ { \mathcal{D}}_{\theta} \hat \Xi (t,\theta) =  { \mathcal{D}}_{t} \hat \Xi (t,\theta) =0$, with
\be
 { \mathcal{D}}_{\theta}= \left [ \frac{\partial}{\partial \theta}+ \frac{t}{2} \frac{\partial \hat \eta}{\partial t} \sigma ^3 +\frac {t}{4} (e^{-i\phi-\theta-\hat \eta}-e^{i\phi+\theta+\hat \eta})\sigma ^+ +\frac {t}{4} (e^{-i\phi-\theta+\hat \eta}-e^{i\phi+\theta-\hat \eta})\sigma ^- \right ]
\label{Dtheta} \, ,
\ee
and
\be
  { \mathcal{D}}_{t}=\left [ t\frac{\partial}{\partial t}- \frac {t}{4} (e^{i\phi+\theta+\hat \eta}+e^{-i\phi-\theta-\hat \eta})\sigma ^+ -\frac {t}{4} (e^{i\phi+\theta-\hat \eta}+e^{-i\phi-\theta+\hat \eta})\sigma ^- \right ] \, .
 \label{Dti2}
\ee
It is straightforward to verify that the compatibility condition $[ { \mathcal{D}}_{\theta},  { \mathcal{D}}_{t}]=0$ is equivalent to the Painlev\'{e} III$_3$ equation
(\ref {pain3}). 

If we now consider equation $ { \mathcal{D}}_{\theta}\hat \Xi (t,\theta) =0$ and write it for the two components 
\be
\hat \Xi = \left ( \begin{array}{c}
\hat \Xi _+ \\
\hat \Xi _-
\end{array} \right ) \, .
\ee
We have
\be 
\frac{d^2  \hat \Xi _{\pm}(\theta)}{d\theta ^2} -\coth (\theta +i\phi \pm \hat \eta )\left[\frac{d \hat \Xi _{\pm}(\theta) }{d \theta }  \pm \frac {t}{2}  \frac {d\hat \eta}{dt} \hat \Xi _{\pm}(\theta)\right ]- \frac {t^2}{4}\left ( \frac {d\hat \eta}{dt}\right )^2 \hat \Xi_{\pm}(\theta)-\frac {t^2}{8}[\cosh (2\theta +2i\phi) - \cosh 2\hat \eta ]\hat \Xi _{\pm}(\theta)=0 \, . \label{Ximode}
\ee
Considering such equations when $t=t_0=4w_0=\frac {8}{3}(-c_0)^{\frac {3}{2}}$ and $\phi =0$, we get by virtue of (\ref {hatXi-Q}) differential equations for the functions $Q_{\pm}\left (\theta + \frac{3i\pi}{2}\right)$. After denoting with $\hat \eta _0$ and $\hat \eta '_0$ the values of $\hat \eta (t)$ and $\frac{d\hat \eta}{dt}$ when $t=4w_0$, we have
\be
\frac{d^2  Q _{\pm}(\theta)}{d\theta ^2} +\tanh (\theta \pm \hat \eta_0)\left[-\frac{d Q _{\pm}(\theta) }{d \theta }  \mp \frac {t_0}{2} \hat \eta '_0 Q _{\pm}(\theta)\right ]- \frac {t_0^2}{4} (\hat \eta '_0 )^2 Q_{\pm}(\theta)+\frac {t_0^2}{8}[\cosh 2\theta +\cosh 2\hat \eta_0 ]Q _{\pm}(\theta)=0 \, . \label{Qpmode}
\ee
Equation (\ref {Qpmode}) is an alternative and convenient way to derive the $Q$-functions in this particular case.

\medskip

Another important circumstance in which a 'rotational symmetry' of $\hat \eta$ occurs \cite {LUK} is the case $p(z)=z^{2N}+c_0$, $c_0<0$, in the limit $N\rightarrow +\infty$, with $l$ and $c_0$ fixed. In this particular limit the apex (\ref {w0sg}) becomes $w_0=-\sqrt {-c_0}$, whilst the polynomial $p(z)=z^{2N}+c_0$ reduces to $c_0$ if $|z|<1$ and diverges as $z^{2N}$ if $|z|>1$.  It will be convenient to use the parametrisation $\sqrt {-c_0}=\frac {r}{4}$. From the form of the modified sinh-Gordon equation 
\be
\partial _z \partial _{\bar z} \eta -e^{2\eta}+|p(z)|^2 e^{-2\eta}=0 \, ,
\label {cl-sinh-3}
\ee
and from the boundary condition $\eta \simeq l\ln z\bar z $ as $z\rightarrow 0$, we can deduce that
in that regime $\eta $ depends on the variable $|z|$ and on the parameters $l$ (present in the boundary condition) and $c_0$ (appearing in (\ref {cl-sinh-3})). Focussing on the case $|z|<1$, we have $P(z=z')=-z^{2N}-c_0\simeq -c_0>0$: then 
\be 
\frac {dw}{dz}\simeq \sqrt {-c_0} \, , 
\ee
which means that $w-w_0 \simeq \sqrt {-c_0}z$; then, $\eta $ depends only on $|w-w_0|$.
Since $P$ is a constant, also $\hat \eta $ depends only on $|w-w_0|$ and on the parameters $l,\sqrt {-c_0}=\frac {r}{4}$. With reference to problem (\ref {lprb}, \ref {Dphi}, \ref {Dti})
it is then convenient to shift $w\rightarrow w-w_0$, to parametrise $w-w_0=\frac {t}{4}e^{i\phi}$, $\bar w-\bar w_0=\frac {t}{4} e^{-i\phi}$ respectively, and to consider the angular and the radial part of the 'shifted' differential operators: they read
\ba
&& [(w-w_0) { \mathcal{D}}_{w-w_0}  -(\bar w-\bar w_0)  { \mathcal{D}}_{\bar w-\bar w_0}] \hat \Xi (w-w_0) \equiv  { \mathcal{D}}_{\theta} \hat \Xi (t,\theta) = 0 \, \label  {lprbtheta} \\ 
&& [(w-w_0)  { \mathcal{D}}_{w-w_0}+(\bar w-\bar w_0)  { \mathcal{D}}_{\bar w-\bar w_0}] \hat \Xi (w-w_0)
 \equiv  { \mathcal{D}}_{t} \hat \Xi (t,\theta) = 0 \, , \label {lprbti}
\ea
with $ { \mathcal{D}}_{\theta}$ and $ { \mathcal{D}}_{t}$ defined as (\ref {Dtheta}, \ref {Dti2}), respectively, in terms of a field $\hat \eta (t=4|w-w_0|)$. It follows that for the two components of the vector $\hat \Xi$ equations (\ref {Ximode}) hold.
Concentrating on these equations, we perform the particular limit, $w-w_0 =\frac {t}{4}e^{i\phi}\rightarrow w_0=-\frac {r}{4}$, which extracts the 
$Q$-functions from $\hat \Xi (w-w_0)$
\be
 \lim \limits _{\substack {t\rightarrow r \\ \phi \rightarrow \pi}}  \hat \Xi (t, \theta )= 
\left ( \begin{array}{c}
e^{\frac{\theta}{2}}Q_+\left(\theta +\frac{i\pi}{2}\right ) \\
e^{-\frac{\theta}{2}}Q_-\left(\theta +\frac{i\pi}{2}\right )
\end{array} \right ) \, . \label{hatXi-Q2}
\ee
Let us now find the behaviour of $\hat \eta$ in this limit.
Using the fact that $\hat \eta $ depends on $|w-w_0|$ when $|z|<1$, we can equivalently study the regime $w-w_0\rightarrow -w_0$, which means the limit $t\rightarrow r$, with $\phi \rightarrow 0$ or for the variable $z$ the limit $z \rightarrow 1$.  Since in this limit $P$ has a zero, we can approximate $\hat \eta \simeq -\frac {1}{4} \ln P\bar P$. 
It is convenient to reach the point $t=r$ from the region $|z|>1$, when $P\simeq -z^{2N}$ and then
$ -\frac {1}{4} \ln P\bar P \simeq -\ln (|z|^N)$. Since the image of the point $z=1$ is $w_1=0$, we have $w \simeq i\frac {z^{N+1}}{N+1}$, which eventually gives (in $w$ variable) $ -\frac {1}{4} \ln P\bar P\simeq -\ln |w|$: this approximation is valid for $|z|>1$ and then also when $z\rightarrow 1^+$, which means $w\rightarrow 0$, {\it i.e.} the limit where, as written before, we have the approximation $\hat \eta \simeq -\frac {1}{4} \ln P\bar P$. Then we conclude that near $w=0$ we have $\hat \eta \simeq -\ln |w|$. Now, going to $w=0$ along the specific direction $t\rightarrow r$ and $\phi =0$, we find $\hat \eta \simeq -\ln |w-w_0+w_0|=-\ln |t-r|$. 
%=In terms of the variables $t$ and $r$, we have 
%$\hat \eta \simeq -\ln |w-w_0+w_0|=-\ln |t-r|$. 

The use of the Painlev\'{e} III$_3$ equation permits to have access to other terms of the expansion around 
the point $t=r$, $\phi =\pi$ of the function $\hat \eta(t)$: one finds
\be 
\hat \eta (t) = -\ln |t-r| +\ln 2- \frac {t-r}{2r}+ \frac {7-16u}{24 r^2}(t-r)^2 + O(t-r)^3 \label
{etalargealfa2}
\ee
with $u$ a constant depending on $c_0,l$. Now, using the approximation (\ref {etalargealfa2}) in equation (\ref {Ximode}) we find that in this limit case it reduces to the modified Mathieu equation
\be 
\frac{\partial ^2  Q _{\pm}(\theta)}{\partial \theta ^2} +\left (\frac {r^2}{8}\cosh 2\theta - u \right ) Q _{\pm}(\theta)=0\, , 
\ee
for the two $Q$-functions defined by (\ref {hatXi-Q2}).

\section {The equivalence of two TBA equations}
\setcounter{equation}{0}
\label{app4}

Let us show how to obtain equation (\ref {TBA-eq-nomoduli-12}) from specialising the general equations (\ref {TBA-eq-nomoduli})  to the case $N=3/2$.  We start from (\ref {TBA-eq-nomoduli}) for $j=1/2$. From (\ref {Ajc_0}) it follows that ${\cal A}_0(\theta, c_0)=0$ and  ${\cal A}_1(\theta, c_0)= {\cal A}_{1/2}(\theta, c_0)$; moreover, since $Y_0=0$, 
$\epsilon _0 (\theta ) \rightarrow -\infty$. Then,  
the equation (\ref {TBA-eq-nomoduli}) for $j=1/2$ reads
\be
\varepsilon _{\frac {1}{2}} (\theta)={\cal A}_{\frac {1}{2}}(\theta , c_0)+\int _{-\infty}^{+\infty} \frac {d\theta '}{2\pi}\frac {3/2}{\cosh \left [\frac {3}{2}(\theta -\theta ')\right ]}\left [ -{\cal A}_{\frac {1}{2}}\left ( \theta  ' , c_0\right )+ \varepsilon_{\frac {1}{2}}\left ( \theta  ' \right )+ \ln \left (1 +e^{-\varepsilon _{j\frac {1}{2}}\left ( \theta '  \right )} \right ) \right ] 
\ee
By 'solving' such equation in terms of $\varepsilon _{\frac {1}{2}} (\theta)$ one immediately finds (\ref {TBA-eq-nomoduli-12}). Remark that this operation changes the integral kernel of the TBA equation.

\medskip

In the general case we prove that (\ref {TBA-final}) can also be obtained by specialising the general equations (\ref 
{TBA-final-general-chi}) to the case $N=3/2$. Let us start from these last equations for $j=1/2$ and observe that, using
(\ref {Y-asy-general}),  
\be
{\cal A}_{1}\left ( \theta  ,  \vec{c} \right ) -  {\cal A}_{0}\left ( \theta  ,  \vec{c} \right )=
{\cal A}_{\frac {1}{2}}\left ( \theta  ,  \vec{c} \right ) \, , \quad {\cal A}_{\frac {1}{2}}\left ( \theta  ,  \vec{c}^R \right )=
{\cal A}_{1}\left ( \theta  ,  \vec{c} \right )
\ee 
Moreover, since $Y_0=0$, $\epsilon _0 (\theta ) \rightarrow -\infty$ and in addition for $N=3/2$, 
$\varepsilon_{1}\left ( \theta  ,  \vec{c} \right )= \varepsilon _{\frac {1}{2}}\left ( \theta  ,  \vec{c}^R \right )$.
Transferring these relations to quantities obtained after discrete Fourier sums, one re-elaborate
equation (\ref 
{TBA-final-general-chi})  for $j=1/2$ as
\be
\chi_{\frac {1}{2},l}(\theta , \vec{c})=\hat {\cal A}_{\frac {1}{2},l}(\theta , \vec{c})+\int _{-\infty}^{+\infty} \frac {d\theta '}{2\pi}\frac {3/2e^{-l (\theta -\theta ') }}{\cosh [3/2(\theta -\theta ')]} \left [ - e^{-{i\pi l}}\hat {\cal A}_{\frac {1}{2},l}(\theta ', \vec{c}) + e^{{i\pi l}}\chi_{\frac {1}{2},l}(\theta ', \vec{c})+e^{{i\pi l}} \tilde \Lambda _{\frac {1}{2},l}(\theta ', \vec{c} ) \right ] \, . 
\ee    
The final step is to 'solve' this equation for $\chi_{\frac {1}{2},l}(\theta , \vec{c})$. One finds then  (\ref {TBA-final}).

\end{document}